\documentclass{aa}  

\usepackage{natbib}
\bibpunct{(}{)}{;}{a}{}{,}
\usepackage{graphicx}
\usepackage{txfonts}
\usepackage{hyperref}
\usepackage{multicol}
\usepackage{subfig}
\usepackage{hyperref}
\usepackage{appendix}
\usepackage[utf8]{inputenc}

\DeclareUnicodeCharacter{2212}{-}
\begin{document} 

\title{Galaxy kinematics across different environments in the RXJ1347-1145 cluster complex
\thanks{Based  on  observations  with  the  European  Southern  Observatory Very Large Telescope (ESO-VLT), observing runs ID 386.A-0688(D) and 087.A-0361(D).}}

\author{J. M. P\'erez-Mart\'inez\inst{1}
           \and B. Ziegler\inst{1}
           \and A. B\"ohm\inst{1}
           \and M. Verdugo\inst{1}
          }
      
\institute{Department of Astrophysics, University of Vienna, T\"urkenschanzstr. 17, A-1180 Vienna, Austria.   \email{jm.perez@univie.ac.at}}

\date{}

\abstract
{}
{In order to understand the role of the different processes that drive galaxy evolution in clusters, we need comprehensive studies that simultaneously examine several of the most important physical properties of galaxies. In this work we  study the interplay between the kinematic state and star formation activity of galaxies in the RXJ1347-1145 cluster complex at z$\sim$0.45.}
{We used VLT/VIMOS to obtain slit spectra for 95 galaxies across the 40$'$x40$'$ area where the RXJ1347-1145 cluster complex resides. We determined the cluster membership of our targets by identifying one or more of the available emission lines within the wavelength range. Our spectroscopy is complemented with archival SUBARU/Suprime-Cam deep photometric observations in five optical bands (B, V, Rc, Ic, z'). We examined the kinematic properties of our sample attending to the degree of distortion of the extracted rotation curves. Regular rotating galaxies were included in our Tully-Fisher analysis while the distorted ones were used to study the role of cluster-specific interactions with respect to star formation and AGN activity.}  
{Our analysis confirmed the cluster membership for approximately half of our targets. We report a higher fraction of galaxies with irregular gas kinematics in the cluster environment than in the field. Cluster galaxies with regular rotation display a moderate brightening in the B-band Tully-Fisher relation compatible with the gradual evolution of the stellar populations with lookback time, and no significant evolution in the stellar-mass Tully-Fisher relation, in line with previous studies at similar redshift. Average specific star formation rate (sSFR) values are slightly lower in our cluster sample (-0.15 dex) with respect to the main sequence of star-forming galaxies, confirming the role of the environment in the early quenching of star formation in clusters. Finally, we carried out an exploratory observational study on the stellar-to-halo mass relation finding that cluster galaxies tend to have slightly lower stellar mass values for a fixed halo mass compared to their field counterparts.} 
{} 


\maketitle


\section{Introduction}
\label{S:Intro} 

The general population of galaxies in the local Universe can be divided into two distinct types: Star-forming galaxies are blue in color, and have disk-like morphologies and  relatively high star formation rates (SFRs), whereas quiescent galaxies are redder, and have more spheroidal shapes and  low levels of star formation. These properties are the product of a series of processes acting over galaxies throughout their lifetime. To give some examples, mass growth (\citealt{Cattaneo11}), morphological transformations (\citealt{Mortlock13}), quenching of star formation (\citealt{Peng10}), and redistribution of angular momentum (\citealt{Swinbank17}) are some of the most important changes that galaxies experience across cosmic time. The scientific community coined the term ``galaxy evolution'' to refer to these processes as a whole, and it has been extensively studied up to z$\sim$2 and beyond during the last decade. 

In recent times, the mass growth and the environment have been revealed as the two main drivers of galaxy evolution (\citealt{Baldry06}). However, both effects act in a similar way over the properties of the general population of galaxies, making it difficult to identify which one is dominant at different cosmic epochs. \cite{Dressler80} was the first to establish that denser environments present higher fractions of quiescent galaxies compared to the field in the local universe. Recently, some studies have linked the under-abundance of star-forming galaxies in clusters to an excess in the population of post-starburst galaxies in clusters at low to intermediate redshifts (\citealt{Socolovsky18}, \citealt{Paccagnella19}), pointing once again towards the influence of the environment in galaxy transformation and the quenching of star formation. However, the exact mechanism causing the stop of star formation is still a matter of debate, with recent studies proposing a two-phase process where a galaxy first slowly consumes most of its gas reservoir in the outskirts of the cluster before being fully quenched due to ram pressure stripping (RPS) in the innermost regions (\citealt{Petropoulou11, Petropoulou12}, \citealt{Wetzel13}, \citealt{Maier19}, \citealt{Ciocan19}). 

This description of galaxy evolution in clusters holds until z$\sim$1, when an increasing fraction of blue galaxies start to populate even the central regions of large-scale structures (\citealt{Butcher78}). At earlier epochs, the star-forming population becomes dominant, and during the cluster assembly even starbursts are common (\citealt{Santos13}, \citealt{Dannerbauer14}, \citealt{Popesso15}, \citealt{Casey17}). On the other hand, \cite{Darvish16} find that the quiescent fraction increases with stellar mass up to z$\sim$3, becoming almost independent of the environment at z>1, with galaxies showing similar SFR and specific SFR (sSFR) values in the field and (proto-)clusters. In addition, \cite{Paulino-Afonso18} reported lower SFR values in dense environments for galaxies below $\log{M_{*}}\leqslant$10.75 with respect to the field at z$\sim$1, while galaxies above that threshold do not show significant differences, which means that mass quenching is only dominant at very high stellar masses at this epoch. 

However, integrated properties such as galaxy luminosity, stellar mass, and star formation activity are not sufficient to understand galaxy evolution in clusters where interactions are frequent. Subtle cluster-specific processes such as starvation might be responsible for the early quenching of star formation for galaxies within massive clusters. However, to achieve a full transformation into passive ellipticals, we still need to comprehend the mechanisms that alter the kinematics of the cluster galaxies and rearrange their 3D  structure during their infalling phase. Processes like merging, harassment, and RPS are probably in play, but also  secondary processes such as the triggering of star formation due to tidal interactions or initial gas disk compression via ram-pressure (\citealt{Ruggiero17})  might intervene in the stronger and accelerated evolution of galaxies living in clusters even though it is still unclear which one predominates. 

The use of galaxy kinematics to study the evolution of star-forming galaxies in clusters has been traditionally linked to scaling relations such as the Tully-Fisher relation (TFR, \citealt{Tully77}), that can only be applied to regular rotating disks. While some authors claimed no difference between the cluster and field TFRs (\citealt{Ziegler03}, \citealt{Nakamura06}), others reported that spiral galaxies were slightly overluminous (\citealt{Bamford05}) and display a larger TFR scatter \citep{Moran07} in the cluster environment at z<1. However, galaxies with irregular kinematics cannot be excluded from a comprehensive environmental study since they represent the majority of the population in clusters (\citealt{Rubin99}, \citealt{Vogt04}). Following this idea, \cite{Bosch1} were able to link RPS events with asymmetries in the gas velocity profile of cluster galaxies that do not show significant distortions in their stellar structure. 

In this work we choose the multicluster system RXJ1347-1145 to investigate the environmental imprints of galaxy evolution for objects displaying regular and irregular gas kinematic behavior, focusing on their Tully-Fisher analysis, star formation, and active galactic nucleus (AGN) activity. The structure of this paper is as follows. In $\text{Sect.}$ 2 we state the target selection, observation conditions, and spectroscopic data reduction. In  $\text{Sect.}$ 3 we describe the methods used during our analysis. We present our results and discussion in $\text{Sect.}$ 4 and $\text{Sect.}$ 5, respectively, followed by our conclusions in $\text{Sect.}$ 6. Throughout this article we assume a \citet{Chabrier03} initial mass function (IMF), and adopt a flat cosmology with $\Omega_{\Lambda}$=0.7, $\Omega_{m}$=0.3, and $H_{0}$=70 km s$^{-1}$Mpc$^{-1}$. All magnitudes quoted in this paper are in the AB system.

\section{Sample selection and observations}
\label{S:Sample}

The galaxy cluster RXJ1347.5-1145 (hereafter RXJ1347) at z$\sim$0.45 is one of the most massive and most X-ray luminous clusters known (\citealt{Schindler95}). In recent years, RXJ1347 has been the subject of intense research, through  spectroscopic (\citealt{Lagana18}, \citealt{Jorgensen17}, \citealt{Fogarty17}), X-ray (\citealt{Ghirardini17}, \citealt{Ueda18}), lensing (\citealt{Chiu18}, \citealt{Umetsu18}), and Sunyaev–Zel’dovich effect (\citealt{Kitayama16}, \citealt{Adam18}) analyses. However, most previous works were focused on the determination of the cluster internal substructures. The presence of two very bright galaxies close to the center of the cluster and the discovery of shocked gas suggest that RXJ1347 is  undergoing a major merger. Furthermore, \cite{Verdugo12} identified a large-scale  cluster complex that extends diagonally across the field for about 20 Mpc and contains $\sim$30 additional group-like structures, including two additional prominent galaxy concentrations: one towards the southeast, coincident with the cluster LCDCS 0825 (\citealt{Gonzalez01}) and another towards the northeast which was named ‘the NE Clump’ by \cite{Verdugo12}.

In this work  we investigate the physical properties of galaxies that belong to this cluster complex, in particular concerning their internal kinematics. We carried out multi-object (MOS) spectroscopy with VIMOS/VLT between March 2011 and September 2012 to obtain spectra for 95 galaxies using two pointings around the RXJ1347 main cluster structure at z$\sim$0.45. Our primary targets were cluster galaxies selected from previous medium-resolution spectroscopic campaigns carried out by our group. 

We used the high-resolution grism, HR-orange, which covers the wavelength range $5200-7600\AA$ and a slit width of 0.8". This configuration yielded a spectral resolution of R$\sim$2500 and an average dispersion of 0.6 \AA/pix with an image scale of 0.205"/pixel. Prior to our observations, we used SExtractor (\citealt{Bertin96}) on CFHT images to compute the position angle of our targets with respect to the orientation of the masks. These values were used during the mask design to align the slits with our targets. However, this and other structural parameters of the galaxies were recomputed using the high-quality SUBARU Suprime-Cam z'-band images after the execution of our observation runs. The reason for this is that the accurate determination of the galaxies' structural parameters plays an important role in the kinematic analysis that will take place at a later stage. Therefore, the CFHT measurements become our slit tilt angles ($\theta$) while the SUBARU measurements are the physical position angles ($PA$) of our objects. 

The slit tilt angles were limited to $\left |{\theta}\right |$ < 45º to ensure a robust sky subtraction and wavelength calibration. However, a small number of our primary targets display PA that exceeded these constraints. In these cases we still analyze the objects by adding the misalignment angle ($\delta$) between $PA$ and $\theta$ as an additional parameter to our analysis (see Sect.\,\ref{SS:StructuralParameters}). The discrepancy between $\theta$ and $PA$ is rather small (<10º) for most galaxies, although a few cases show larger values due to the lower imaging quality of the CFHT observations. The total integration time slightly varies between observing runs, being 2.1h for targets observed during period P86 and 1.85h for targets in P87. Our observing program was conducted with average seeing conditions of 0.8" FWHM and airmass $\sim $1.1 during both observing runs. The spectroscopic data reduction was carried out using the ESO-REFLEX pipeline for VIMOS. The main reduction steps were bias subtraction, flat normalization, and wavelength calibration. We co-added the 2D spectra exposures using an IRAF sigma-clipping algorithm that performs  bad pixel and cosmic ray rejection.

\begin{figure}
\centering
\includegraphics[width=\columnwidth]{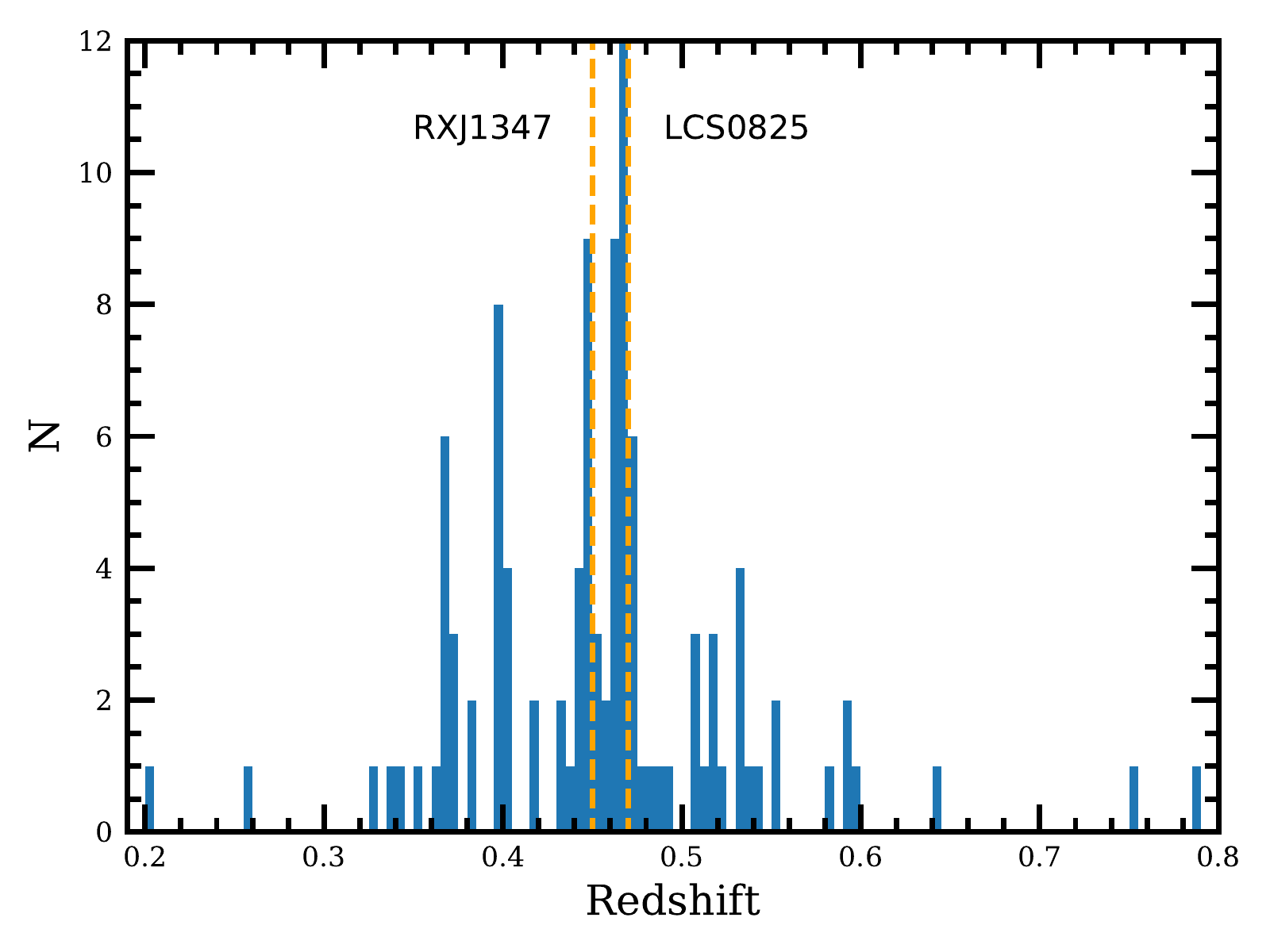}
\caption{Distribution of our targets in redshift space. The two dashed orange lines at z$\sim$0.45 and z$\sim$0.47 correspond to the two major structures of the cluster complex identified as RXJ1347.5-1145 and LCDCS 0825 by \citealt{Verdugo12}.}
\label{F:redshift}
\end{figure}

\begin{figure}
\centering

\includegraphics[width=\columnwidth]{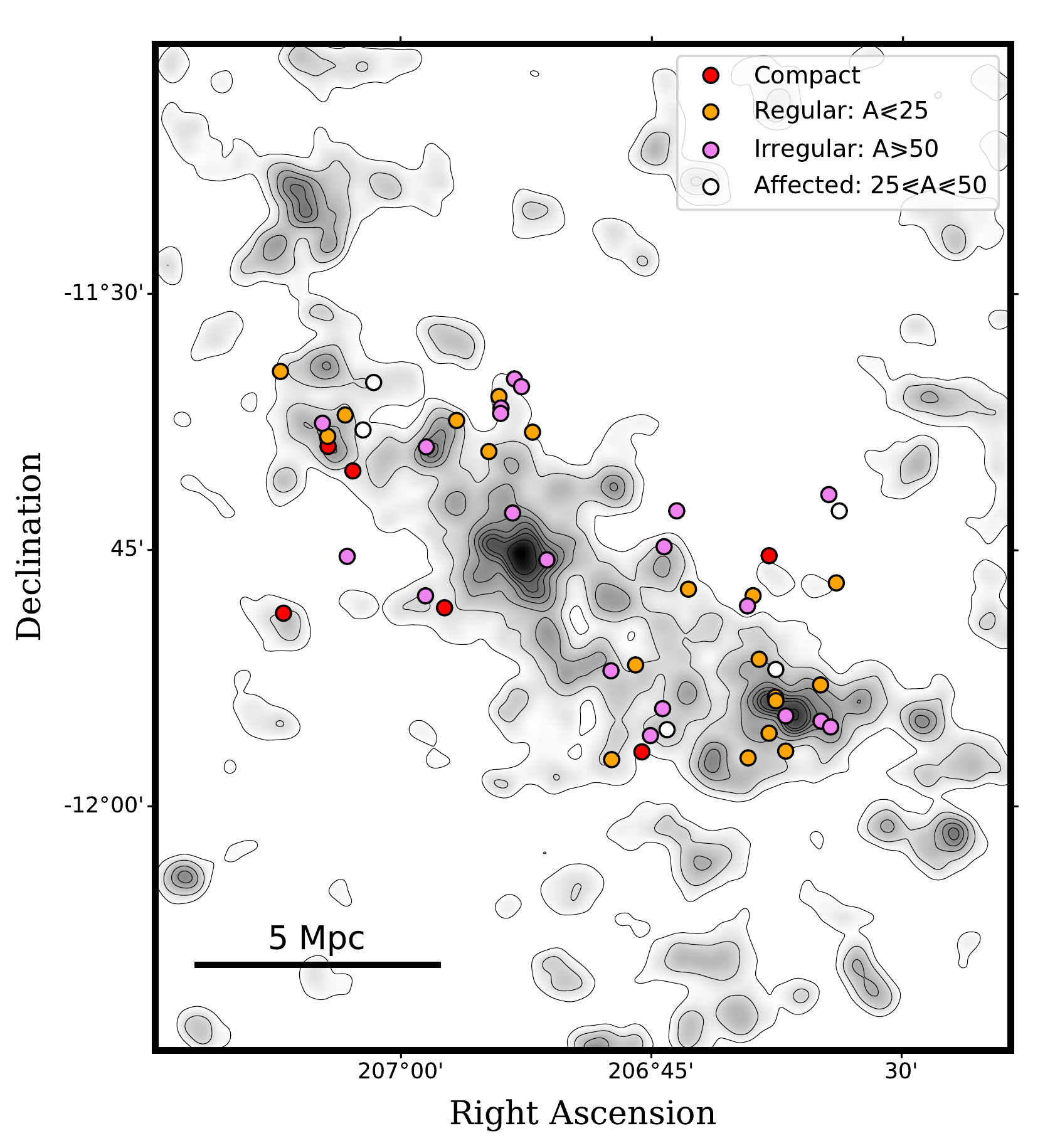}
\caption{Spatial distribution of our cluster sample over the galaxy number density map shown in \cite{Verdugo12} using the nearest-neighbor counting technique. The figure covers an approximate area of 50$\times$50\,arcmin$^2$ around the center of the RXJ1347 cluster complex. Density contours start 1$\sigma$ above the value of the general field at the redshift of the cluster, and gradually increase   to 200 Mpc$^{-1}$ in the inner regions of the cluster complex. The orange, purple, and white circles respectively indicate the galaxies classified as regular, irregular, and affected according to their gas kinematics (see Sect.\,\ref{SS:Asymmetry}). The red circles are galaxies with compact gas emission; these galaxies are not included in our kinematic analysis.}
\label{F:density}
\end{figure}

We use several prominent emission lines ([OII] 3727\AA, H$\beta$ 4861\AA, [OIII] 4959,5007\AA) to measure the redshift of our targets and determine their cluster membership. The distribution of our targets in redshift space is shown in the top panel of Fig.\,\ref{F:redshift}. Two peaks are clearly visible at z$\sim$0.45 and z$\sim$0.47, and correspond to the two main structures of the cluster complex, RXJ1347 and LCDCS 0825. However, more than thirty additional smaller group-like structures have been previously reported to be part of the same large-scale structure (\citealt{Verdugo12}). To encompass most of these structures in redshift space, our cluster membership window is defined as 0.415<z<0.485. Observed galaxies outside this redshift range are considered to be part of the general field population and will form our field comparison sample. Figure\,\ref{F:density} shows the distribution of our cluster targets over the density map of the cluster structure presented in \cite{Verdugo12}. The contours define the galaxy number density of a given area in units of Mpc$^{-2}$. The first contour starts at 14 Mpc$^{-2}$ which is 1$\sigma$ above the mean density value in the field and gradually increases up to 200 Mpc$^{-2}$ in the innermost regions of the cluster complex. Most of our cluster sample is located in the low- to intermediate-density areas. In this figure, our cluster objects have been split into four different subgroups. Three of them (regular, affected, and irregular) depend on the degree of distortion found in their kinematics, which is measured by the asymmetry index $A$ (see Sect.\,\ref{SS:Asymmetry}). The fourth subgroup (compact) contains galaxies whose gas emission is mostly concentrated in the inner parts of the disk. It was not feasible to analyze the spatially resolved kinematics of these objects to a large galactocentric radii, and therefore they  were not included in our kinematic analysis. Further, our spectroscopic campaign benefits from complementary archival SUBARU Suprime-Cam wide-field imaging in five bands (B, V, Rc, Ic, z$'$) and CFHT/MEGACAM g'-band (\citealt{Umetsu14}). The depth and seeing of our co-added mosaic images are shown in Table\,\ref{T:imaging}. The coordinates, redshifts, rest-frame colors, and magnitudes of our final sample are summarized in Appendix \,\ref{appendix:a}. The combination of the large field of view of Suprime-Cam (34$'$x 27$'$), its image quality, depth, and the wealth of our VLT/VIMOS spectroscopic programs allow us to present a comprehensive picture of the physical properties of galaxies in clusters at intermediate redshift.

\begin{table}
\caption{Summary of the imaging data used in this work}
\centering
\begin{tabular}{llcccc}
\hline
\noalign{\vskip 0.1cm}
Telescope &  Filter   & Exp. Time &  FWHM \\ 
          &           & (s)       &   (")   \\ \hline 
\noalign{\vskip 0.1cm}
SUBARU/Suprime-Cam      & B         &  1\,440  &  2.20 \\ 
\ldots      & V         &   2\,160  &  0.75 \\
\ldots      & Rc         &   2\,880  &  0.74 \\ 
\ldots      & Ic         &   3\,240  &  1.14 \\  
\ldots      & z$'$         &  4\,860   &  0.72\\
CFHT/MEGACAM      & g$'$         &  4\,200   &  1.01\\ \hline 

\end{tabular}
\label{T:imaging}
\end{table}
  
\section{Methods} 
\label{SS:Methods}
  
\subsection{Rest-frame magnitudes and stellar masses}
\label{SS:Imaging}

We used the publicly available photometric catalogs produced by the CLASH team (\citealt{Umetsu14}) using SExtractor (\citealt{Bertin96}) over PSF matched SUBARU images in five bands (B, V, Rc, Ic, z$'$) to obtain the observed magnitudes of our targets. Stellar masses and rest-frame magnitudes were computed by using {\sc Lephare} (\citealt{Ilbert2006}, \citealt{Arnouts2011}). This code applies a $\chi^2$ minimization algorithm to match stellar population synthesis models (\citealt{BC03}) to the spectral energy distribution (SED) derived from the photometry available assuming a Chabrier IMF (\citealt{Chabrier03}). We constrained the possible ages to values lower than the age of the Universe at z$\sim$0.45 (i.e., $\sim$9 Gyrs), and applied Calzetti's attenuation law (\citealt{Calzetti2000}) with extinction values of $E(B-V)=0 - 0.5$\,mag in steps of 0.1\,mag. We estimate the total calibration for all bands to have an accuracy of 0.1 magnitudes and $\sim$0.15 dex for the logarithmic stellar masses.

  \begin{figure}
   \centering
   \includegraphics[width=\columnwidth]{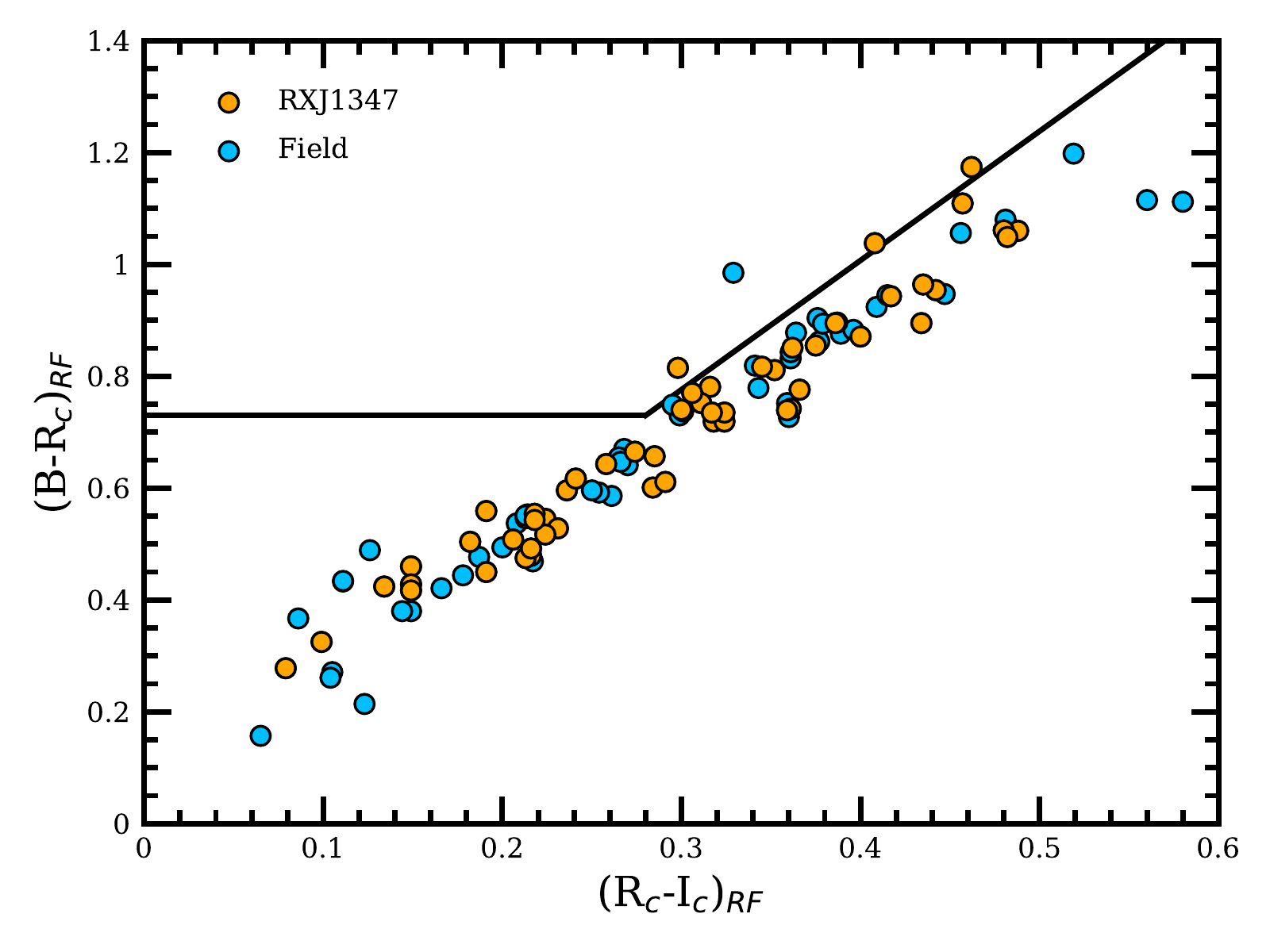}
   \caption{Color-color diagram for our cluster (orange) and field (blue) sample of galaxies. Both samples show very similar distributions in the star-forming region of the diagram with only a few galaxies in the exclusion (passive) area.}
   \label{F:colorcolor}
  \end{figure}

To put our sample in context we plotted our galaxies into the rest-frame $(B-R)$ versus $(R-I)$ color-color diagram (Fig. \ref{F:colorcolor}). This diagram splits the galaxies into two different groups, an old-age sequence of quiescent galaxies (upper left corner) and a star-forming sequence of galaxies with stronger star formation rate and higher dust content (\citealt{Whitaker13}). These regions are empirically delimited by previous studies so that the passive population is easily distinguished from the star-forming one. \cite{Kuchner17} recently applied the BRI diagram for this purpose in another cluster at a similar redshift and compared their results with those obtained by using the popular UVJ diagram (see \citealt{Whitaker13} and \citealt{Vanderwel14}) finding a high degree of consistency between the two  classification schemes. In our study the vast majority of galaxies lie within the star-forming region with very similar distributions between the cluster and field samples.

\subsection{Structural parameters}
\label{SS:StructuralParameters}  

We use the z$'$-band SUBARU Suprime-Cam images to measure the structural parameters of our galaxies. There are two reasons for this choice: the very good seeing conditions (FWHM$\sim$0.7") achieved during the observations in this band, and the redder filters that trace the structure of the disk more accurately and to larger galactocentric radii, avoiding the contamination coming from prominent star formation features that are usually visible in bluer wavelength regimes. This makes the z$'$-band photometry the best available option for computing the structural parameters of our targets.

We used the GALFIT package \citep{Peng02} to model the surface brightness profile of our targets and extract their main structural parameters. For every object we first compute an exponential profile ($n_{s}=1$) to model the disk component of the galaxy. This single-component model is subtracted from the original image and the residuals are visually inspected. If the model is not able to subtract most of the emission in the central area of the galaxy we infer the presence of a bulge component. In these cases we use the single-component parameters as initial guess values in a two-component surface brightness profile with $n_{s}=4$ for the bulge and $n_{s}=1$ for the disk component. Otherwise, we keep the modeled parameters from the one-component fit as our final result. The most important parameters for the analysis presented in this work are the inclination ($i$), the position angle ($PA$), and the effective radius ($R_{e}$). The inclination is computed from the ratio between the apparent major and minor axis ($b/a$) following \citet{Heidmann72} and assuming that the ratio of the disk scale length to the scale height is consistent with the observed value for typical spirals in the local Universe (i.e., q=0.2, \citealt{Tully98}). In a few cases the GALFIT $PA$ values may significantly differ from those computed before our observations using SExtractor on CFHT images, and that were used in the first place to align the slits with the galaxies' major axis. The lower S/N and the poorer seeing conditions of the CFHT imaging are responsible for this misalignment. We compute the angular difference between the orientation of the slit ($\theta$) and the major axis of the galaxy ($PA$) measured by GALFIT and include this additional parameter in our kinematic analysis. The values of $i$, $PA$, $\theta$, and $R_e$ of our objects can be found in Appendix \,\ref{appendix:a}. 

\subsection{Determining the maximum intrinsic velocity (V$_{max}$)}
   
The rotation-curve extraction and determination of V$_{max}$ from 2D spectra is explained in full detail in several previous publications within our own group (see \citealt{Boehm04}, \citealt{Bosch2}, \citealt{Boehm16}). In the following  we provide a brief summary of our approach to obtain V$_{max}$.  

Typically, [OII] and H$\beta$ are the brightest spectral features within the wavelength range of our observations, and thus the sources from which we extract our rotation curves. We enhance the signal-to-noise ratio by averaging over up to 3 pixels (i.e., $\sim$0.6" in the spatial axis) and examine the red- and blueshifts of the emission line under scrutiny as a function of galactocentric radius. These shifts are later converted into positive and negative velocity values with respect to the kinematic center of the galaxy, where by definition $v=0$. To determine this position, we first average each 2D spectrum over $\sim80\AA$ around the emission line and extract the luminous spatial profile of the galaxy in its vicinity. The width of this window varies to avoid contamination by nearby sky-line residuals. Then we fit it with a Gaussian function and provisionally assign the galaxy center to the peak of the luminous spatial profile. However, we allow for a small shift of up to $\pm 1$ pixel to minimize possible asymmetries caused by offsets between the luminous and kinematic centers of rotating galaxies. This shift has a maximum value of $\sim$1.2 kpc in spatial scale at the redshift of our targets. Large offsets hint the presence of interactions causing tidal or ram pressure tails (\citealt{Kronberger08b}). Thus, we avoid matching the kinematic and luminous centers beyond this limit.

Finally, we compute a simulated velocity field that takes into account all observational, geometrical, and instrumental effects: seeing, disk inclination, scale length, misalignment angle, and slit width. We assume the intrinsic rotational law introduced by \citealt{Courteau}, which is defined by a linear rise in the rotation velocity up to the turnover radius and a convergence into a constant value, $V_{max}$, at large galactocentric radii. By extracting the simulated rotation curve we obtain the intrinsic maximum rotation velocity $V_{max}$. The typical error on $V_{max}$ is $\leq 20$ km/s depending on the accuracy of the structural parameters and the quality and extent of the rotation curve. The synthetic velocity fields and simulated and observed rotation curves for our sample can be found in Appendix \,\ref{appendix:a}.

\subsection{Rotation-curve asymmetry}
\label{SS:Asymmetry}

Throughout their lifetimes galaxies may undergo interactions of different kinds, either with other objects or with the medium where they reside. These interactions alter the motion of the gas and stars orbiting around the center of the galaxy, introducing a certain degree of distortion in the kinematics. To quantify these disturbances \cite{Dale01} introduced an asymmetry index ($A$) that measures the difference between the area under the approaching and receding arms of a rotation curve as a function of galactocentric radius. This parameter is particularly sensitive to disturbances affecting the outer parts of the rotation curves and to large offsets between the emission line and the kinematic center of the galaxy. It has been applied successfully to identify distorted galaxies at z$\sim$0.2 by \cite{Bosch1} using the following prescription: 
\begin{equation}
A = \sum_{i} \frac{\left | v(r_{i})+v(-r_i)  \right | }{\sqrt{\sigma_{v}^{2}\left (r_{i} \right )  +  \sigma_{v}^{2}\left (-r_{i} \right )}} \left [ \frac{1}{2} \sum_{i} \frac{\left | v\left (r_{i} \right ) \right | + \left | v\left (-r_{i} \right ) \right |}{\sqrt{\sigma_{v}^{2}\left (r_{i} \right )  +  \sigma_{v}^{2}\left (-r_{i} \right )}} \right ]^{-1}
\label{EQ_asy}
\end{equation}

\begin{figure}
\centering
\includegraphics[width=\columnwidth]{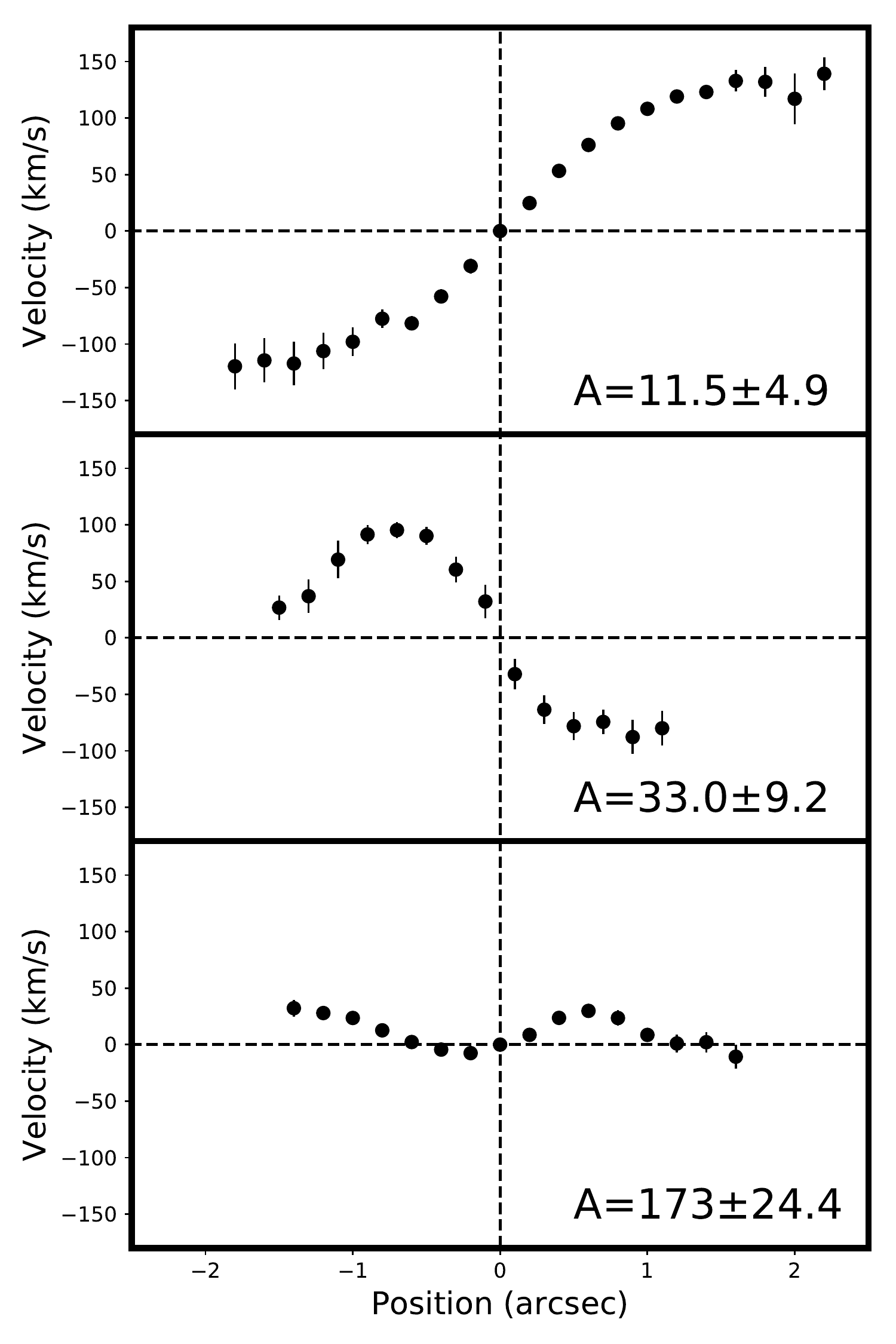}
\caption{Three examples of the asymmetry classification scheme applied to our sample of galaxies. From top to bottom: Regular ($A     \leqslant25$), affected ($25\leqslant A\leqslant50$), and irregular ($A\geqslant50$) kinematics.}
\label{F:asymmetry}
\end{figure}

\cite{Dale01} defined the asymmetry index as a percentage quantity. However, we multiply the result of Eq.\,\ref{EQ_asy} by a factor of 100 throughout this work to obtain absolute $A$ values. The pairs $(v(r_{i}), v(−r_{i}))$ represent the velocity of the two wings of the rotation curve weighted by their errors $(\sigma_{v}(r_{i}), \sigma_{v}(−r_{i}))$. Thus, the asymmetry index cannot  be computed for incomplete pairs of velocity values. This means that if the receding side of the rotation curve is more extended than the approaching side, the asymmetry index will only be calculated within the part of the rotation curve that comprises measurements from both sides. This also implies that to compute $A$, any shift in the kinematic center should be discrete (i.e., in steps of $\pm 0.5$ or $\pm 1$ pixel), preserving the symmetrical spatial distribution of the velocity measurements with respect to $r=0$. Considering this constraint, we set the kinematic center position that provides us with the lowest possible asymmetry value for the subsequent analyses. For undisturbed galaxies, we expect $\left | v(r_{i})+v(-r_{i})  \right |$ to be close to zero, which translates into a very low asymmetry index value, while those galaxies with asymmetric rotation curve wings, significant distortions in one side of the galaxy, or with completely chaotic kinematics will yield higher $A$ values. We apply error propagation in Eq.\,\ref{EQ_asy} to compute the error of $A$. 

The asymmetry index is particularly sensitive to distortions in the outer parts of the rotation curves; here the absolute velocity values are higher and the relative velocity difference between the receding and approaching sides of the rotation curve can also be larger. Furthermore, interactions are more likely to disturb the gas component in the outer parts of the galaxy disks where the gas is less gravitationally bound. In that sense the objects with measured gas kinematics up to large galactocentric radii have a greater chance to show some distortions and get higher $A$ values. However, they also provide the most complete kinematic information. Taking all these effects into account we conclude that our $A$ values provide us with a good estimate for the degree of asymmetry within the ionized gas region of the galaxy, although in some cases they may only represent a lower limit with respect to the asymmetry index of fully extended rotation curves, which can only be measured by neutral gas observations. 

Based on the experience of \cite{Bosch1} with this index we created three categories according to the degree of asymmetry displayed by our galaxies. Those objects with $A\leqslant25$ are labeled as regular rotators. Galaxies displaying intermediate $A$ values such as $25\leqslant A\leqslant50$ are considered to be affected by recent interactions even if they still show signs of their former regular rotational status. Finally, galaxies with $A\geqslant50$ are classified as irregulars. However, we cannot discard the misclassification of some objects between adjacent categories given the intrinsic error of $A$, the diverse spatial extension of our rotation curves, and the presence of small systemic velocities that cannot be accounted for by our kinematic center correction. We show examples of these three categories in Fig.\,\ref{F:asymmetry}.

\section{Results}

In this section we study the link between the kinematic state of our galaxies, the environment, and some of their most important physical properties such as the SFR and the AGN activity. We  investigate the behavior of our sample in several scaling relations. Our sample is initially comprised of 95 spectroscopically detected galaxies, 50 of them in the cluster and 45 in the field. However, the additional requirements imposed to obtain the physical quantities previously mentioned will progressively diminish the size of our sample. 

\subsection{Kinematic state and environment}

The first step in our analysis consists in the identification of the kinematic state of our cluster and field samples. To achieve this we first inspect the spectra of our targets and extract position-velocity diagrams such as those shown in Fig.\,\ref{F:asymmetry}. Those objects with kinematic information up to a sufficiently large galactocentric radii will be classified as regulars, affected, and irregulars according to their asymmetry index values. We find that the fraction of irregular galaxies in the cluster environment is higher than in the field  (40.0$\%$ and 23.3$\%,$ respectively; see Table \ref{T:asymmetry}). However, the total fraction of galaxies that have suffered some kind of disturbance (affected + irregulars) is similar between the cluster and field environment (50$\%$ and 41.8$\%$). This suggests that while the field population of galaxies at 0.3<z<0.6 already contains a significant fraction of galaxies showing some degree of distortion, cluster-specific interactions contribute to enhance their asymmetry index, increasing the fraction of irregular galaxies according to our gas kinematics asymmetry criterion. These results are in line with previous studies  that reported significant fractions of field galaxies with perturbed kinematics at intermediate redshifts (\citealt{Yang08}, \citealt{Kutdemir10}).

\begin{table}
\caption{Kinematic state fractions}
\centering
\begin{tabular}{ccllll}
\hline
\noalign{\vskip 0.1cm}
             &  Cluster   & Field \\ \hline 
\noalign{\vskip 0.1cm}
Regular       & 38.0$\%$ (19/50) &   41.8$\%$ (18/45)  \\ 
Affected      & 10.0$\%$ (5/50)   &  18.6$\%$ (8/45)  \\
Irregular     & 40.0$\%$ (20/50) & 23.3$\%$ (10/45) \\
Compact       & 12.0$\%$ (6/50)  & 16.3$\%$ (7/45) \\
\hline 
\end{tabular}
\label{T:asymmetry}
\end{table}

\subsection{The Tully-Fisher relation}

We used galaxies with regular  kinematics to extract a reliable value for $V_{max}$ for our Tully-Fisher diagrams. Objects classified as regular rotators using the asymmetry index criterion, and those labeled as affected but with sufficiently extended kinematics were included in our analysis. In this work, we chose the B-band and the stellar-mass ($M_{*}$) TFR to look for imprints of environmental effects in our cluster galaxies. The B band is dominated by the light of massive young stars, and is therefore  very sensitive to recent episodes of star formation. On the other hand, the stellar mass acts as a proxy to trace the weight of the overall underlying population of old stars within the galaxy. In summary, these two complementary representations of the TFR provide a way to examine the recent and cumulative star formation history of the galaxies through their kinematics.  

 \begin{figure*}
    \begin{multicols}{2}
      \includegraphics[width=\linewidth]{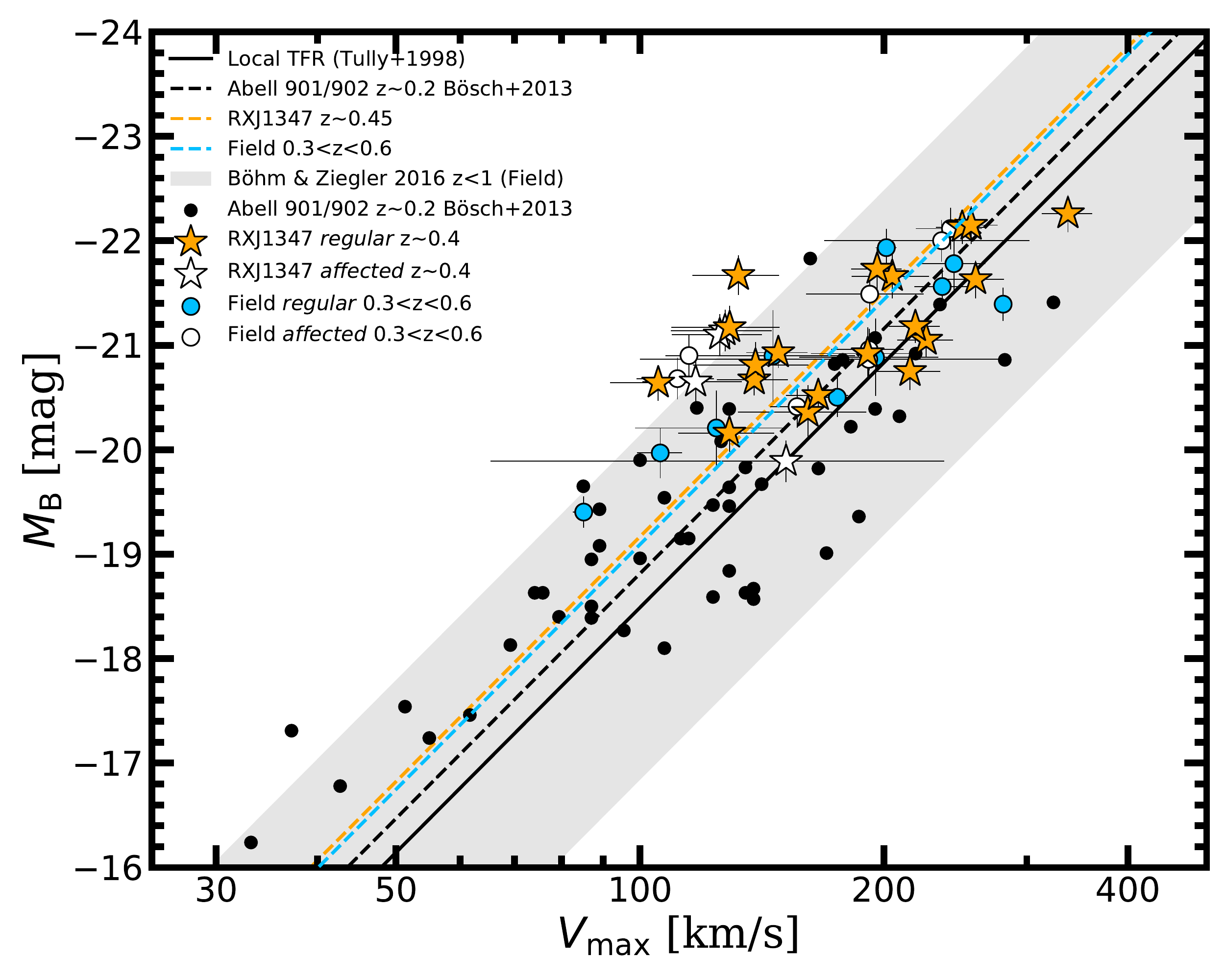}\par 
      \includegraphics[width=\linewidth]{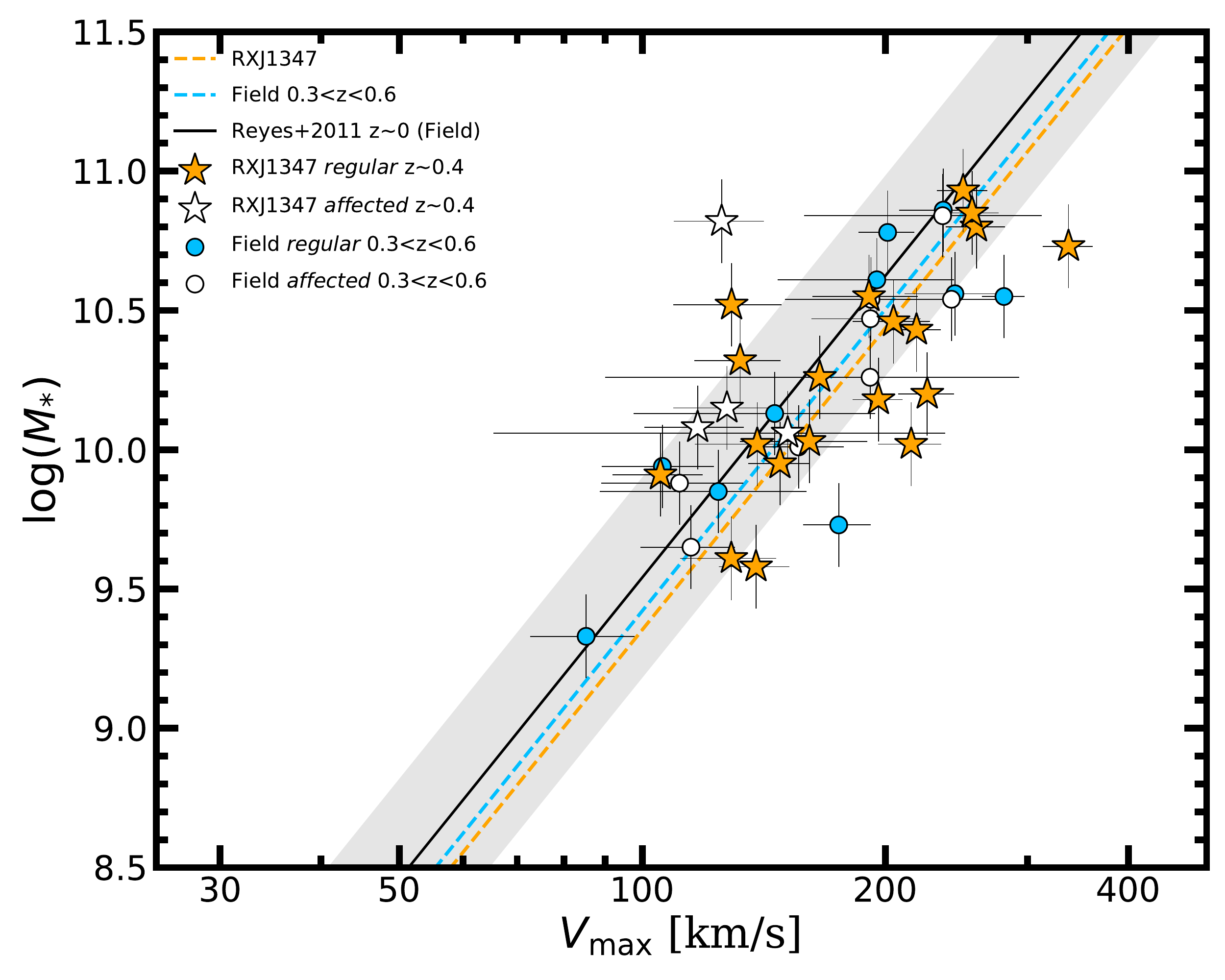}\par 
      \end{multicols}
      \caption{\textit{Left}: Tully-Fisher B-band diagram. In both diagrams the  orange and white stars represent regular and affected cluster objects, respectively, while the blue and white circles represent regular and affected field galaxies. Black circles represent cluster galaxies at z$\sim$0.2 from \cite{Bosch2}. The solid black line shows the local B-band TFR (\cite{Tully98}) with a 3$\sigma$ scatter area around reported by \cite{Boehm16} for galaxies at 0<z<1 (gray area). The orange, blue, and black dashed lines represent the best fit for the cluster and field sample of this study and the cluster sample from \cite{Bosch2}, respectively.  \textit{Right}:  Stellar-mass Tully-Fisher diagram. The symbols and their colors follow the same description as in the left-hand panel. The solid black line shows the local $M_{*}$-TFR from \cite{Reyes11}, with a 3$\sigma$ scatter gray area around it.}
         \label{F:TFR}
      \end{figure*}

 \begin{figure*}
    \begin{multicols}{2}
      \includegraphics[width=\linewidth]{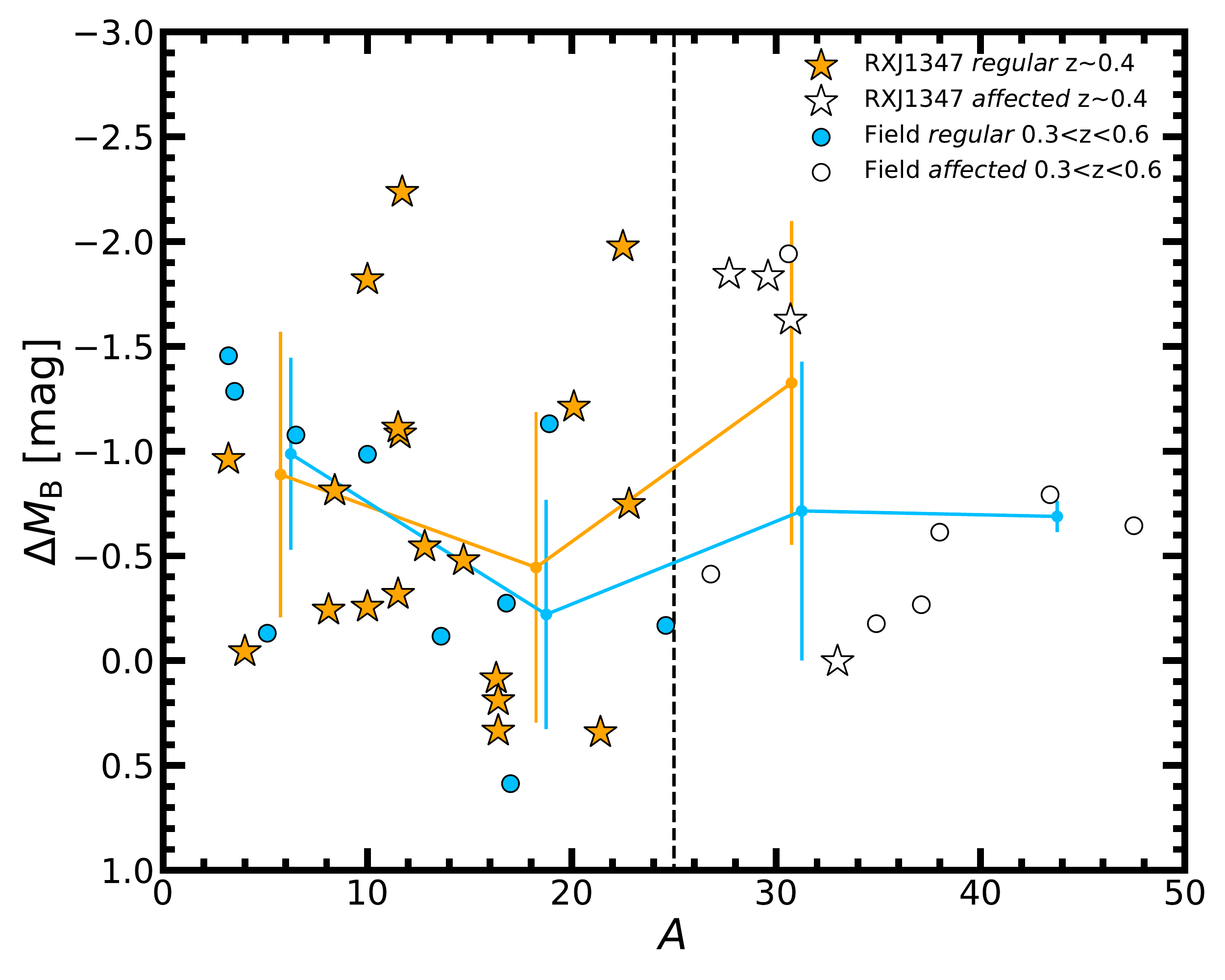}\par 
      \includegraphics[width=\linewidth]{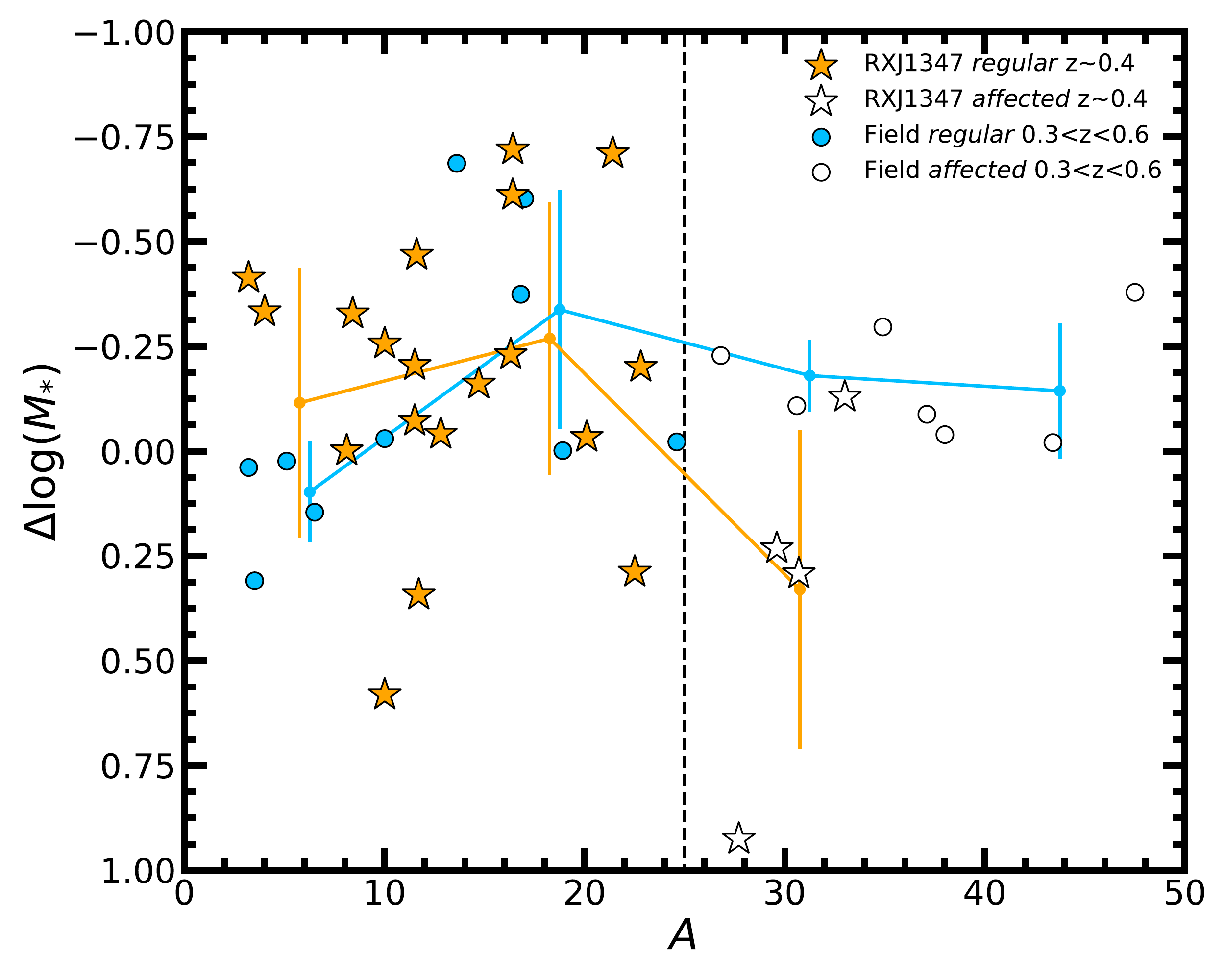}\par 
      \end{multicols}
      \caption{\textit{Left}: B-band Tully-Fisher offsets compared to the asymmetry index values. \textit{Right}: Stellar-mass (M$_{*}$) Tully-Fisher offsets compared to the asymmetry index values. In both diagrams the orange and white stars represent regular and affected cluster objects, respectively, while the blue and white circles represent regular and affected field galaxies. The orange and blue dots joined by lines of the same color show the mean values and standard deviation of our cluster and field sample divided into four bins according to their asymmetry index ($A\leqslant12.5$, $12.5\leqslant A\leqslant25$, $25\leqslant A\leqslant37.5$, and $37.5\leqslant A\leqslant50$). The vertical dashed black line at A=25 shows the limit between the kinematically regular and affected categories.}
         \label{F:TFR2}
      \end{figure*}

\begin{figure}
\centering
\includegraphics[width=\columnwidth]{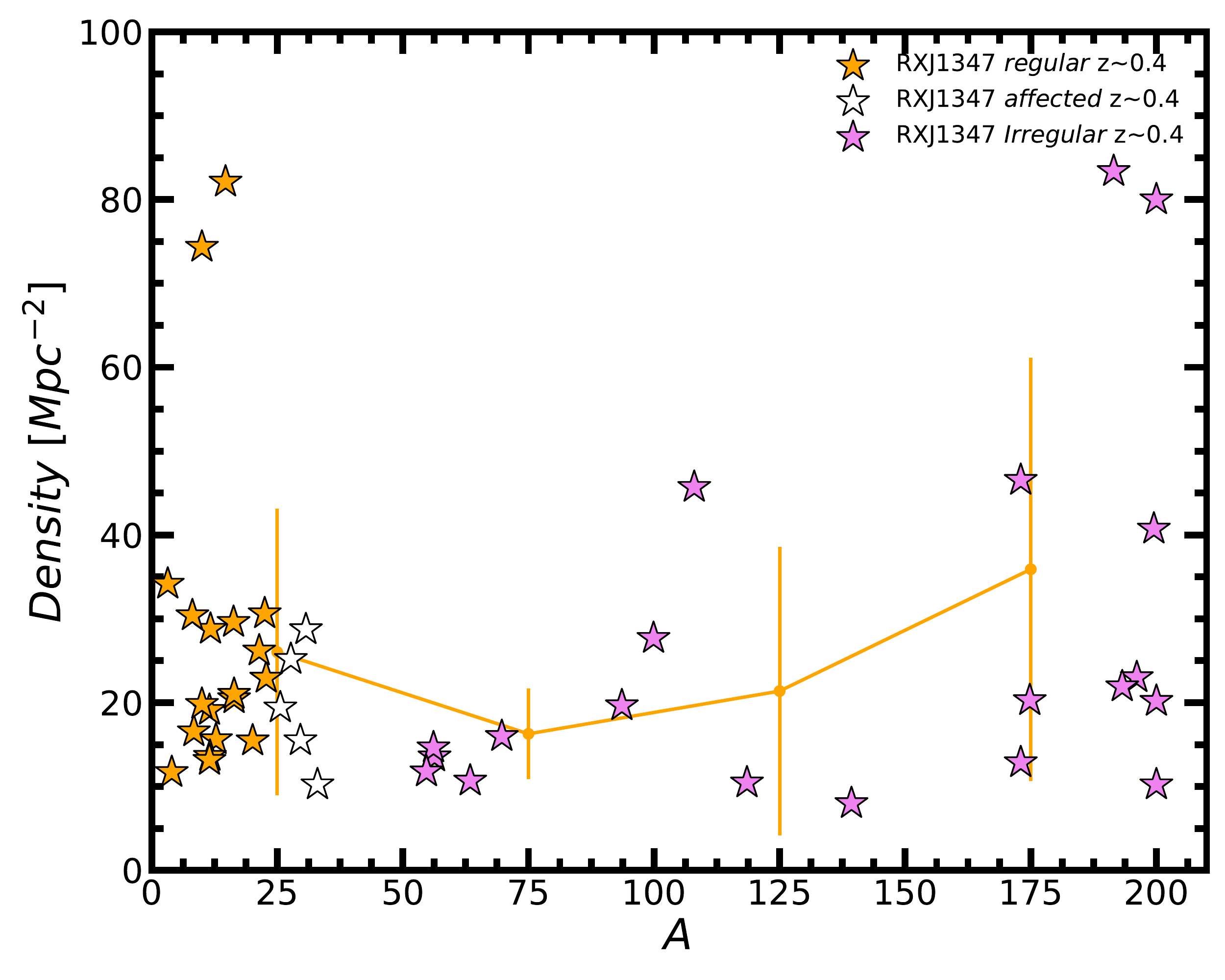}\par 
\caption{Galaxy number density of the area where our cluster objects lie (\citealt{Verdugo12}) with respect to their asymmetry index. The orange, white, and violet stars represent regular, affected, and irregular cluster objects, respectively. The orange dots joined by lines of the same color show the mean values and standard deviation of our cluster sample divided into four bins according to their asymmetry index ($A\leqslant50$, $50\leqslant A\leqslant100$, $100\%\leqslant A\leqslant150$, and $150\leqslant A\leqslant200$).}
\label{F:TFR3}
\end{figure}

Before presenting our results on the B-band TFR, we must emphasize the importance of correcting our absolute magnitudes from extinction. In general, edge-on spiral galaxies show higher values of extinction than their face-on counterparts, the reason is that the light coming from the stars goes through a larger portion of the disk when the galaxy is edge-on with respect to the line of sight. Moreover, more massive disks are dustier than lower-mass disks \citep{Giovanelli95}. We take into account these two effects following the prescription given by \citet{Tully98} to correct rest-frame B-band absolute magnitudes for intrinsic dust absorption. This correction diverges for completely edge-on galaxies (i.e., $i=90º$), and for this reason we exclude from our sample one cluster and four field galaxies. After applying this correction, the typical errors for the B-band absolute magnitude values in the TFR are $\sim$0.2-0.3 mag. Two more galaxies were excluded (one in the cluster and one in the field)  after finding that their mismatch angles were $\delta\geqslant 45º$. Finally, four field galaxies lie beyond the edge of the SUBARU images and were excluded due to the lack of enough photometric bands to compute reliable rest-frame magnitudes and stellar masses. 

We present our B-band TFR in Fig.\,\ref{F:TFR} (left). Our final TFR cluster sample is composed of 19 regular and 4 affected objects (orange and white stars, respectively). However, we     only use the regular objects to study the evolution of the Tully-Fisher relation. To find the best fit for our sample we keep the slope of the local relation by \citet{Tully98}, while we let the intercept vary. We find an average deviation of $\Delta M_B$=-0.7$\pm$0.8 mag for our cluster sample. We use three different auxiliary samples for comparison. First, we make use of the ten remaining field regular galaxies observed by our program, finding that $\Delta M_B$=-0.6$\pm$0.7 mag. We also include a sample of 50 cluster star-forming galaxies at z=0.16 studied by \cite{Bosch2}, who reported $\Delta M_B$=-0.3$\pm$0.7 mag with respect to the local TFR. Finally, we compare our results against a sample made of 124 field star-forming galaxies at 0<z<1 that is representative of the typical scatter of this scaling relation in the given redshift range (gray area). In all these data sets we note that  $V_{max}$ and $M_B$ were computed using the same methods presented in this study, which makes them ideal for a direct comparison. The scatter of our cluster sample at z$\sim$0.45 is consistent with what has been previously found in the field (\citealt{Boehm16}), while the reported offset in B-band luminosity is larger than that of the \cite{Bosch2} sample at lower redshift, but in line with what has been found by previous observational studies (\citealt{Bamford05}) and predictions from semianalytical models (\citealt{Dutton11}). We repeated our analysis for the $M_{*}$-TFR (Fig.\,\ref{F:TFR}, right side) finding a mild offset ($\Delta M_{*}$=0.2$\pm$0.4) between our targets at z$\sim$0.45 and the local relation (\citealt{Reyes11}). In the case of our field sample the offset is even smaller ($\Delta M_{*}$=0.1$\pm$0.3). This supports previous results claiming no significant evolution on the $M_{*}$-TFR up to z$\sim$1 (\citealt{Pelliccia17}, \citealt{Harrison17}) and points towards a small influence of the environment in the M$_{*}$-TFR at this redshift.

In Fig.\,\ref{F:TFR2} we analyze the possible relation between the asymmetry index (A) and the residuals from the B-band and M$_{*}$-TFR. We bin our objects in four bins according to their asymmetry index value ($A\leqslant12.5$, $12.5\leqslant A\leqslant25$, $25\leqslant A\leqslant37.5$, and $37.5\leqslant A\leqslant50$) and compute the average of the residuals in these bins and its standard deviation. We restrict our study to only the first three bins of the cluster sample due to the lack of galaxies in the fourth. Our results show that the cluster and field samples are similarly distributed with no clear trends. In Fig.\,\ref{F:TFR3} we investigate the influence of the local galaxy number density on the asymmetry index of our objects. Within our cluster sample we find that most objects included in our Tully-Fisher analysis are located in moderate- to low-density regions of the structure, while only two regular galaxies are found in the densest areas of the cluster complex. In an environment-based quenching scenario we expect that most star-forming galaxies increase their asymmetry index (becoming kinematically irregular; \citealt{Bosch1}) and gradually stop their star formation during their infalling path towards the central and densest areas of the cluster (\citealt{Haines15}). The only two regular objects lying in these dense regions show $\log{M_*}$=10.52 and $\log{M_*}$=10.85, which may indicate that only the most massive star-forming galaxies would still show significant star formation activity across their disks and maintain their regular rotation once they reach the central areas of massive clusters. On the other hand, most galaxies with irregular gas kinematics within our sample also lie in low- to intermediate-density regions, which suggests that a significant fraction of field galaxies infalling into cluster structures already carry gas kinematic distortions before being affected by environmental effects.  

\subsection{Star formation activity}  

The most reliable and most commonly used SFR calibrator is H$\alpha$. However, the observation of intermediate- to high-redshift targets makes it difficult to get access to this emission line using optical spectroscopy. This is the case of our VIMOS programs for which we can only detect spectral features for cluster galaxies between $3600-5200\AA$ in the rest frame. Thus, we rely on the [OII]$\lambda$3727 doublet to estimate the SFR of our targets. Due to the slit positioning of our objects with respect to the center of the VIMOS mask, the wavelength range of some objects is slightly offset towards redder or bluer wavelengths. In the former case, this may shift the [OII] line out of the visible wavelength range, making it impossible to determine the SFR using this method. This reduces our sample to 31 cluster galaxies split into the same three groups described in Sec.\,\ref{SS:Asymmetry} according to their asymmetry index. We apply the  prescription given by \cite{Gilbank11} to compute reliable $SFR$ values,
\begin{equation}
SFR_{emp,corr}/(M_{\odot} {yr}^{-1})=\frac{L([OII])/3.80\times 10^{40}erg s^{-1}}{a \tanh [(x-b)/c]+d}
,\end{equation}
where $a$ = −1.424, $b$ = 9.827, $c$ = 0.572, $d$ = 1.700, and $x$=$\log{(M_{*}/M_{\odot})}$. This approach includes an empirical mass-dependent correction that takes into account the effects of metallicity and dust extinction over the SFR. However, \cite{Gilbank11} assume a Kroupa IMF, while all the quantities in this paper have been computed following a Chabrier IMF. To maintain consistency, we multiply $SFR_{emp,corr}$ by a factor of 0.9, which accounts for the stellar mass transformation between the Kroupa and Chabrier IMFs.

\begin{figure}
   \centering
   \includegraphics[width=\columnwidth]{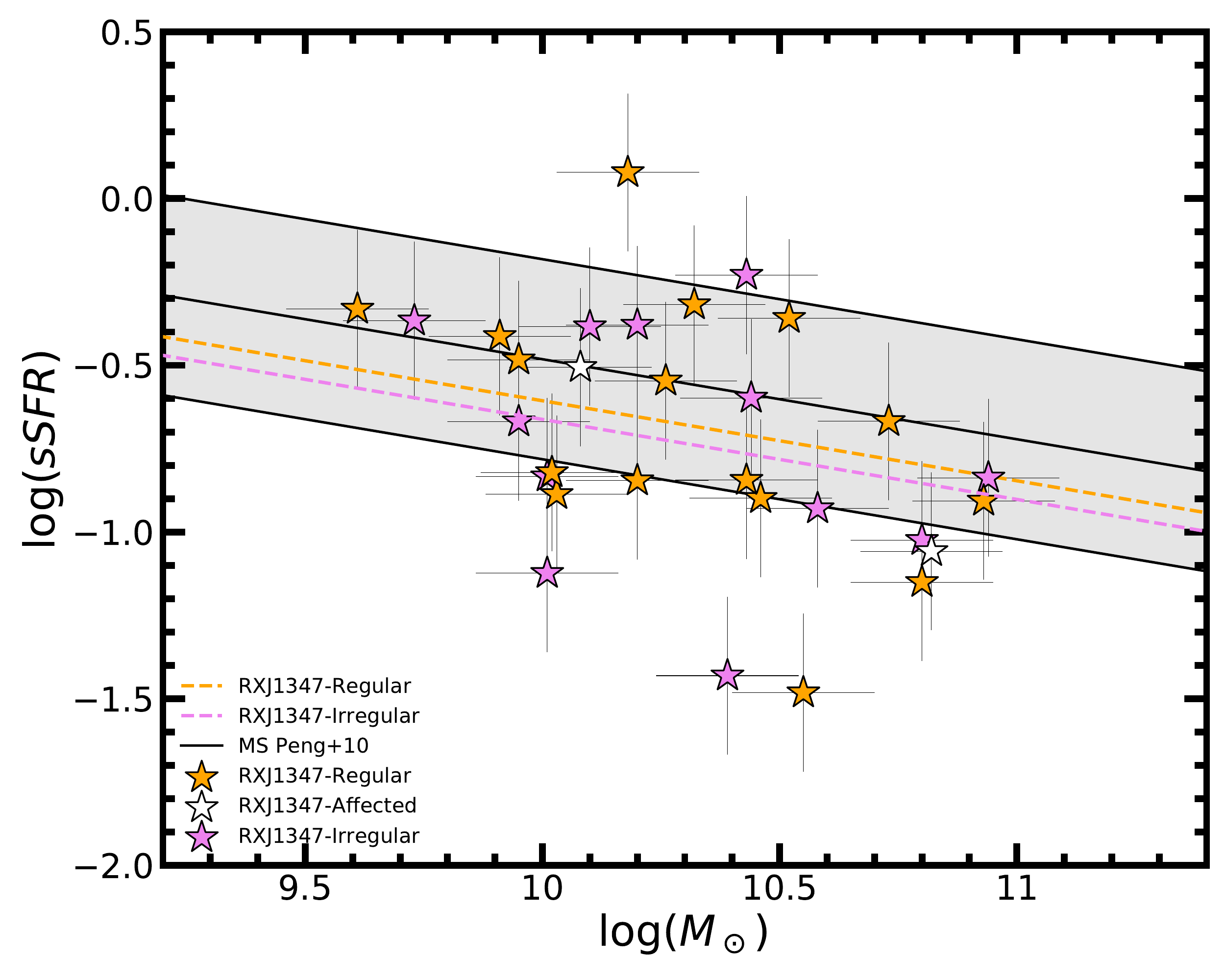}
   \includegraphics[width=\columnwidth]{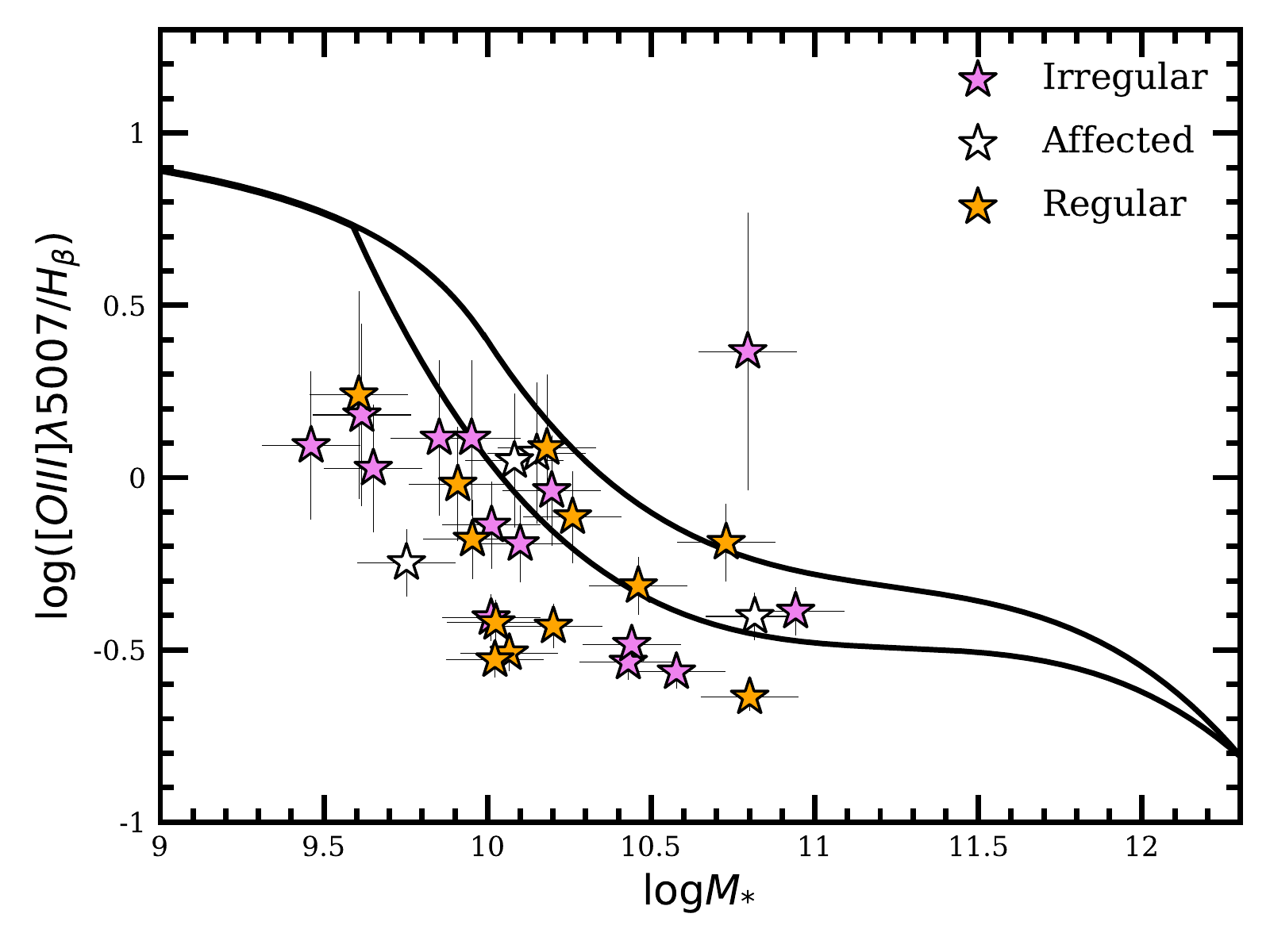}
   \caption{\textit{Top}: sSFR-log(M$_*$) diagram. Orange, white, and violet stars respectively represent cluster galaxies classified as regular, affected, and irregular according to their kinematic asymmetry index value ($A$). The black solid line shows the main sequence of star-forming galaxies at z$\sim$0.45 given by \cite{Peng10} with a 3$\sigma$ gray area region. The orange and violet dashed lines represent the best fit linear regressions to our sample of regular and irregular galaxies, respectively, assuming the slope given by the main sequence and a Chabrier IMF. \textit{Bottom}: Mass-excitation diagram (symbol shapes and colors  as  in the top panel). The diagram is divided into three different regions according to the dominating source of gas excitation: star-forming  (bottom left), AGN (top right), and composite (central stripe).}
   \label{F:SSFR}
\end{figure}

To study the star formation activity of our cluster galaxies we present the sSFR-mass relation in Fig.\,\ref{F:SSFR} (upper panel). We  use the main sequence (Eq. 1 in \citealt{Peng10}) at z$\sim$0.45 as a reference for the expected star formation activity in the field. The goal of this analysis is to study the environmental imprints on the star formation activity of our galaxies, and the connection with their kinematic state. We find that regular cluster galaxies lie on average 0.1$\pm$0.3 dex below the main sequence, while irregular galaxies show a slightly larger offset of 0.2$\pm$0.4 dex. On average, the specific star formation of kinematically irregular galaxies is slightly more suppressed than in their regular counterparts. However, this difference becomes statistically insignificant once we take the errors into account.

\subsection{Gas excitation diagnostics} 

In this section we aim to investigate the ionizing source of the interstellar medium (ISM) in our galaxies, and if it is related to cluster-specific interactions that influence their degree of gas kinematic asymmetry at the same time. The two candidate processes are the products of star formation (i.e., hot young stars), and the presence of a supermassive black hole in the center of the galaxy injecting a large amount of energy in the ISM. Given the wavelength constraints of our spectroscopic observations, we are unable to apply the often used BPT diagram (\citealt{Baldwin81}) for this purpose, and thus we are forced to use other diagnostics that require emission lines in the bluer part of the spectrum. One of these representations is the mass-excitation (MEx) diagram introduced by \cite{Juneau11}. This diagram takes the [OIII]$\lambda$5007/H$\beta$ ratio from the BPT diagram and substitutes the [NII]/H$\alpha$ ratio for the stellar mass. It has been tested up to z$\sim$2 showing a good degree of consistency with respect to BPT analyses at z<1 (\citealt{Juneau14}). 

We present our results in Fig.\,\ref{F:SSFR} (bottom panel), where we divide our sample according to their kinematic state in the same way that we described in the previous section. We find that most of our galaxies lie within the star-forming or composite regions, independently of their kinematic classification. There are only two galaxies within the AGN region,  one  classified as irregular and the other as regular according to their asymmetry indexes. Our results suggest that AGN activity is not connected with kinematic gas distortions, and thus interactions between galaxies or with the intracluster medium are not likely to trigger a strong AGN response by channeling gas towards the central regions of the galaxies on our spatial scales.
        
\subsection{Halo masses} 

Dark matter  halos are key to understanding the formation of the first galaxies and the hierarchical growth of structures in the universe. For most individual galaxies, dark matter represents more than 80\%\ of their total mass, yet due to the non-interacting nature of dark matter little can be said about its properties. For this reason, galaxy evolution studies are usually focused on the study of the baryonic component of the different populations of galaxies across cosmic time. However, measurements of the halo mass are fundamental in order to achieve a comprehensive understanding of galaxy formation and evolution since the gravitational potential, dominated by the dark component, drives most of the interactions that a galaxy undergoes during its lifetime. 

The presence of dark matter within galaxies is ideally inferred from observations of baryonic matter if possible. The traditional method for measuring the dark matter content of galaxies requires  their dynamical masses to be derived. However, the available baryonic information is usually limited by observational constraints, and the use of models and simulations is needed to translate our observables into the parameters required to compute the dynamical mass. Recently, \citealt{Conselice18} adopted several combined observational and theoretical approaches to derive the halo mass of field galaxies up to z$\sim$3. In the next sections, we follow a method introduced by \cite{Lampichler17} and tested by \cite{Conselice18} to derive the halo mass values for our sample of galaxies. This method is only valid for objects with v/$\sigma$>1. While we did not carry out a velocity dispersion analysis for our samples, it is reasonable to assume that galaxies labeled as regulars (A<25) according to their asymmetry index comply with this requirement. The halo mass can be defined as
\begin{equation}
M_{h}=\frac{v_{h}^{2}R_{h}}{G}
,\end{equation}
where R$_{h}$ is the virial radius of the halo, v$_{h}$ is the rotation velocity at R$_{h}$, and G is Newton's gravitational constant. However, R$_{h}$ and v$_{h}$ cannot be directly obtained from our observational data. Thus, we need to find a way to compute these quantities from the effective radius R$_e$ and the maximum rotation velocity V$_{max}$ measured in our study. \citealt{Kravtsov13} established a relationship between the half-mass radius (R$_m$) and the virial radius assuming that the relation between the total mass of halos, M$_h$, and stellar mass of galaxies they host, M$_\ast$, is approximately monotonic, and the cumulative abundances of halos and galaxies match (n$_h$($>$M) = n$_g$($>$M$_{\ast}$)): 
\begin{equation} \label{eq2}
R_{m}\approx0.015 R_h
\end{equation}
In the case of disk galaxies, we can convert the half-mass radius into optical half-light radius by using the empirical relation presented in \cite{Szomoru13} for galaxies at 0.5<z<2.5. In this work, the authors used deep HST data in several fields to derive accurate stellar-mass surface density profiles, from which R$_m$ can be extracted by assuming a certain M/L ratio dependent on the galaxies' properties. In the low-redshift regime, they found that R$_{m}$ is on average 25$\%$ smaller than the rest-frame optical R$_e$: 
\begin{equation}\label{eq3}
R_e\approx1.33R_m
\end{equation}
For a more in-depth discussion of the methods used we refer to \cite{Szomoru13}. Equations \ref{eq2} and \ref{eq3} provide us with a relationship between the virial radius of the dark matter halo (R$_{h}$) and the half-light optical radius (R$_e$). Finally, to compute the halo mass we need to connect V$_{max}$ with v$_h$. Several works (\citealt{Dutton10}, \citealt{Papastergis11}, \citealt{Cattaneo14}) have investigated this relation by comparing the rotation velocity measured at several scale lengths from the center of the galaxy (v$_{opt}$) with theoretical models that take into account the contribution of different dark matter halo profiles to obtain the rotation velocity at the virial radius (v$_h$). Given the description of v$_{opt}$ in those studies we can assume that in our work V$_{max} \approx v_{opt}$ in the following. However, the ratio v$_{opt}$/v$_h$ is strongly dependant on the model used, ranging from v$_{opt}$/v$_h$=1.1 to v$_{opt}$/v$_h$=1.5 between different studies. We here adopt the mean value, v$_{opt}$/v$_h$=1.3, to compute our halo masses. Taking into account these approximations we can estimate the halo mass of our targets in the following way:
\begin{equation}
M_{h}\approx\frac{29.7V_{max}^{2}R_{e}}{G}
\end{equation}
In Fig.\,\ref{F:Halo} we present the stellar-to-halo mass relation for our sample of cluster and field galaxies at intermediate redshift in comparison with the theoretical relation derived by \cite{Moster13}, which in this case has been  parameterized for z$\sim$0.45. The distribution of both our cluster and field samples follow the theoretical relation with significant systematic errors inherent to the computation of M$_{h}$. Interestingly, most cluster galaxies with $\log{M_*}$<10.5 lie below the theoretical prediction, while this effect is not seen for field galaxies with similar stellar mass. However, the scatter of the sample, the low number of objects, and the fact that this $M_{h}$ estimation does not take into account the effects of the cluster environment on the dark matter halos of satellite galaxies make it difficult to draw conclusions about the origin of this difference.

\begin{figure}
   \centering
   \includegraphics[width=\columnwidth]{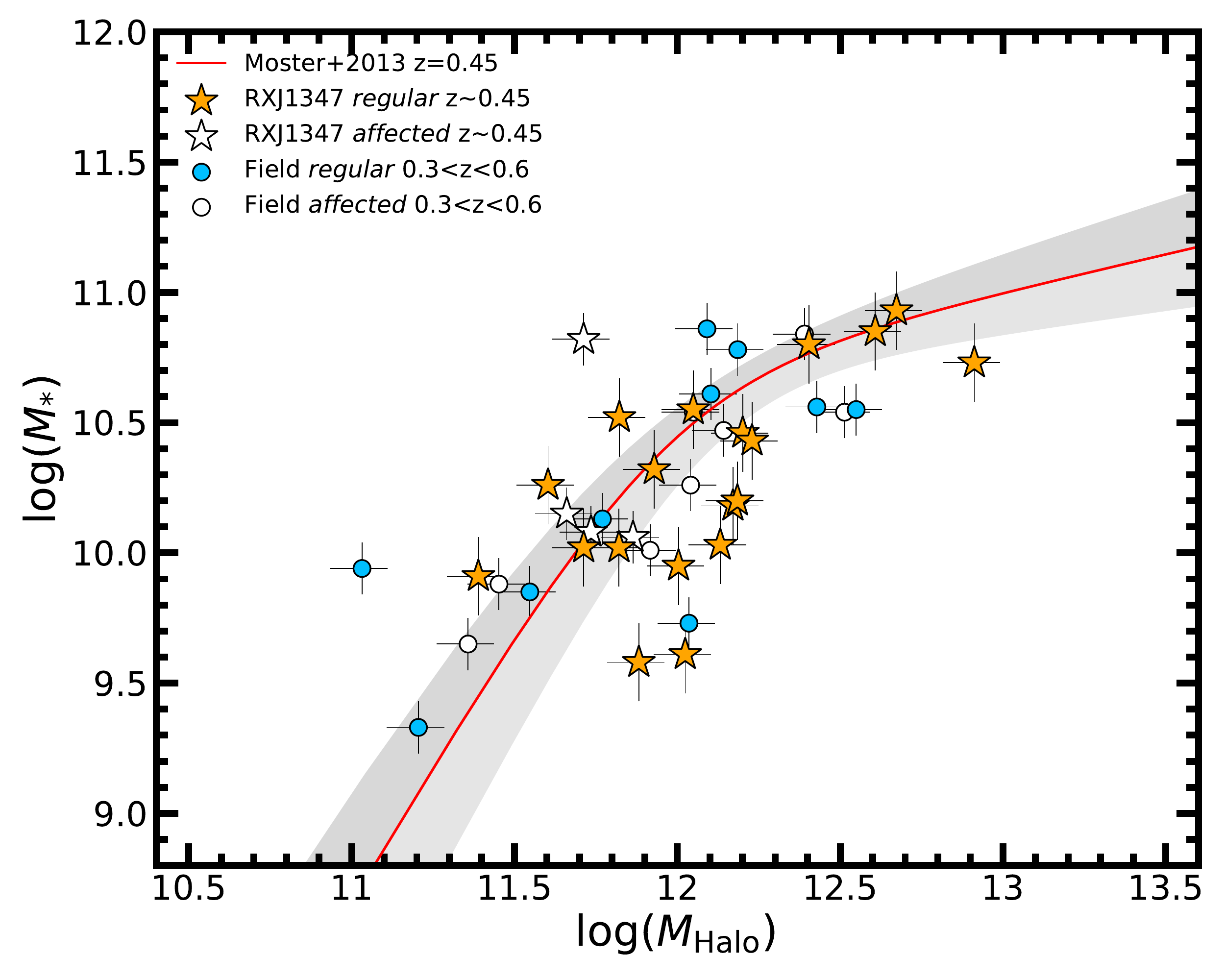}
   \caption{Stellar-to-halo mass diagram. The red solid line represents the expectations for this relation given by \citealt{Moster13} at z$\sim$0.45 with a 1$\sigma$ gray area. All the other symbols have the same meaning as in Fig.\,\ref{F:TFR}.}
   \label{F:Halo}
\end{figure}

\section{Discussion}
  
The study of different galaxy properties provides us with diverse pieces of the galaxy evolution picture. This work is focused on the study of galaxy kinematics, its connection with the environment, and the star formation and AGN activity. The kinematics analysis yielded a higher fraction of irregular galaxies in the cluster environment than in the field according to our asymmetry index criteria. This can be explained by the influence of cluster-specific interactions on the population of galaxies that are progressively infalling towards the central regions of the cluster complex. On the other hand, we speculate that the relatively high fraction (41.9$\%$) of field galaxies displaying signs of disturbances (irregular + affected) may be caused by the continuous mass growth of the galaxies in the field via accretion and minor merging events at intermediate redshifts. Our results suggest that field galaxies do not live in complete isolation and are subject to minor interactions with relative frequency, disturbing their gas kinematics to some extent. This scenario was already introduced by \cite{Puech08} and \cite{Kutdemir10} in the past.  

The Tully-Fisher relation and its evolution for cluster galaxies have been a subject of debate during the last two decades. Currently the dominant view is that there are no significant differences in the slope and the zero point of the relation between field and cluster galaxies at a fixed redshift, but larger scatter has been reported in cluster samples. However, the combination of different representations of the TFR can be useful to obtain information about the evolutionary stage of the stellar populations that are part of the studied galaxies. Some attempts in this direction were made by group in \cite{JM17} for a small sample of cluster galaxies at z$\sim$1.4. However, we do not find remarkable effects for our cluster galaxies at z$\sim$0.45. The cluster B-band TFR yielded a moderate brightening of $\Delta M_B$=-0.7$\pm$0.8 mag with respect to the local B-band TFR, which can be explained by the gradual evolution of stellar populations with lookback time and the intrinsic scatter of our sample. These results agree within the errors with previous works in the cluster environment (\citealt{Bamford05}). In parallel, we report no significant evolution in the $M_{*}$-TFR ($\Delta M_{*}$=-0.2$\pm$0.4), in line with previous observational studies in the field (\citealt{Ubler17}, \citealt{Tiley19}) and cluster environment (\citealt{Pelliccia19}), with semianalytical models (\citealt{Dutton11}) and with recent results from the EAGLE hydrodynamical simulations at z$\sim$0 (\citealt{Ferrero17}). The lack of evolution in the $M_{*}$-TFR across environment and cosmic time points towards the presence of a tight link between the stellar mass growth and the changes in dynamical mass (for which $V_{max}$ acts as a proxy) of a given galaxy during its lifetime. Moreover, we do not see a clear correlation between the kinematic degree of asymmetry $A$ of our objects and the residuals with respect to the local B-band and stellar-mass TFR. However, our TFR cluster and field samples are mainly composed of regular galaxies (i.e., with $A\leq25$ values), and thus we lack the statistics to investigate the possible offsets caused by galaxies that have been likely affected by some kind of interactions (i.e., $25\leq A\leq50$ values) in these two scaling relations. Similarly, we do not find a correlation between the local galaxy number density and the degree of asymmetry of the cluster sample. Most of our regular, affected, and irregular galaxies populate regions with intermediate local density within the cluster. This suggests that interactions in dense environments are already common during the infalling phase. Moreover, the lack of objects with extensive gas disk emission in the densest regions of the cluster also suggests that star formation is more efficiently suppressed in the cluster core.

Dense environments accelerate and strengthen the quenching of star formation in galaxies (\citealt{Maier16}, \citealt{Bruno17}) and their morphological transformation (\citealt{Kuchner17}). In this work, we  used this information to investigate the star formation activity for cluster galaxies displaying different kinematic behaviors. Our cluster sample agrees with previous results (\citealt{Bruno17}) showing lower sSFR values compared to the field main sequence of star-forming galaxies at intermediate redshift. Furthermore, galaxies classified as irregulars ($A\geqslant50$) display a slightly lower average sSFR value (-0.2 dex) than their regular counterparts (-0.1 dex), although this difference becomes insignificant when taking into account the scatter of our sample.

Some authors have speculated that gravitational interactions such as close encounters or mergers that heavily disturb the gas kinematics of galaxies may trigger starburst events, enhancing momentarily the SFR of the galaxies involved (\citealt{Schweizer05}, \citealt{Teyssier10}, \citealt{Hoyos16}). However, the outcome of these interactions strongly depends on the intrinsic properties of the interacting objects, and in most cases it appears to have only a mild effect on the SFR. For example, \cite{Knapen15} found an increase in the SFRs by an average factor of 1.9 for extreme interacting field galaxies in the local universe compared to their non-interacting control sample. However, approximately 50\% of these extreme interaction events also yielded SFR values consistent with or even slightly lower than those of the non-interacting control sample. Similar results were found by \cite{Pearson19} using different surveys at 0<z<4. This supports the scenario where gravitational interactions may cause starburst events, albeit not in a systematic way. The similarity between the regular and the irregular gas kinematic samples in the sSFR-M$_{*}$ relation found in this work may imply that star formation activity is (in general) independent of the galaxy kinematics in clusters at intermediate redshift.

We consider two different explanations for these results. Interactions, such as starvation, act gradually over the gas reservoir on a relatively long timescale, increasing in strength towards the cluster core regions, and acting as a cumulative effect over the physical properties of a given galaxy. In a similar way, RPS may become an important effect when the galaxy approaches the innermost regions of the cluster, acting for several hundred million years. On the other hand, mergers and close encounters can be considered   instantaneous events, and their effects are strongly dependant on the previous properties of the objects involved. Cluster complexes as big as RXJ1347 grow by accreting infalling groups of galaxies that have been subject to a certain degree of pre-processing, causing a partial depletion of their gas reservoir, and thus making it more difficult to trigger a strong starburst after a merging event. Furthermore, starbursts are very short-lived events, lasting no more than a few hundred million years at most. In a context where the gas reservoir of the hosts has  already been partially depleted due to its interaction with the ICM for some time, the duration of a starburst event could be even shorter. This means that at a given time, very few objects may be experiencing this phase, making their detection difficult. 

Recent results by the STAGES collaboration show signs of star formation enhancement in some cases of RPS detected thanks to the analysis of deep HST photometry and narrow-band H$_{\alpha}$ imaging in the Abell 901/902 cluster complex at z$\sim$0.16 (\citealt{Roman-Oliveira19}). In that work the average $\log{(sSFR)}$ value of RPS-galaxies is $\sim$0.2 dex higher than that measured for their cluster star-forming mother-sample at a fixed M$_\ast$. If such cases are present in our study, the kinematic analysis we carried out does not allow us to unambiguously identify them, which may contribute to the scatter of our cluster sample in the sSFR-log(M$_*$) relation.

We also examine the influence of galaxy kinematics on the AGN activity. Recent results by \citealt{Poggianti17} suggest that some interactions such as RPS feed the central black hole of massive disk galaxies, triggering AGN activity. Due to the characteristics of our sample, we choose a more simplistic approach by studying the frequency of AGNs in galaxies that display regular and distorted gas kinematics in the cluster environment. Our results, though limited due to the size of our sample, show that the fraction of AGNs is very similar (and very low) for both classes. It seems that in general, interactions in the cluster environment are not likely to channel gas from the outskirts of the galaxy towards its central regions. Thus, the appearance of AGN is probably dominated by the effects of the mass growth, as  happens in the field, though attenuated by the partial depletion of the cold gas reservoir due to the influence of the ICM.

Finally, we carried out an exploratory analysis of the stellar-to-halo mass relation using the method outlined by our group (\citealt{Lampichler17}) and in \cite{Conselice18}. We find that our cluster and field samples follow the theoretical predictions proposed by \cite{Moster13} at the redshift of our targets. Similarly, \cite{Niemiec18} studied this relation for satellite cluster galaxies using the Illustris simulations. The authors follow the evolutionary path of every satellite galaxy since it is accreted by the cluster gravitational potential well, finding that satellite cluster galaxies are shifted towards lower halo masses compared to the results for central galaxies. We do not see this trend in our observational study, although several circumstances do not allow us to discard its existence. First, the method outlined to compute $M_{h}$ relies on a series of approximations that have  only been tested for field galaxies and in the local universe for the most part. The dark matter halos of satellite galaxies in clusters interact with each other and with the main halo of the cluster (i.e., the central galaxy halo) during their infalling period. The consequences of these interactions may result in the partial stripping or merging of the dark matter component of satellite galaxies, which in turn may invalidate the conversion between ($R_h$,$V_h$) and ($R_e$, $V_{max}$) in the cluster. Second, the conversion itself is model dependent, which significantly increases the systematic uncertainties in the determination of $M_{h}$, and contributes to the scatter of the sample. The conversion between V$_{max}$ and v$_h$ ranges between 1.1 and 1.5 for different studies (\citealt{Dutton10}, \citealt{Papastergis11}, \citealt{Cattaneo14}). In the same way, the conversion between the half-light and the half-mass radius of late-type galaxies has different values in the literature (e.g., R$_e$=R$_m$ in \citealt{Lanyon-Foster12}, while R$_e$=1.33R$_m$ in \citealt{Szomoru13}). We estimate that these model-dependent uncertainties add 0.1-0.3 dex to the error budget computed in the determination of halo masses. Third, the small number statistics of our sample combined with cluster-specific processes that affect the distribution of matter within the galaxy and its stellar-mass budget can introduce additional biases when comparing the cluster and field population. For example, the bulge growth of late-type galaxies (\citealt{Kuchner17}) or the effect of additional subtle interactions in clusters add further uncertainties on the transformation between R$_e$ and R$_m$ in comparison with the field. Due to the non-trivial implications that the combinations of these effects may have for our sample, we do not attempt to explain the possible offset between cluster and field galaxies visible in our $M_*-M_h$ diagram and conclude that statistically significant studies in clusters at different epochs are required to shed light onto the stellar-to-halo mass evolution of cluster galaxies.

\section{Conclusions} 

In this work, we   used the VIMOS/VLT spectrograph to investigate the kinematics of a sample of galaxies in the RXJ1347 cluster complex. In particular, we   studied the possible link between the kinematic asymmetries, the star formation rate, and the gas excitation of the gas disk component. Our kinematic analysis used the asymmetry index $A$ (\citealt{Dale01} and \citealt{Bosch1}) to measure the degree of disturbance of the gas component of our galaxies. Those objects with regular enough kinematics according to this index were included in our Tully-Fisher and stellar-to-halo mass analysis, while those that show significant distortions were the focus of a subsequent star formation and AGN activity analysis. We compared our results with reference samples in the local universe and at intermediate redshift. Our main findings can be summarized as follows:

\begin{enumerate}
    \item The fraction of galaxies that display strong kinematic asymmetries in the cluster (42.0$\%$) is higher than in the field (23.3$\%$). A possible explanation for this difference is the influence of cluster-specific interactions. However, this fraction rises to 41.9$\%$ in the field when we combine galaxies with strong irregularities (irregulars) and those with mild but perceptible disturbances (affected). This may be caused by a higher accretion activity and minor merger frequency than expected in the field at intermediate redshifts. This scenario has been proposed by some authors in the past (\citealt{Puech08}, \citealt{Yang08}, and \citealt{Kutdemir10}).
    
    \item Cluster galaxies with sufficiently regular rotation curves ($A\leqslant50$) display a moderate albeit non-significant brightening in the B-band TFR ($\Delta M_B$=-0.7$\pm$0.8 mag) and non-significant evolution in the $M_{*}$-TFR ($\Delta M_{*}$=-0.2$\pm$0.4 mag). In the field, we find very similar results in both scaling relations at intermediate redshift $\Delta M_B$=-0.6$\pm$0.7 mag and $\Delta M_{*}$=-0.1$\pm$0.3 mag. These results suggest that cluster and field galaxies behave similarly in the different representations of the TFR at this redshift. The reported B-band evolution can be explained by the successively younger stellar populations towards longer lookback time, while our results in the $M_{*}$-TFR agree with recent observational studies in the field and cluster environment that reported no significant evolution up to z=1 (\citealt{Tiley19} and \citealt{Pelliccia19}, respectively).
    
    \item We report average lower sSFR values for our cluster sample compared to the field expectations given by the main sequence of star-forming galaxies (\citealt{Peng10}) at z$\sim$0.45. In particular, we find slightly lower sSFR values for those galaxies classified as irregulars according to their asymmetry index ($A\geqslant50$) with respect to those classified as fully regular ($A\leqslant25$). We do not see signs of a star formation burst for galaxies that may have suffered an interaction in their recent past in clusters at intermediate redshift.
    
    \item There is no correlation between the kinematic classification of our galaxies and AGN activity measured through the MEx diagram (\citealt{Juneau11}).
    
    \item We explored the stellar-to-halo mass relation for our sample of cluster and field galaxies at intermediate redshift. Our results agree with the theoretical prediction proposed by \citealt{Moster13} parameterized for z=0.45. However, some cluster galaxies display smaller stellar masses for a given halo mass compared to the field and in contrast with results from hydrodynamics simulations (\citealt{Niemiec18}). However, the assumptions made to estimate $M_{h}$ do not take into account the effects of the cluster environment over the dark matter halos of satellite galaxies. This, together with the low number of objects studied and their scatter, does not allow us to investigate the origin of this trend in a systematic way. Additional observations are required to improve our understanding of the stellar-to-halo mass relation in clusters. 
\end{enumerate}

After several decades of environmental studies, many aspects of galaxy evolution are still not well understood by the astronomical community, even at low to intermediate redshift. We emphasize the importance of carrying out comprehensive studies that investigate galaxy evolution from different perspectives (i.e., with respect to stellar population properties, morphologies, kinematics, etc.) and making use of large data sets. In particular, the use of IFU observations in comparison with high-resolution simulations will be of key importance in order to disentangle the influence of different cluster-specific interactions over the physical properties of galaxies in the near future.

\begin{acknowledgements}

The authors thank the anonymous referee for providing useful and constructive feedback that helped us to improve this manuscript during the reviewing process.

\end{acknowledgements}

\bibliographystyle{aa.bst} 
\bibliography{references.bib} 

\begin{thebibliography}{97}
\expandafter\ifx\csname natexlab\endcsname\relax\def\natexlab#1{#1}\fi

\bibitem[{{Adam} {et~al.}(2018){Adam}, {Hahn}, {Ruppin}, {Ade}, {Andr{\'e}},
  {Arnaud}, {Bartalucci}, {Beelen}, {Beno{\^\i}t}, {Bideaud}, {Billot},
  {Bourrion}, {Calvo}, {Catalano}, {Coiffard}, {Comis}, {D'Addabbo},
  {D{\'e}sert}, {Doyle}, {Ferrari}, {Goupy}, {Kramer}, {Lagache}, {Leclercq},
  {Lestrade}, {Mac{\'\i}as-P{\'e}rez}, {Martinez Aviles}, {Martizzi},
  {Maurogordato}, {Mauskopf}, {Mayet}, {Monfardini}, {Pajot}, {Pascale},
  {Perotto}, {Pisano}, {Pointecouteau}, {Ponthieu}, {Pratt}, {Rev{\'e}ret},
  {Ricci}, {Ritacco}, {Rodriguez}, {Romero}, {Roussel}, {Schuster}, {Sievers},
  {Triqueneaux}, {Tucker}, {Wu}, \& {Zylka}}]{Adam18}
{Adam}, R., {Hahn}, O., {Ruppin}, F., {et~al.} 2018, \aap, 614, A118

\bibitem[{{Arnouts} \& {Ilbert}(2011)}]{Arnouts2011}
{Arnouts}, S. \& {Ilbert}, O. 2011, Astrophysics Source Code Library
  [\eprint[ascl]{1108.009}]

\bibitem[{{Baldry} {et~al.}(2006){Baldry}, {Balogh}, {Bower}, {Glazebrook},
  {Nichol}, {Bamford}, \& {Budavari}}]{Baldry06}
{Baldry}, I.~K., {Balogh}, M.~L., {Bower}, R.~G., {et~al.} 2006, \mnras, 373,
  469

\bibitem[{{Baldwin} {et~al.}(1981){Baldwin}, {Phillips}, \&
  {Terlevich}}]{Baldwin81}
{Baldwin}, J.~A., {Phillips}, M.~M., \& {Terlevich}, R. 1981, \pasp, 93, 5

\bibitem[{{Bamford} {et~al.}(2005){Bamford}, {Milvang-Jensen},
  {Arag{\'o}n-Salamanca}, \& {Simard}}]{Bamford05}
{Bamford}, S.~P., {Milvang-Jensen}, B., {Arag{\'o}n-Salamanca}, A., \&
  {Simard}, L. 2005, \mnras, 361, 109

\bibitem[{{Bertin} \& {Arnouts}(1996)}]{Bertin96}
{Bertin}, E. \& {Arnouts}, S. 1996, \aaps, 117, 393

\bibitem[{{B{\"o}hm} \& {Ziegler}(2016)}]{Boehm16}
{B{\"o}hm}, A. \& {Ziegler}, B.~L. 2016, \aap, 592, A64

\bibitem[{{B{\"o}hm} {et~al.}(2004){B{\"o}hm}, {Ziegler}, {Saglia}, {Bender},
  {Fricke}, {Gabasch}, {Heidt}, {Mehlert}, {Noll}, \& {Seitz}}]{Boehm04}
{B{\"o}hm}, A., {Ziegler}, B.~L., {Saglia}, R.~P., {et~al.} 2004, \aap, 420, 97

\bibitem[{{B{\"o}sch} {et~al.}(2013{\natexlab{a}}){B{\"o}sch}, {B{\"o}hm},
  {Wolf}, {Arag{\'o}n-Salamanca}, {Barden}, {Gray}, {Ziegler}, {Schindler}, \&
  {Balogh}}]{Bosch1}
{B{\"o}sch}, B., {B{\"o}hm}, A., {Wolf}, C., {et~al.} 2013{\natexlab{a}}, \aap,
  549, A142

\bibitem[{{B{\"o}sch} {et~al.}(2013{\natexlab{b}}){B{\"o}sch}, {B{\"o}hm},
  {Wolf}, {Arag{\'o}n-Salamanca}, {Ziegler}, {Barden}, {Gray}, {Balogh},
  {Meisenheimer}, \& {Schindler}}]{Bosch2}
{B{\"o}sch}, B., {B{\"o}hm}, A., {Wolf}, C., {et~al.} 2013{\natexlab{b}}, \aap,
  554, A97

\bibitem[{{Bruzual} \& {Charlot}(2003)}]{BC03}
{Bruzual}, G. \& {Charlot}, S. 2003, \mnras, 344, 1000

\bibitem[{{Butcher} \& {Oemler}(1978)}]{Butcher78}
{Butcher}, H. \& {Oemler}, Jr., A. 1978, \apj, 219, 18

\bibitem[{{Calzetti} {et~al.}(2000){Calzetti}, {Armus}, {Bohlin}, {Kinney},
  {Koornneef}, \& {Storchi-Bergmann}}]{Calzetti2000}
{Calzetti}, D., {Armus}, L., {Bohlin}, R.~C., {et~al.} 2000, \apj, 533, 682

\bibitem[{{Casey} {et~al.}(2017){Casey}, {Cooray}, {Killi}, {Capak}, {Chen},
  {Hung}, {Kartaltepe}, {Sanders}, \& {Scoville}}]{Casey17}
{Casey}, C.~M., {Cooray}, A., {Killi}, M., {et~al.} 2017, \apj, 840, 101

\bibitem[{{Cattaneo} {et~al.}(2011){Cattaneo}, {Mamon}, {Warnick}, \&
  {Knebe}}]{Cattaneo11}
{Cattaneo}, A., {Mamon}, G.~A., {Warnick}, K., \& {Knebe}, A. 2011, \aap, 533,
  A5

\bibitem[{{Cattaneo} {et~al.}(2014){Cattaneo}, {Salucci}, \&
  {Papastergis}}]{Cattaneo14}
{Cattaneo}, A., {Salucci}, P., \& {Papastergis}, E. 2014, \apj, 783, 66

\bibitem[{{Chabrier}(2003)}]{Chabrier03}
{Chabrier}, G. 2003, \pasp, 115, 763

\bibitem[{{Chiu} {et~al.}(2018){Chiu}, {Umetsu}, {Sereno}, {Ettori},
  {Meneghetti}, {Merten}, {Sayers}, \& {Zitrin}}]{Chiu18}
{Chiu}, I.~N., {Umetsu}, K., {Sereno}, M., {et~al.} 2018, \apj, 860, 126

\bibitem[{{Ciocan} {et~al.}(2019){Ciocan}, {Maier}, {Ziegler}, \&
  {Verdugo}}]{Ciocan19}
{Ciocan}, B.~I., {Maier}, C., {Ziegler}, B.~L., \& {Verdugo}, M. 2019, arXiv
  e-prints, arXiv:1909.07988

\bibitem[{{Conselice} {et~al.}(2018){Conselice}, {Twite}, {Palamara}, \&
  {Hartley}}]{Conselice18}
{Conselice}, C.~J., {Twite}, J.~W., {Palamara}, D.~P., \& {Hartley}, W. 2018,
  \apj, 863, 42

\bibitem[{{Courteau}(1997)}]{Courteau}
{Courteau}, S. 1997, \aj, 114, 2402

\bibitem[{{Dale} {et~al.}(2001){Dale}, {Giovanelli}, {Haynes}, {Hardy}, \&
  {Campusano}}]{Dale01}
{Dale}, D.~A., {Giovanelli}, R., {Haynes}, M.~P., {Hardy}, E., \& {Campusano},
  L.~E. 2001, \aj, 121, 1886

\bibitem[{{Dannerbauer} {et~al.}(2014){Dannerbauer}, {Kurk}, {De Breuck},
  {Wylezalek}, {Santos}, {Koyama}, {Seymour}, {Tanaka}, {Hatch}, \&
  {Altieri}}]{Dannerbauer14}
{Dannerbauer}, H., {Kurk}, J.~D., {De Breuck}, C., {et~al.} 2014, \aap, 570,
  A55

\bibitem[{{Darvish} {et~al.}(2016){Darvish}, {Mobasher}, {Sobral}, {Rettura},
  {Scoville}, {Faisst}, \& {Capak}}]{Darvish16}
{Darvish}, B., {Mobasher}, B., {Sobral}, D., {et~al.} 2016, \apj, 825, 113

\bibitem[{{Dressler}(1980)}]{Dressler80}
{Dressler}, A. 1980, \apj, 236, 351

\bibitem[{{Dutton} {et~al.}(2010){Dutton}, {Conroy}, {van den Bosch}, {Prada},
  \& {More}}]{Dutton10}
{Dutton}, A.~A., {Conroy}, C., {van den Bosch}, F.~C., {Prada}, F., \& {More},
  S. 2010, \mnras, 407, 2

\bibitem[{{Dutton} {et~al.}(2011){Dutton}, {van den Bosch}, {Faber}, {Simard},
  {Kassin}, {Koo}, {Bundy}, {Huang}, {Weiner}, {Cooper}, {Newman}, {Mozena}, \&
  {Koekemoer}}]{Dutton11}
{Dutton}, A.~A., {van den Bosch}, F.~C., {Faber}, S.~M., {et~al.} 2011, \mnras,
  410, 1660

\bibitem[{{Ferrero} {et~al.}(2017){Ferrero}, {Navarro}, {Abadi}, {Sales},
  {Bower}, {Crain}, {Frenk}, {Schaller}, {Schaye}, \& {Theuns}}]{Ferrero17}
{Ferrero}, I., {Navarro}, J.~F., {Abadi}, M.~G., {et~al.} 2017, \mnras, 464,
  4736

\bibitem[{{Fogarty} {et~al.}(2017){Fogarty}, {Postman}, {Larson}, {Donahue}, \&
  {Moustakas}}]{Fogarty17}
{Fogarty}, K., {Postman}, M., {Larson}, R., {Donahue}, M., \& {Moustakas}, J.
  2017, \apj, 846, 103

\bibitem[{{Ghirardini} {et~al.}(2017){Ghirardini}, {Ettori}, {Amodeo},
  {Capasso}, \& {Sereno}}]{Ghirardini17}
{Ghirardini}, V., {Ettori}, S., {Amodeo}, S., {Capasso}, R., \& {Sereno}, M.
  2017, \aap, 604, A100

\bibitem[{{Gilbank} {et~al.}(2010){Gilbank}, {Baldry}, {Balogh}, {Glazebrook},
  \& {Bower}}]{Gilbank11}
{Gilbank}, D.~G., {Baldry}, I.~K., {Balogh}, M.~L., {Glazebrook}, K., \&
  {Bower}, R.~G. 2010, \mnras, 405, 2594

\bibitem[{{Giovanelli} {et~al.}(1995){Giovanelli}, {Haynes}, {Salzer},
  {Wegner}, {da Costa}, \& {Freudling}}]{Giovanelli95}
{Giovanelli}, R., {Haynes}, M.~P., {Salzer}, J.~J., {et~al.} 1995, \aj, 110,
  1059

\bibitem[{{Gonzalez} {et~al.}(2001){Gonzalez}, {Zaritsky}, {Dalcanton}, \&
  {Nelson}}]{Gonzalez01}
{Gonzalez}, A.~H., {Zaritsky}, D., {Dalcanton}, J.~J., \& {Nelson}, A. 2001,
  \apjs, 137, 117

\bibitem[{{Haines} {et~al.}(2015){Haines}, {Pereira}, {Smith}, {Egami},
  {Babul}, {Finoguenov}, {Ziparo}, {McGee}, {Rawle}, \& {Okabe}}]{Haines15}
{Haines}, C.~P., {Pereira}, M.~J., {Smith}, G.~P., {et~al.} 2015, \apj, 806,
  101

\bibitem[{{Harrison} {et~al.}(2017){Harrison}, {Johnson}, {Swinbank}, {Stott},
  {Bower}, {Smail}, {Tiley}, {Bunker}, {Cirasuolo}, {Sobral}, {Sharples},
  {Best}, {Bureau}, {Jarvis}, \& {Magdis}}]{Harrison17}
{Harrison}, C.~M., {Johnson}, H.~L., {Swinbank}, A.~M., {et~al.} 2017, \mnras,
  467, 1965

\bibitem[{{Heidmann} {et~al.}(1972){Heidmann}, {Heidmann}, \& {de
  Vaucouleurs}}]{Heidmann72}
{Heidmann}, J., {Heidmann}, N., \& {de Vaucouleurs}, G. 1972, \memras, 75, 85

\bibitem[{{Hoyos} {et~al.}(2016){Hoyos}, {Arag{\'o}n-Salamanca}, {Gray},
  {Wolf}, {Maltby}, {Bell}, {B{\"o}hm}, \& {Jogee}}]{Hoyos16}
{Hoyos}, C., {Arag{\'o}n-Salamanca}, A., {Gray}, M.~E., {et~al.} 2016, \mnras,
  455, 295

\bibitem[{{Ilbert} {et~al.}(2006){Ilbert}, {Arnouts}, {McCracken},
  {Bolzonella}, {Bertin}, {Le F{\`e}vre}, {Mellier}, {Zamorani}, {Pell{\`o}},
  {Iovino}, {Tresse}, {Le Brun}, {Bottini}, {Garilli}, {Maccagni}, {Picat},
  {Scaramella}, {Scodeggio}, {Vettolani}, {Zanichelli}, {Adami}, {Bardelli},
  {Cappi}, {Charlot}, {Ciliegi}, {Contini}, {Cucciati}, {Foucaud}, {Franzetti},
  {Gavignaud}, {Guzzo}, {Marano}, {Marinoni}, {Mazure}, {Meneux}, {Merighi},
  {Paltani}, {Pollo}, {Pozzetti}, {Radovich}, {Zucca}, {Bondi}, {Bongiorno},
  {Busarello}, {de La Torre}, {Gregorini}, {Lamareille}, {Mathez}, {Merluzzi},
  {Ripepi}, {Rizzo}, \& {Vergani}}]{Ilbert2006}
{Ilbert}, O., {Arnouts}, S., {McCracken}, H.~J., {et~al.} 2006, \aap, 457, 841

\bibitem[{{J{\o}rgensen} {et~al.}(2017){J{\o}rgensen}, {Chiboucas}, {Berkson},
  {Smith}, {Takamiya}, \& {Villaume}}]{Jorgensen17}
{J{\o}rgensen}, I., {Chiboucas}, K., {Berkson}, E., {et~al.} 2017, \aj, 154,
  251

\bibitem[{{Juneau} {et~al.}(2014){Juneau}, {Bournaud}, {Charlot}, {Daddi},
  {Elbaz}, {Trump}, {Brinchmann}, {Dickinson}, {Duc}, {Gobat}, {Jean-Baptiste},
  {Le Floc'h}, {Lehnert}, {Pacifici}, {Pannella}, \& {Schreiber}}]{Juneau14}
{Juneau}, S., {Bournaud}, F., {Charlot}, S., {et~al.} 2014, \apj, 788, 88

\bibitem[{{Juneau} {et~al.}(2011){Juneau}, {Dickinson}, {Alexander}, \&
  {Salim}}]{Juneau11}
{Juneau}, S., {Dickinson}, M., {Alexander}, D.~M., \& {Salim}, S. 2011, \apj,
  736, 104

\bibitem[{{Kitayama} {et~al.}(2016){Kitayama}, {Ueda}, {Takakuwa}, {Tsutsumi},
  {Komatsu}, {Akahori}, {Iono}, {Izumi}, {Kawabe}, {Kohno}, {Matsuo}, {Ota},
  {Suto}, {Takizawa}, \& {Yoshikawa}}]{Kitayama16}
{Kitayama}, T., {Ueda}, S., {Takakuwa}, S., {et~al.} 2016, Publications of the
  Astronomical Society of Japan, 68, 88

\bibitem[{{Knapen} {et~al.}(2015){Knapen}, {Cisternas}, \&
  {Querejeta}}]{Knapen15}
{Knapen}, J.~H., {Cisternas}, M., \& {Querejeta}, M. 2015, \mnras, 454, 1742

\bibitem[{{Kravtsov}(2013)}]{Kravtsov13}
{Kravtsov}, A.~V. 2013, \apjl, 764, L31

\bibitem[{{Kronberger} {et~al.}(2008){Kronberger}, {Kapferer},
  {Unterguggenberger}, {Schindler}, \& {Ziegler}}]{Kronberger08b}
{Kronberger}, T., {Kapferer}, W., {Unterguggenberger}, S., {Schindler}, S., \&
  {Ziegler}, B.~L. 2008, \aap, 483, 783

\bibitem[{{Kuchner} {et~al.}(2017){Kuchner}, {Ziegler}, {Verdugo}, {Bamford},
  \& {H{\"a}u{\ss}ler}}]{Kuchner17}
{Kuchner}, U., {Ziegler}, B., {Verdugo}, M., {Bamford}, S., \&
  {H{\"a}u{\ss}ler}, B. 2017, \aap, 604, A54

\bibitem[{{Kutdemir} {et~al.}(2010){Kutdemir}, {Ziegler}, {Peletier}, {Da
  Rocha}, {B{\"o}hm}, \& {Verdugo}}]{Kutdemir10}
{Kutdemir}, E., {Ziegler}, B.~L., {Peletier}, R.~F., {et~al.} 2010, \aap, 520,
  A109

\bibitem[{{Lagan{\'a}} \& {Ulmer}(2018)}]{Lagana18}
{Lagan{\'a}}, T.~F. \& {Ulmer}, M.~P. 2018, \mnras, 475, 523

\bibitem[{{Lampichler} {et~al.}(2017){Lampichler}, {Maier}, \&
  {Ziegler}}]{Lampichler17}
{Lampichler}, N., {Maier}, C., \& {Ziegler}, B. 2017, arXiv e-prints
  [\eprint[arXiv]{1707.09838}]

\bibitem[{{Lanyon-Foster} {et~al.}(2012){Lanyon-Foster}, {Conselice}, \&
  {Merrifield}}]{Lanyon-Foster12}
{Lanyon-Foster}, M.~M., {Conselice}, C.~J., \& {Merrifield}, M.~R. 2012,
  \mnras, 424, 1852

\bibitem[{{Maier} {et~al.}(2016){Maier}, {Kuchner}, {Ziegler}, {Verdugo},
  {Balestra}, {Girardi}, {Mercurio}, {Rosati}, {Fritz}, {Grillo}, {Nonino}, \&
  {Sartoris}}]{Maier16}
{Maier}, C., {Kuchner}, U., {Ziegler}, B.~L., {et~al.} 2016, \aap, 590, A108

\bibitem[{{Maier} {et~al.}(2019){Maier}, {Ziegler}, {Haines}, \&
  {Smith}}]{Maier19}
{Maier}, C., {Ziegler}, B.~L., {Haines}, C.~P., \& {Smith}, G.~P. 2019, \aap,
  621, A131

\bibitem[{{Moran} {et~al.}(2007){Moran}, {Miller}, {Treu}, {Ellis}, \&
  {Smith}}]{Moran07}
{Moran}, S.~M., {Miller}, N., {Treu}, T., {Ellis}, R.~S., \& {Smith}, G.~P.
  2007, \apj, 659, 1138

\bibitem[{{Mortlock} {et~al.}(2013){Mortlock}, {Conselice}, {Hartley},
  {Ownsworth}, {Lani}, {Bluck}, {Almaini}, {Duncan}, {van der Wel},
  {Koekemoer}, {Dekel}, {Dav{\'e}}, {Ferguson}, {de Mello}, {Newman}, {Faber},
  {Grogin}, {Kocevski}, \& {Lai}}]{Mortlock13}
{Mortlock}, A., {Conselice}, C.~J., {Hartley}, W.~G., {et~al.} 2013, \mnras,
  433, 1185

\bibitem[{{Moster} {et~al.}(2013){Moster}, {Naab}, \& {White}}]{Moster13}
{Moster}, B.~P., {Naab}, T., \& {White}, S.~D.~M. 2013, \mnras, 428, 3121

\bibitem[{{Nakamura} {et~al.}(2006){Nakamura}, {Arag{\'o}n-Salamanca},
  {Milvang-Jensen}, {Arimoto}, {Ikuta}, \& {Bamford}}]{Nakamura06}
{Nakamura}, O., {Arag{\'o}n-Salamanca}, A., {Milvang-Jensen}, B., {et~al.}
  2006, \mnras, 366, 144

\bibitem[{{Niemiec} {et~al.}(2018){Niemiec}, {Jullo}, {Giocoli}, {Limousin}, \&
  {Jauzac}}]{Niemiec18}
{Niemiec}, A., {Jullo}, E., {Giocoli}, C., {Limousin}, M., \& {Jauzac}, M.
  2018, arXiv e-prints [\eprint[arXiv]{1811.04996}]

\bibitem[{{Paccagnella} {et~al.}(2019){Paccagnella}, {Vulcani}, {Poggianti},
  {Moretti}, {Fritz}, {Gullieuszik}, \& {Fasano}}]{Paccagnella19}
{Paccagnella}, A., {Vulcani}, B., {Poggianti}, B.~M., {et~al.} 2019, \mnras,
  482, 881

\bibitem[{{Papastergis} {et~al.}(2011){Papastergis}, {Martin}, {Giovanelli}, \&
  {Haynes}}]{Papastergis11}
{Papastergis}, E., {Martin}, A.~M., {Giovanelli}, R., \& {Haynes}, M.~P. 2011,
  \apj, 739, 38

\bibitem[{{Paulino-Afonso} {et~al.}(2018){Paulino-Afonso}, {Sobral}, {Darvish},
  {Ribeiro}, {Stroe}, {Best}, {Afonso}, \& {Matsuda}}]{Paulino-Afonso18}
{Paulino-Afonso}, A., {Sobral}, D., {Darvish}, B., {et~al.} 2018, \aap, 620,
  A186

\bibitem[{{Pearson} {et~al.}(2019){Pearson}, {Wang}, {Alpaslan}, {Baldry},
  {Bilicki}, {Brown}, {Grootes}, {Holwerda}, {Kitching}, {Kruk}, \& {van der
  Tak}}]{Pearson19}
{Pearson}, W.~J., {Wang}, L., {Alpaslan}, M., {et~al.} 2019, \aap, 631, A51

\bibitem[{{Pelliccia} {et~al.}(2019){Pelliccia}, {Lemaux}, {Tomczak}, {Lubin},
  {Shen}, {Epinat}, {Wu}, {Gal}, {Rumbaugh}, {Kocevski}, {Tresse}, \&
  {Squires}}]{Pelliccia19}
{Pelliccia}, D., {Lemaux}, B.~C., {Tomczak}, A.~R., {et~al.} 2019, \mnras, 482,
  3514

\bibitem[{{Pelliccia} {et~al.}(2017){Pelliccia}, {Tresse}, {Epinat}, {Ilbert},
  {Scoville}, {Amram}, {Lemaux}, \& {Zamorani}}]{Pelliccia17}
{Pelliccia}, D., {Tresse}, L., {Epinat}, B., {et~al.} 2017, \aap, 599, A25

\bibitem[{{Peng} {et~al.}(2002){Peng}, {Ho}, {Impey}, \& {Rix}}]{Peng02}
{Peng}, C.~Y., {Ho}, L.~C., {Impey}, C.~D., \& {Rix}, H.-W. 2002, \aj, 124, 266

\bibitem[{{Peng} {et~al.}(2010){Peng}, {Lilly}, {Kova{\v c}}, {Bolzonella},
  {Pozzetti}, {Renzini}, {Zamorani}, {Ilbert}, {Knobel}, {Iovino}, {Maier},
  {Cucciati}, {Tasca}, {Carollo}, {Silverman}, {Kampczyk}, {de Ravel},
  {Sanders}, {Scoville}, {Contini}, {Mainieri}, {Scodeggio}, {Kneib}, {Le
  F{\`e}vre}, {Bardelli}, {Bongiorno}, {Caputi}, {Coppa}, {de la Torre},
  {Franzetti}, {Garilli}, {Lamareille}, {Le Borgne}, {Le Brun}, {Mignoli},
  {Perez Montero}, {Pello}, {Ricciardelli}, {Tanaka}, {Tresse}, {Vergani},
  {Welikala}, {Zucca}, {Oesch}, {Abbas}, {Barnes}, {Bordoloi}, {Bottini},
  {Cappi}, {Cassata}, {Cimatti}, {Fumana}, {Hasinger}, {Koekemoer},
  {Leauthaud}, {Maccagni}, {Marinoni}, {McCracken}, {Memeo}, {Meneux}, {Nair},
  {Porciani}, {Presotto}, \& {Scaramella}}]{Peng10}
{Peng}, Y.-j., {Lilly}, S.~J., {Kova{\v c}}, K., {et~al.} 2010, \apj, 721, 193

\bibitem[{{P{\'e}rez-Mart{\'{\i}}nez}
  {et~al.}(2017){P{\'e}rez-Mart{\'{\i}}nez}, {Ziegler}, {Verdugo}, {B{\"o}hm},
  \& {Tanaka}}]{JM17}
{P{\'e}rez-Mart{\'{\i}}nez}, J.~M., {Ziegler}, B., {Verdugo}, M., {B{\"o}hm},
  A., \& {Tanaka}, M. 2017, \aap, 605, A127

\bibitem[{{Petropoulou} {et~al.}(2012){Petropoulou}, {V{\'\i}lchez}, \&
  {Iglesias-P{\'a}ramo}}]{Petropoulou12}
{Petropoulou}, V., {V{\'\i}lchez}, J., \& {Iglesias-P{\'a}ramo}, J. 2012, \apj,
  749, 133

\bibitem[{{Petropoulou} {et~al.}(2011){Petropoulou}, {V{\'\i}lchez},
  {Iglesias-P{\'a}ramo}, {Papaderos}, {Magrini}, {Cedr{\'e}s}, \&
  {Reverte}}]{Petropoulou11}
{Petropoulou}, V., {V{\'\i}lchez}, J., {Iglesias-P{\'a}ramo}, J., {et~al.}
  2011, \apj, 734, 32

\bibitem[{{Poggianti} {et~al.}(2017){Poggianti}, {Jaff{\'e}}, {Moretti},
  {Gullieuszik}, {Radovich}, {Tonnesen}, {Fritz}, {Bettoni}, {Vulcani},
  {Fasano}, {Bellhouse}, {Hau}, \& {Omizzolo}}]{Poggianti17}
{Poggianti}, B.~M., {Jaff{\'e}}, Y.~L., {Moretti}, A., {et~al.} 2017, \nat,
  548, 304

\bibitem[{{Popesso} {et~al.}(2015){Popesso}, {Biviano}, {Finoguenov}, {Wilman},
  {Salvato}, {Magnelli}, {Gruppioni}, {Pozzi}, {Rodighiero}, {Ziparo}, {Berta},
  {Elbaz}, {Dickinson}, {Lutz}, {Altieri}, {Aussel}, {Cimatti}, {Fadda},
  {Ilbert}, {Le Floch}, {Nordon}, {Poglitsch}, \& {Xu}}]{Popesso15}
{Popesso}, P., {Biviano}, A., {Finoguenov}, A., {et~al.} 2015, \aap, 574, A105

\bibitem[{{Puech} {et~al.}(2008){Puech}, {Flores}, {Hammer}, {Yang}, {Neichel},
  {Lehnert}, {Chemin}, {Nesvadba}, {Epinat}, {Amram}, {Balkowski}, {Cesarsky},
  {Dannerbauer}, {di Serego Alighieri}, {Fuentes-Carrera}, {Guiderdoni},
  {Kembhavi}, {Liang}, {{\"O}stlin}, {Pozzetti}, {Ravikumar}, {Rawat},
  {Vergani}, {Vernet}, \& {Wozniak}}]{Puech08}
{Puech}, M., {Flores}, H., {Hammer}, F., {et~al.} 2008, \aap, 484, 173

\bibitem[{{Reyes} {et~al.}(2011){Reyes}, {Mandelbaum}, {Gunn}, {Pizagno}, \&
  {Lackner}}]{Reyes11}
{Reyes}, R., {Mandelbaum}, R., {Gunn}, J.~E., {Pizagno}, J., \& {Lackner},
  C.~N. 2011, \mnras, 417, 2347

\bibitem[{{Rodr{\'{\i}}guez del Pino} {et~al.}(2017){Rodr{\'{\i}}guez del
  Pino}, {Arag{\'o}n-Salamanca}, {Chies-Santos}, {Weinzirl}, {Bamford}, {Gray},
  {B{\"o}hm}, {Wolf}, \& {Maltby}}]{Bruno17}
{Rodr{\'{\i}}guez del Pino}, B., {Arag{\'o}n-Salamanca}, A., {Chies-Santos},
  A.~L., {et~al.} 2017, \mnras, 467, 4200

\bibitem[{{Roman-Oliveira} {et~al.}(2019){Roman-Oliveira}, {Chies-Santos},
  {Rodr{\'{\i}}guez del Pino}, {Arag{\'o}n-Salamanca}, {Gray}, \&
  {Bamford}}]{Roman-Oliveira19}
{Roman-Oliveira}, F.~V., {Chies-Santos}, A.~L., {Rodr{\'{\i}}guez del Pino},
  B., {et~al.} 2019, \mnras, 484, 892

\bibitem[{{Rubin} {et~al.}(1999){Rubin}, {Waterman}, \& {Kenney}}]{Rubin99}
{Rubin}, V.~C., {Waterman}, A.~H., \& {Kenney}, J. D.~P. 1999, \aj, 118, 236

\bibitem[{{Ruggiero} \& {Lima Neto}(2017)}]{Ruggiero17}
{Ruggiero}, R. \& {Lima Neto}, G.~B. 2017, \mnras, 468, 4107

\bibitem[{{Santos} {et~al.}(2013){Santos}, {Altieri}, {Popesso}, {Strazzullo},
  {Valtchanov}, {Berta}, {B{\"o}hringer}, {Conversi}, {Demarco}, {Edge},
  {Lidman}, {Lutz}, {Metcalfe}, {Mullis}, {Pintos-Castro},
  {S{\'a}nchez-Portal}, {Rawle}, {Rosati}, {Swinbank}, \& {Tanaka}}]{Santos13}
{Santos}, J.~S., {Altieri}, B., {Popesso}, P., {et~al.} 2013, \mnras, 433, 1287

\bibitem[{{Schindler} {et~al.}(1995){Schindler}, {Guzzo}, {Ebeling},
  {Boehringer}, {Chincarini}, {Collins}, {de Grandi}, {Neumann}, {Briel},
  {Shaver}, \& {Vettolani}}]{Schindler95}
{Schindler}, S., {Guzzo}, L., {Ebeling}, H., {et~al.} 1995, \aap, 299, L9

\bibitem[{{Schweizer}(2005)}]{Schweizer05}
{Schweizer}, F. 2005, Astrophysics and Space Science Library, Vol. 329,
  {Merger-Induced Starbursts}, ed. R.~{de Grijs} \& R.~M. {Gonz{\'a}lez
  Delgado}, 143

\bibitem[{{Socolovsky} {et~al.}(2018){Socolovsky}, {Almaini}, {Hatch}, {Wild},
  {Maltby}, {Hartley}, \& {Simpson}}]{Socolovsky18}
{Socolovsky}, M., {Almaini}, O., {Hatch}, N.~A., {et~al.} 2018, \mnras, 476,
  1242

\bibitem[{{Swinbank} {et~al.}(2017){Swinbank}, {Harrison}, {Trayford},
  {Schaller}, {Smail}, {Schaye}, {Theuns}, {Smit}, {Alexander}, {Bacon},
  {Bower}, {Contini}, {Crain}, {de Breuck}, {Decarli}, {Epinat}, {Fumagalli},
  {Furlong}, {Galametz}, {Johnson}, {Lagos}, {Richard}, {Vernet}, {Sharples},
  {Sobral}, \& {Stott}}]{Swinbank17}
{Swinbank}, A.~M., {Harrison}, C.~M., {Trayford}, J., {et~al.} 2017, \mnras,
  467, 3140

\bibitem[{{Szomoru} {et~al.}(2013){Szomoru}, {Franx}, {van Dokkum}, {Trenti},
  {Illingworth}, {Labb{\'e}}, \& {Oesch}}]{Szomoru13}
{Szomoru}, D., {Franx}, M., {van Dokkum}, P.~G., {et~al.} 2013, \apj, 763, 73

\bibitem[{{Teyssier} {et~al.}(2010){Teyssier}, {Chapon}, \&
  {Bournaud}}]{Teyssier10}
{Teyssier}, R., {Chapon}, D., \& {Bournaud}, F. 2010, \apjl, 720, L149

\bibitem[{{Tiley} {et~al.}(2019){Tiley}, {Bureau}, {Cortese}, {Harrison},
  {Johnson}, {Stott}, {Swinbank}, {Smail}, {Sobral}, {Bunker}, {Glazebrook},
  {Bower}, {Obreschkow}, {Bryant}, {Jarvis}, {Bland-Hawthorn}, {Magdis},
  {Medling}, {Sweet}, {Tonini}, {Turner}, {Sharples}, {Croom}, {Goodwin},
  {Konstantopoulos}, {Lorente}, {Lawrence}, {Mould}, {Owers}, \&
  {Richards}}]{Tiley19}
{Tiley}, A.~L., {Bureau}, M., {Cortese}, L., {et~al.} 2019, \mnras, 482, 2166

\bibitem[{{Tully} \& {Fisher}(1977)}]{Tully77}
{Tully}, R.~B. \& {Fisher}, J.~R. 1977, \aap, 54, 661

\bibitem[{{Tully} {et~al.}(1998){Tully}, {Pierce}, {Huang}, {Saunders},
  {Verheijen}, \& {Witchalls}}]{Tully98}
{Tully}, R.~B., {Pierce}, M.~J., {Huang}, J.-S., {et~al.} 1998, \aj, 115, 2264

\bibitem[{{{\"U}bler} {et~al.}(2017){{\"U}bler}, {F{\"o}rster Schreiber},
  {Genzel}, {Wisnioski}, {Wuyts}, {Lang}, {Naab}, {Burkert}, {van Dokkum},
  {Tacconi}, {Wilman}, {Fossati}, {Mendel}, {Beifiori}, {Belli}, {Bender},
  {Brammer}, {Chan}, {Davies}, {Fabricius}, {Galametz}, {Lutz}, {Momcheva},
  {Nelson}, {Saglia}, {Seitz}, \& {Tadaki}}]{Ubler17}
{{\"U}bler}, H., {F{\"o}rster Schreiber}, N.~M., {Genzel}, R., {et~al.} 2017,
  \apj, 842, 121

\bibitem[{{Ueda} {et~al.}(2018){Ueda}, {Kitayama}, {Oguri}, {Komatsu},
  {Akahori}, {Iono}, {Izumi}, {Kawabe}, {Kohno}, {Matsuo}, {Ota}, {Suto},
  {Takakuwa}, {Takizawa}, {Tsutsumi}, \& {Yoshikawa}}]{Ueda18}
{Ueda}, S., {Kitayama}, T., {Oguri}, M., {et~al.} 2018, \apj, 866, 48

\bibitem[{{Umetsu} {et~al.}(2014){Umetsu}, {Medezinski}, {Nonino}, {Merten},
  {Postman}, {Meneghetti}, {Donahue}, {Czakon}, {Molino}, {Seitz}, {Gruen},
  {Lemze}, {Balestra}, {Ben{\'{\i}}tez}, {Biviano}, {Broadhurst}, {Ford},
  {Grillo}, {Koekemoer}, {Melchior}, {Mercurio}, {Moustakas}, {Rosati}, \&
  {Zitrin}}]{Umetsu14}
{Umetsu}, K., {Medezinski}, E., {Nonino}, M., {et~al.} 2014, \apj, 795, 163

\bibitem[{{Umetsu} {et~al.}(2018){Umetsu}, {Sereno}, {Tam}, {Chiu}, {Fan},
  {Ettori}, {Gruen}, {Okumura}, {Medezinski}, {Donahue}, {Meneghetti}, {Frye},
  {Koekemoer}, {Broadhurst}, {Zitrin}, {Balestra}, {Ben{\'\i}tez}, {Higuchi},
  {Melchior}, {Mercurio}, {Merten}, {Molino}, {Nonino}, {Postman}, {Rosati},
  {Sayers}, \& {Seitz}}]{Umetsu18}
{Umetsu}, K., {Sereno}, M., {Tam}, S.-I., {et~al.} 2018, \apj, 860, 104

\bibitem[{{van der Wel} {et~al.}(2014){van der Wel}, {Franx}, {van Dokkum},
  {Skelton}, {Momcheva}, {Whitaker}, {Brammer}, {Bell}, {Rix}, {Wuyts},
  {Ferguson}, {Holden}, {Barro}, {Koekemoer}, {Chang}, {McGrath},
  {H{\"a}ussler}, {Dekel}, {Behroozi}, {Fumagalli}, {Leja}, {Lundgren},
  {Maseda}, {Nelson}, {Wake}, {Patel}, {Labb{\'e}}, {Faber}, {Grogin}, \&
  {Kocevski}}]{Vanderwel14}
{van der Wel}, A., {Franx}, M., {van Dokkum}, P.~G., {et~al.} 2014, \apj, 788,
  28

\bibitem[{{Verdugo} {et~al.}(2012){Verdugo}, {Lerchster}, {B{\"o}hringer},
  {Hildebrandt}, {Ziegler}, {Erben}, {Finoguenov}, \& {Chon}}]{Verdugo12}
{Verdugo}, M., {Lerchster}, M., {B{\"o}hringer}, H., {et~al.} 2012, \mnras,
  421, 1949

\bibitem[{{Vogt} {et~al.}(2004){Vogt}, {Haynes}, {Giovanelli}, \&
  {Herter}}]{Vogt04}
{Vogt}, N.~P., {Haynes}, M.~P., {Giovanelli}, R., \& {Herter}, T. 2004, \aj,
  127, 3325

\bibitem[{{Wetzel} {et~al.}(2013){Wetzel}, {Tinker}, {Conroy}, \& {van den
  Bosch}}]{Wetzel13}
{Wetzel}, A.~R., {Tinker}, J.~L., {Conroy}, C., \& {van den Bosch}, F.~C. 2013,
  \mnras, 432, 336

\bibitem[{{Whitaker} {et~al.}(2013){Whitaker}, {van Dokkum}, {Brammer},
  {Momcheva}, {Skelton}, {Franx}, {Kriek}, {Labb{\'e}}, {Fumagalli},
  {Lundgren}, {Nelson}, {Patel}, \& {Rix}}]{Whitaker13}
{Whitaker}, K.~E., {van Dokkum}, P.~G., {Brammer}, G., {et~al.} 2013, \apjl,
  770, L39

\bibitem[{{Yang} {et~al.}(2008){Yang}, {Flores}, {Hammer}, {Neichel}, {Puech},
  {Nesvadba}, {Rawat}, {Cesarsky}, {Lehnert}, {Pozzetti}, {Fuentes-Carrera},
  {Amram}, {Balkowski}, {Dannerbauer}, {di Serego Alighieri}, {Guiderdoni},
  {Kembhavi}, {Liang}, {{\"O}stlin}, {Ravikumar}, {Vergani}, {Vernet}, \&
  {Wozniak}}]{Yang08}
{Yang}, Y., {Flores}, H., {Hammer}, F., {et~al.} 2008, \aap, 477, 789

\bibitem[{{Ziegler} {et~al.}(2003){Ziegler}, {B{\"o}hm}, {J{\"a}ger}, {Heidt},
  \& {M{\"o}llenhoff}}]{Ziegler03}
{Ziegler}, B.~L., {B{\"o}hm}, A., {J{\"a}ger}, K., {Heidt}, J., \&
  {M{\"o}llenhoff}, C. 2003, \apjl, 598, L87

\end{thebibliography}

\begin{appendix}
\section{Additional material}
\label{appendix:a}
In this section we present the data tables containing all the relevant parameters of the cluster and field galaxies that were included in our TFR analysis, i.e., those galaxies classified as regular or affected attending to our gas kinematics asymmetry index criterion. We also display the observed and computed rotation curves for the same objects (Figs. \ref{foot} and \ref{foot2}). In addition, we add three examples of irregular galaxies ($A\geq50$) with varying degrees of distortion ($50\geq A\geq200$) to show the typical cases that this class encompass (Fig. \ref{foot3}).
\suppressfloats[t]
\begin{sidewaystable*}[h!]
\centering
\caption{General properties of the cluster galaxies included in our TFR analysis. Regular galaxies have identification names (ID) C1-C19, while affected galaxies are labeled  C20-C23. Column description: IDs, J2000 coordinates, redshift, BRI absolute magnitudes in the AB system, extinction in B-band, effective radius in the z-band, inclination angle of the galaxy, position angle of the galaxy with respect to the vertical axis, position angle of the slit with respect to the vertical axis, logarithmic stellar mass, star formation rate, maximum rotation velocity, and asymmetry index.} 
\begin{tabular}{cccccccccccccccccc}
\hline
\noalign{\vskip 0.1cm}
ID & RA & DEC & $z$ & $M_B$ & $M_{Rc}$ & $M_{Ic}$ & $A_B$ & $R_e$ & $i$ & $PA$ & $\theta$ & $\log{M_{\ast}}$ & SFR & $V_{max}$ & $V_{max,err}$ & $A$ & $A_{err}$   \\
   & (hh:mm:ss)  & (dd:mm:ss.s) &     & (mag) & (mag) & (mag) & (mag) & (kpc) & (º) & (º) & (º)  & & ($M_\odot/yr$) & (km/s) & (km/s) &  & \\
\noalign{\vskip 0.1cm}
\hline 
\hline 
\noalign{\vskip 0.2cm}
C1 & 13:48:29.2 & -11:34:39.1 & 0.4195 & -21.92 & -22.82 & -23.25 & -0.21 & 10.94 & 38.3 & -60.9 & -31.9 & 10.93 & 10.6 & 249.7 & 7.9  & 12.8 & 4.8  \\
C2 & 13:48:13.6 & -11:37:12.3 & 0.4457 & -20.62 & -21.57 & -22.01 & -0.67 & 5.49  & 78.0 & 14.5 & 20.9 &  10.46 & 3.6  & 204.7 & 12.4 & 22.8 & 9.0  \\
C3 & 13:48:17.9 & -11:38:27.4 & 0.4609 & -20.58 & -21.17 & -21.41 & -1.22 & 5.58  & 85.6 & -31.9 & -33.3 &  10.18 & 18.1 & 196.1 & 4.0  & 3.2 & 2.6   \\
C4 & 13:47:36.9 & -11:36:07.5 & 0.4551 & -20.01 & -20.43 & -20.58 & -0.16 & 5.79  & 39.7 & -28.9 & -30.2 &  9.61  & 1.9  & 128.9 & 7.5  & 8.4 & 9.1   \\
C5 & 13:47:47.0 & -11:37:31.7 & 0.4488 & -21.90 & -22.64 & -23.00 & -0.36 & 10.40 & 45.4 & 47.3 & 48.0 &  10.73 & 11.5 & 337.2 & 14.1 & 21.4 & 5.9 \\
C6 & 13:47:28.9 & -11:38:12.7 & 0.4700 & -20.42 & -20.75 & -20.85 & -0.25 & 5.77  & 49.5 & -39.7 & -42.0 &  9.58  & -    & 138.3 & 3.9  & 11.6 & 1.8 \\
C7 & 13:47:39.3 & -11:39:20.7 & 0.4521 & -20.45 & -21.05 & -21.33 & -0.59 & 6.66  & 64.0 & 11.1 & 13.0 &  9.95  & 2.9  & 148.1 & 2.8  & 11.5 & 4.9 \\ 
C8  & 13:47:04.2 & -11:51:50.2 & 0.4619 & -21.33 & -22.18 & -22.56 & -0.32 & 5.46 & 45.9 & -46.6 & -45.0 &  10.8  & 4.5  & 259.4 & 11.9 & 16.3 & 10.3  \\
C9 & 13:47:09.9 & -11:57:22.9 & 0.4306 & -20.77 & -21.51 & -21.81 & -0.28 & 4.36 & 44.9 & -26.9 & -39.0 &  10.2  & 2.3  & 225.4 & 7.9  & 16.4 & 2.8  \\
C10 & 13:46:34.7 & -11:51:30.2 & 0.4615 & -20.96 & -21.62 & -21.90 & -0.70 & 7.04 & 76.8 & -37.5 & -44.0 &  10.32 & 10.1 & 132.2 & 6.2  & 11.7 & 7.5  \\
C11 & 13:46:19.9 & -11:53:00.0 & 0.4683 & -20.32 & -21.38 & -21.86 & -0.60 & 4.45  & 64.8 & 30.4 & 30.5 &  10.55 & 1.2  & 190.9 & 18.5 & 8.1 & 5.1   \\
C12 & 13:46:30.8 & -11:53:43.0 & 0.4802 & -20.91 & -21.67 & -21.98 & -0.28 & 5.78  & 53.2 & 9.9 & 1.0 &  10.52 & 14.5 & 129.0 & 9.7  & 10.0 & 11.6  \\
C13 & 13:46:30.6 & -11:53:55.8 & 0.4733 & -21.26 & -22.15 & -22.54 & -0.56 & 8.95  & 72.4 & 29.7 & 45.0 &  10.85 & -    & 256.1 & 10.2 & 14.7 & 4.8 \\
C14 & 13:46:32.3 & -11:55:49.9 & 0.4555 & -20.17 & -20.73 & -20.94 & -0.47 & 3.20  & 71.8 & -10.2 & -5.1 &  9.91  & 3.14 & 105.3 & 3.5  & 22.5 & 3.0  \\
C15 & 13:46:28.3 & -11:56:52.7 & 0.4461 & -20.05 & -20.83 & -21.19 & -0.33 & 6.38  & 50.6 & -13.1 & -8.5 &  10.03 & 1.39 & 161.1 & 19.0 & 10.0 & 4.7 \\
C16 & 13:46:37.3 & -11:57:16.6 & 0.4473 & -20.06 & -20.80 & -21.12 & -0.31 & 5.28  & 71.0 & 32.8 & 36.2 &  10.02 & 1.58 & 167.2 & 7.4  & 16.4 & 6.1 \\
C17 & 13:46:51.6 & -11:47:24.1 & 0.4482 & -20.54 & -21.15 & -21.44 & -0.17 & 5.69  & 54.3 & 66.7 & 45.0 &  10.02 & -    & 114.6 & 10.5 & 20.1 & 6.5 \\
C18 & 13:46:16.2 & -11:47:01.6 & 0.4604 & -20.36 & -21.13 & -21.44 & -0.16 & 2.11  & 38.3 & 15.0 & 33.4 &  10.26 & 5.17 & 165.9 & 4.6  & 11.5 & 13.5 \\
C19 & 13:46:36.1 & -11:47:46.7 & 0.4695 & -20.65 & -21.39 & -21.70 & -0.56 & 5.14  & 60.9 &  58.0 & 45.0 &  10.43 & 3.85 & 218.6 & 5.8  & 4.0 & 10.8   \\
\noalign{\vskip 0.2cm}
\hline 
\noalign{\vskip 0.2cm}
C20 & 13:48:09.4 & -11:38:04.8 & 0.4690 & -20.69 & -21.24 & -21.46 & -0.45 & 4.09  & 65.3 & -4.0 & -16.5 &  10.15 & -    & 127.3 & 8.1  & 29.6 & 2.0 \\
C21 & 13:46:56.7 & -11:55:37.5 & 0.4435 & -20.92 & -21.98 & -22.46 & -0.18 & 4.75  & 44.3 & -1.3 & -7.0 &  10.82 & 5.78 & 125.4 & 6.0  & 27.7 & 10.1 \\
C22 & 13:46:30.7 & -11:52:06.0 & 0.4717 & -20.24 & -20.85 & -21.09 & -0.41 & 5.74  & 64.9 & 15.9 & 13.9 &  10.08 & 3.75 & 117.1 & 6.5  & 30.7 & 25.7  \\
C23 & 13:46:15.5 & -11:42:48.8 & 0.4538 & -19.72 & -20.54 & -20.88 & -0.17 & 4.62  & 40.3 & 22.3 & 17.7 &  10.06 & -    & 151.4 & 76.0 & 33.0 & 9.2  \\
\noalign{\vskip 0.2cm}
\hline 
\noalign{\vskip 0.2cm}
\label{T:tabtop}
\end{tabular}
\end{sidewaystable*}

\begin{sidewaystable*}
\centering
\caption{General properties of the field galaxies included in our TFR analysis. Regular galaxies have identification names (ID) F1-F10, while affected galaxies are labeled  F11-F18. Columns are labeled as in Table \ref{T:tabtop}.}

\begin{tabular}{cccccccccccccccccc}
\hline
\noalign{\vskip 0.1cm}
ID & RA & DEC & $z$ & $M_B$ & $M_{Rc}$ & $M_{Ic}$ & $A_B$ & $R_e$ & $i$ & $PA$ & $\theta$ & $\log{M_{\ast}}$ &  SFR & $V_{max}$ & $V_{max,err}$ & $A$ & $A_{err}$   \\
   & (hh:mm:ss)  & (dd:mm:ss.s) &     & (mag) & (mag) & (mag) & (mag) & (kpc) & (º) & (º) & (º)  & & ($M_\odot/yr$) & (km/s) & (km/s) &  & \\
\noalign{\vskip 0.1cm}
\hline 
\hline 
\noalign{\vskip 0.2cm}
F1   & 13:48:18.9 & -11:44:21.9 & 0.3834 & -19.60 & -20.50 & -20.88 & -0.37 & 1.39 & 66.0 & -53.8 & -45.0 &  9.94  & 1.1  & 105.9 & 6.8   & 3.5 & 15.4  \\
F2   & 13:48:16.2 & -11:35:09.5 & 0.5326 & -21.12 & -22.00 & -22.36 & -0.44 & 3.21 & 54.6 & -16.6 & -25.0 &  10.86 & 14.6 & 235.9 & 18.0  & 24.6 & 5.8 \\
F3   & 13:48:11.7 & -11:35:45.2 & 0.7878 & -20.29 & -21.24 & -21.65 & -0.59 & 4.82 & 64.5 & -15.7 & -37.5 &  10.61 & -    & 195.2 & 38.0  & 5.1 & 12.6  \\
F4   & 13:47:42.1 & -11:44:46.0 & 0.5380 & -21.37 & -21.92 & -22.14 & -0.41 & 6.53 & 51.9 & 30.7 & 11.8 &  10.56 & 14.4 & 243.9 & 22.7  & 16.8 & 8.8 \\
F5   & 13:46:51.7 & -11:53:48.8 & 0.4935 & -21.08 & -21.98 & -22.36 & -0.85 & 5.48 & 73.2 & -2.0 & -1.6 &  10.78 & -    & 201.3 & 6.1   & 6.5 & 2.6  \\
F6   & 13:46:15.3 & -11:52:24.2 & 0.4007 & -19.72 & -20.31 & -20.57 & -0.78 & 5.13 & 55.7 & -37.3 & -39.0 &  9.73  & -    & 175.1 & 6.8   & 13.6 & 5.9  \\
F7   & 13:46:28.9 & -11:53:19.6 & 0.3701 & -20.45 & -19.90 & -20.93 & -0.31 & 3.31 & 57.0 & -46.2 & 0.0 &  9.85  & 0.6  & 124.2 & 25.6  & 10.0 & 9.1  \\ 
F8   & 13:47:13.6 & -11:45:51.9 & 0.3998 & -20.34 & -21.17 & -21.53 & -0.56 & 4.01 & 68.0 & 64.6 & 20.0 &  10.13 & 2.1  & 145.9 & 38.4  & 18.9 & 9.3 \\
F9   & 13:46:57.5 & -11:47:00.3 & 0.3284 & -19.27 & -19.74 & -19.96 & -0.13 & 3.20 & 47.8 & -33.1 & -27.2 &  9.33  & -    &  85.2 & 2.6   & 3.2 & 3.3  \\
F10  & 13:46:32.5 & -11:46:19.1 & 0.3676 & -21.15 & -22.02 & -22.39 & -0.24 & 6.52 & 39.4 & -22.8 & -20.6 &  10.55 & 3.3  & 280.5 & 7.2   & 17.0 & 6.6  \\
\noalign{\vskip 0.2cm}
\hline 
\noalign{\vskip 0.2cm}
F11  & 13:48:21.5 & -11:46:54.8 & 0.5240 & -20.73 & -21.48 & -21.77 & -1.39 & 8.10 & 85.8 & -17.6 & -18.5 &  10.15 & 8.5  & 241.5 & 22.6  & 47.5 & 6.5  \\
F12  & 13:48:32.2 & -11:47:51.9 & 0.5339 & -21.32 & -21.87 & -22.09 & -0.17 & 4.40 & 36.9 & 55.7 & 45.0 &  10.82 & 24.4 & 192.0 & 31.8  & 43.4 & 10.3  \\
F13  & 13:47:46.7 & -11:38:03.4 & 0.3696 & -20.10 & -20.83 & -21.19 & -0.31 & 4.90 & 52.4 & -52.6 & -45.0 &  10.08 & -    & 156.2 & 11.4  & 26.8 & 5.2  \\
F14  & 13:47:31.5 & -11:43:24.7 & 0.5337 & -20.55 & -20.82 & -20.93 & -0.35 & 2.50 & 61.5 & 12.0 & 30.8 &  10.06 & 3.6  & 114.9 & 5.4   & 30.6 & 17.2 \\
F15  & 13:46:50.1 & -11:42:04.2 & 0.3994 & -20.80 & -21.57 & -21.92 & -0.07 & 4.34 & 24.8 & 25.7 & -13.0 &  10.15 & -    & 191.5 & 91.6  & 34.9 & 5.4  \\
F16  & 13:47:02.9 & -11:46:38.6 & 0.3679 & -20.38 & -21.26 & -21.64 & -0.58 & 5.47 & 66.3 & 27.7 & 21.7 &  10.82 & 1.6  & 191.7 & 19.6  & 37.1 & 10.6  \\
F17  & 13:46:31.5 & -11:46:42.8 & 0.3515 & -20.37 & -20.96 & -21.22 & -0.30 & 3.32 & 59.5 & 36.4 & 31.3 &  10.08 & 1.8  & 111.2 & 12.2  & 25.6 & 3.1 \\
F18  & 13:46:56.0 & -11:53:10.7 & 0.3990 & -21.29 & -22.35 & -22.80 & -0.71 & 6.42 & 65.7 & 22.5 & 21.7 &  10.06 & -    & 235.5 & 66.9  & 38.0 & 1.9\\
\noalign{\vskip 0.2cm}
\hline 
\noalign{\vskip 0.2cm}
\label{T:tabtop2}
\end{tabular}
\end{sidewaystable*}

\begin{figure*}[h!]
\centering
\includegraphics[width=\textwidth]{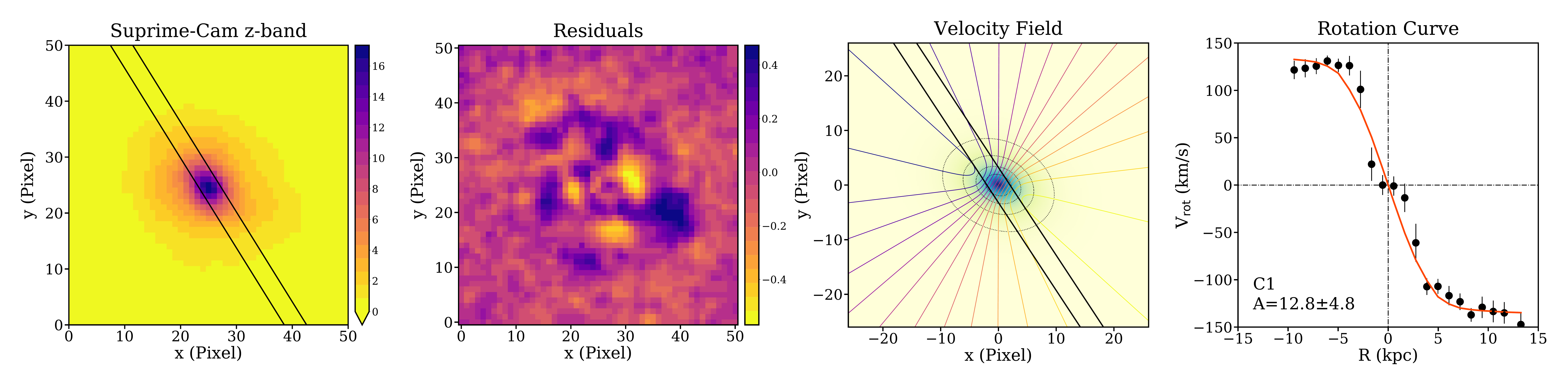}
\includegraphics[width=\textwidth]{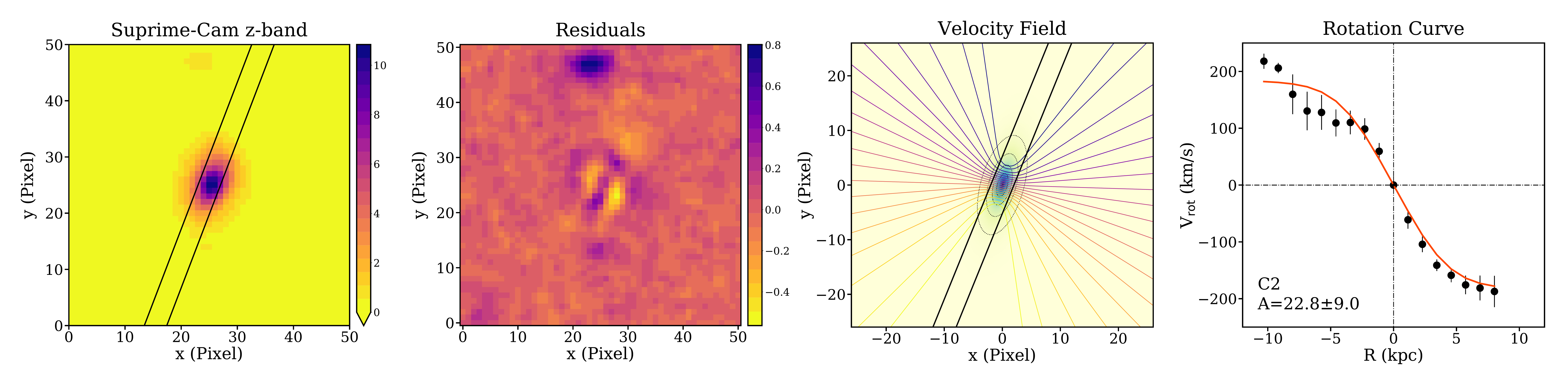}
\includegraphics[width=\textwidth]{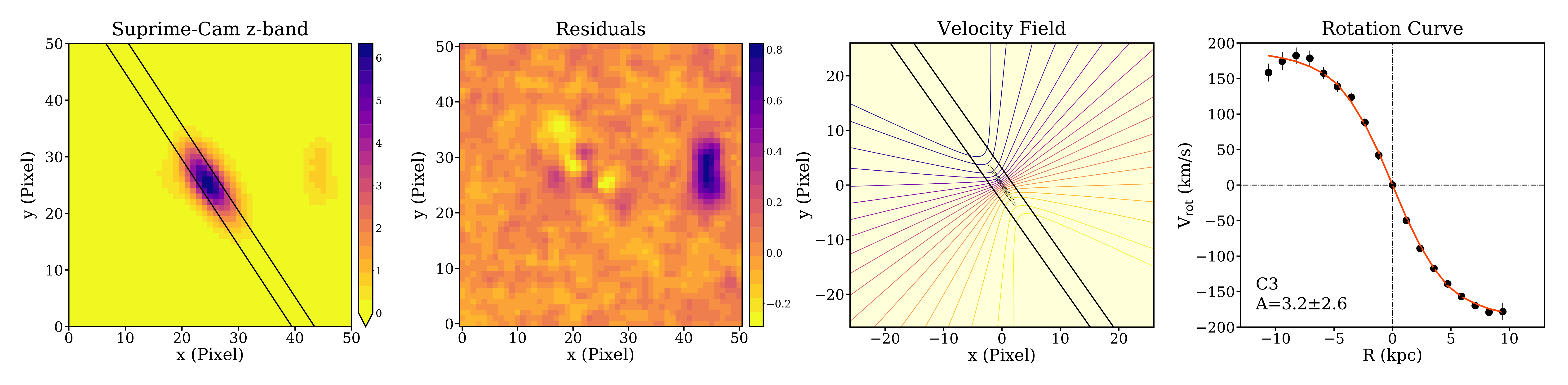}
\includegraphics[width=\textwidth]{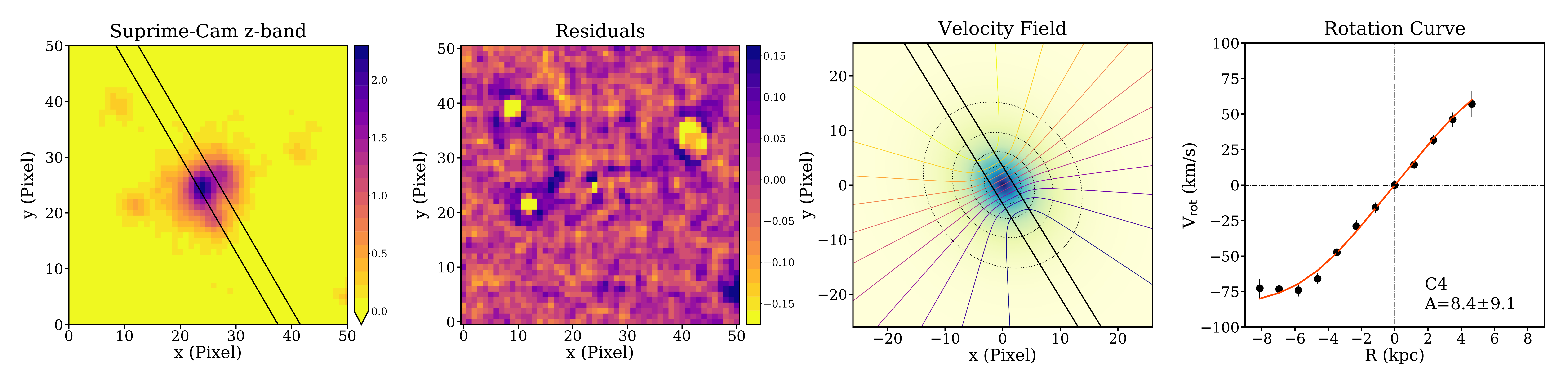}
\includegraphics[width=\textwidth]{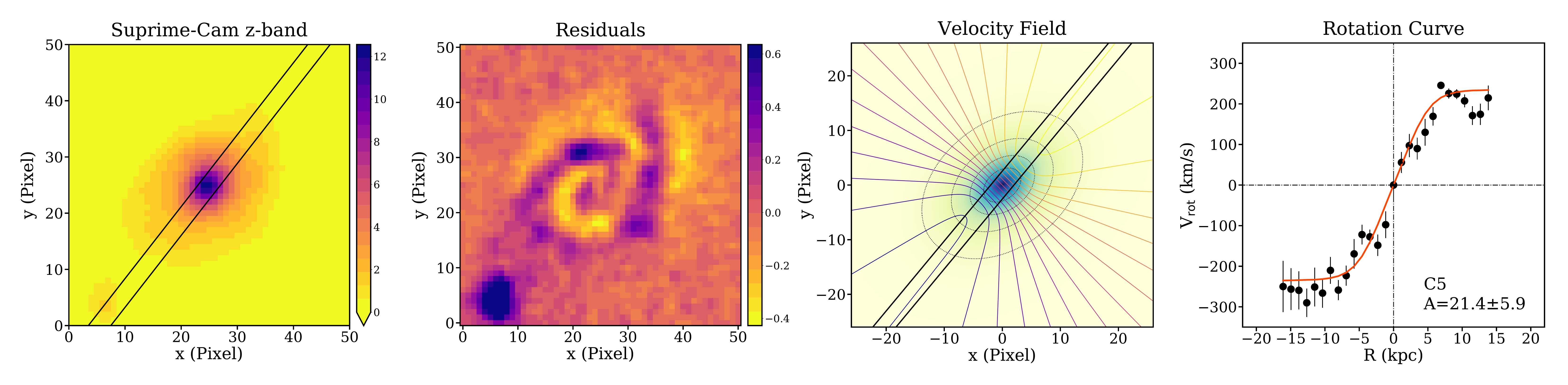}
\caption{Our sample of cluster galaxies studied following the methods explained in Sect. \ref{SS:Methods} and presented in the same order as in Table \ref{T:tabtop}. Column 1:  z-band Suprime-Cam image centered on the target; Column 2:  residuals after subtracting the 2D model of the galaxy; Column 3:  synthetic velocity field after fitting the simulated rotation curve to the observed curve; Column 4:  rotation curve (black dots) in the observed frame and the simulated rotation curve (red line). The black solid parallel lines in the panels in Cols. 1 and 3  depict the position of the edges of the slit.}
\ContinuedFloat
\label{foot}
\end{figure*}

\begin{figure*}[h!]
\centering
\includegraphics[width=\textwidth]{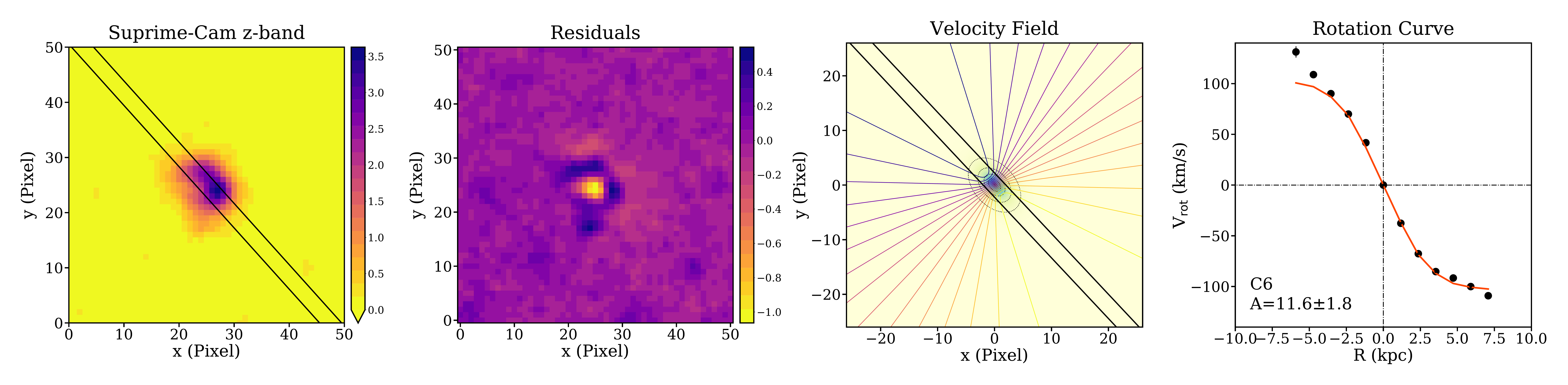}
\includegraphics[width=\textwidth]{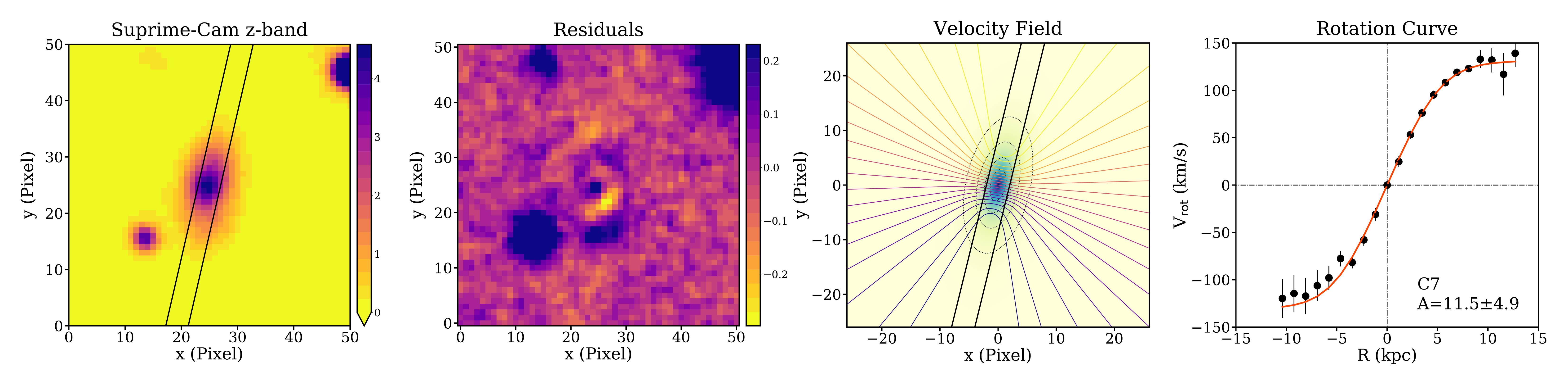}
\includegraphics[width=\textwidth]{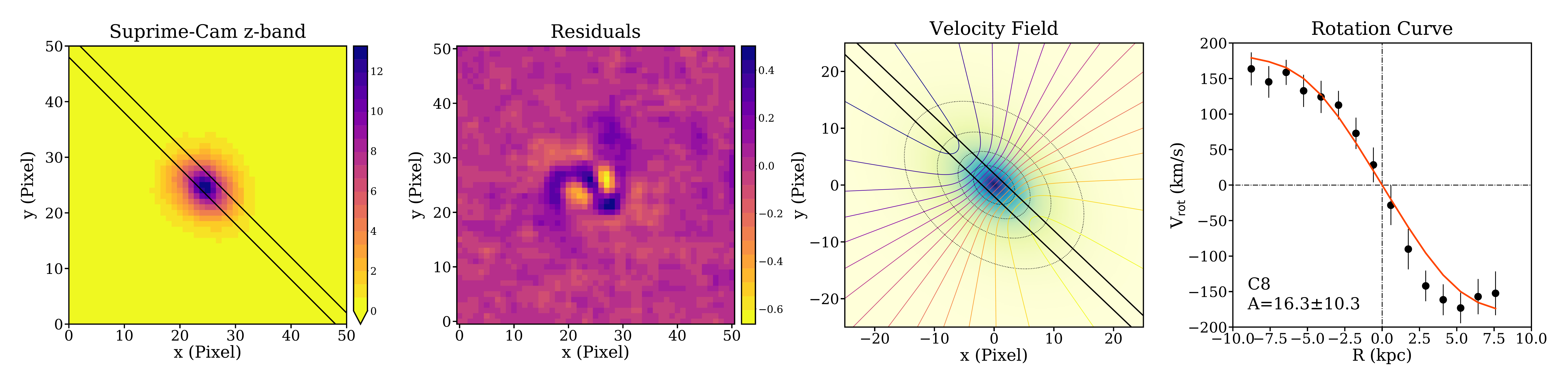}
\includegraphics[width=\textwidth]{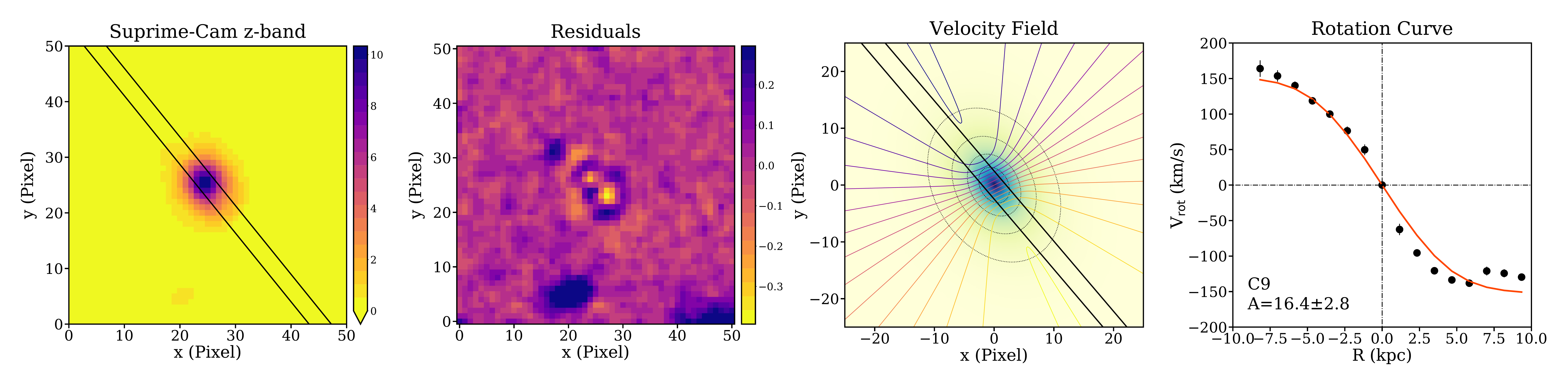}
\includegraphics[width=\textwidth]{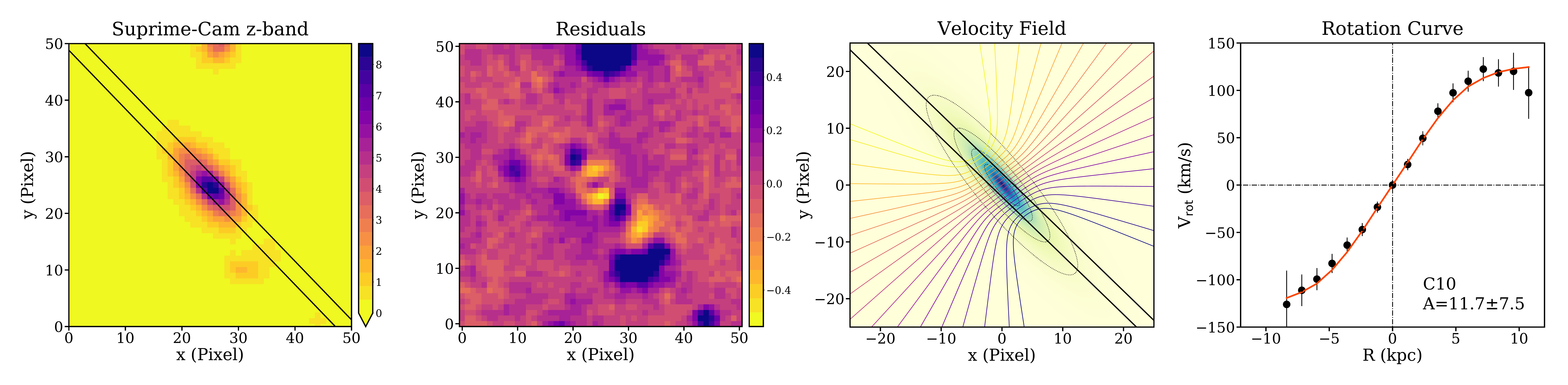}
\caption[]{(Continued)}
\ContinuedFloat

\end{figure*}

\begin{figure*}[h!]
\centering
\includegraphics[width=\textwidth]{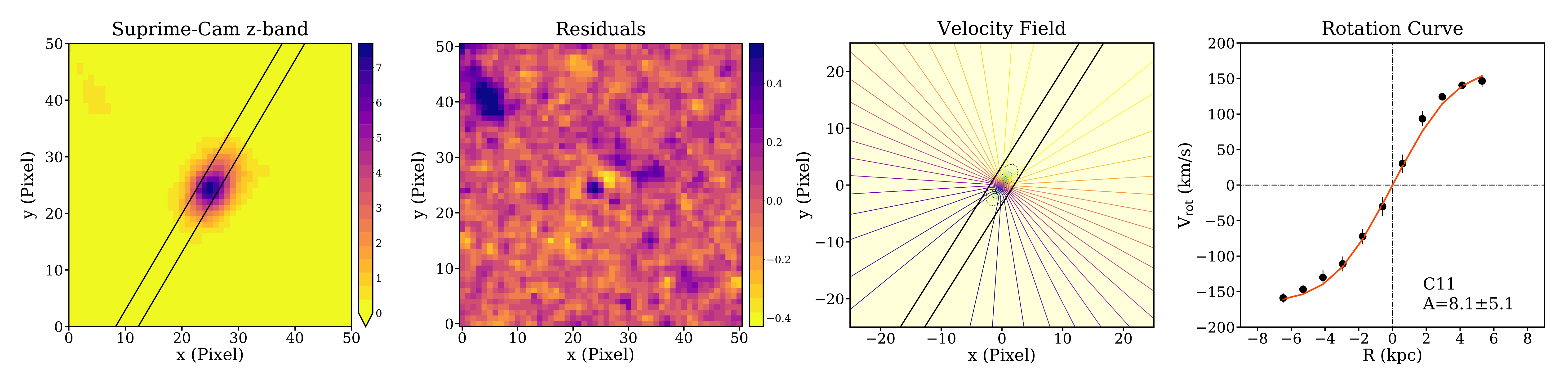}
\includegraphics[width=\textwidth]{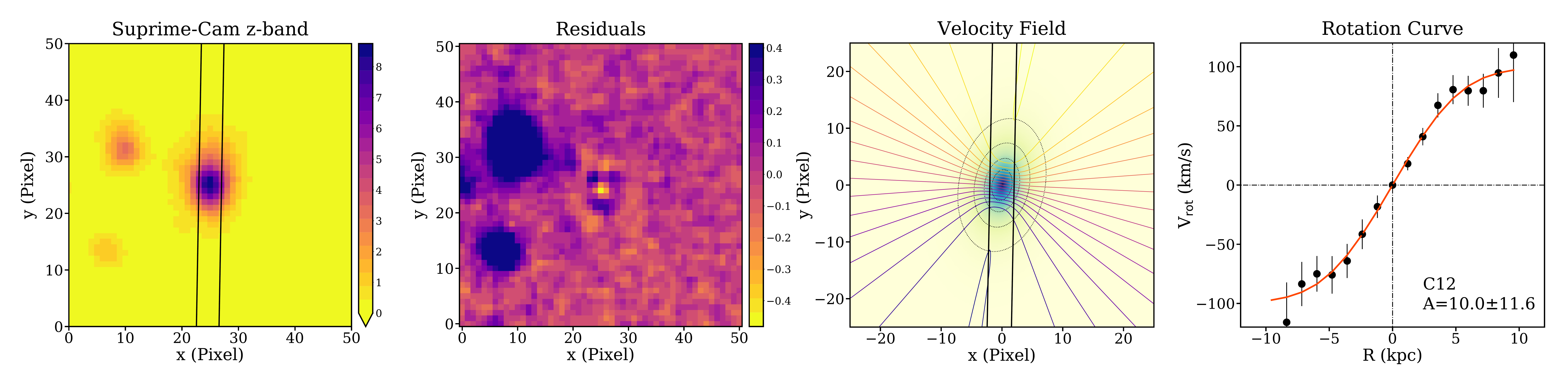}
\includegraphics[width=\textwidth]{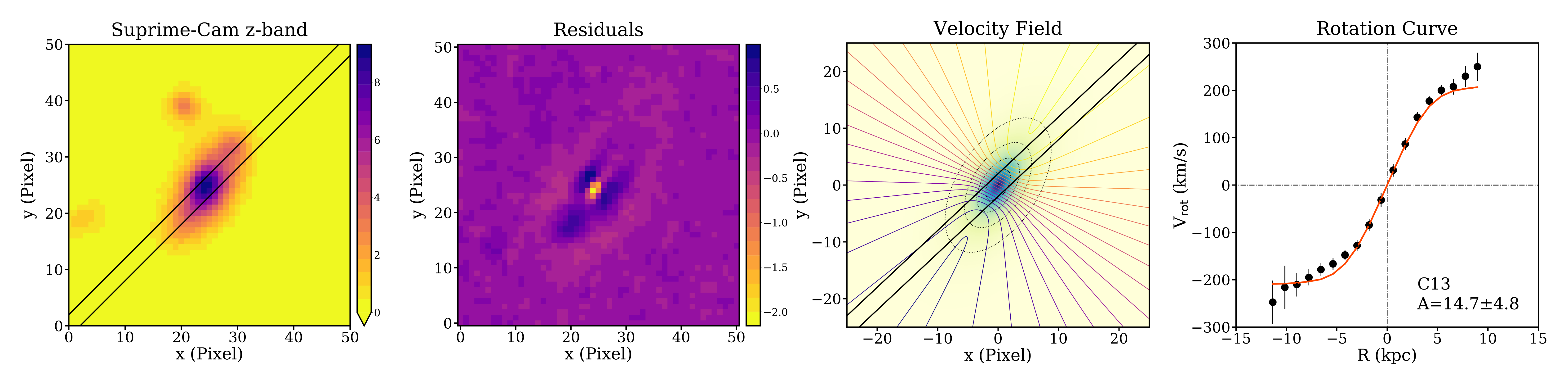}
\includegraphics[width=\textwidth]{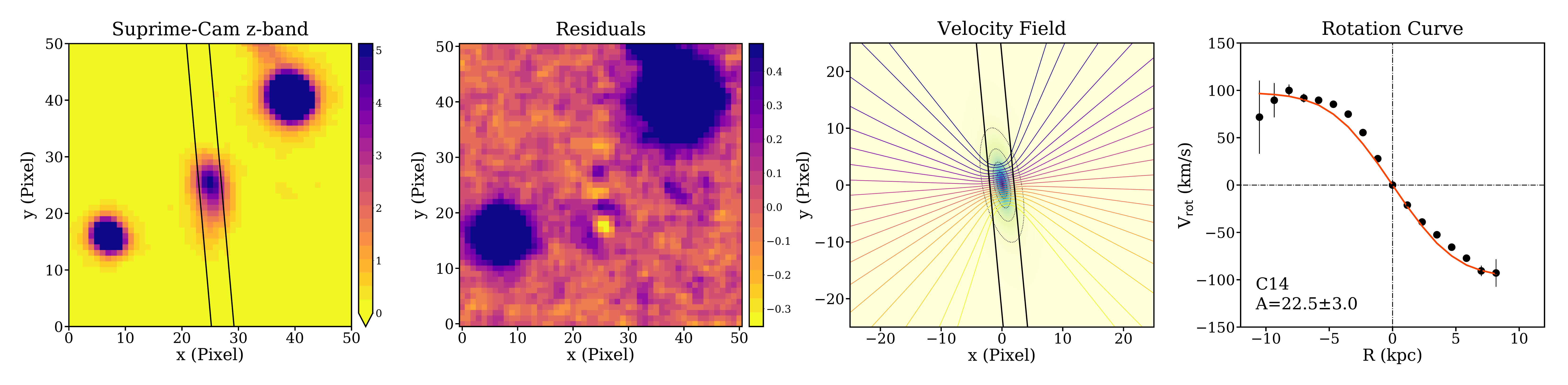}
\includegraphics[width=\textwidth]{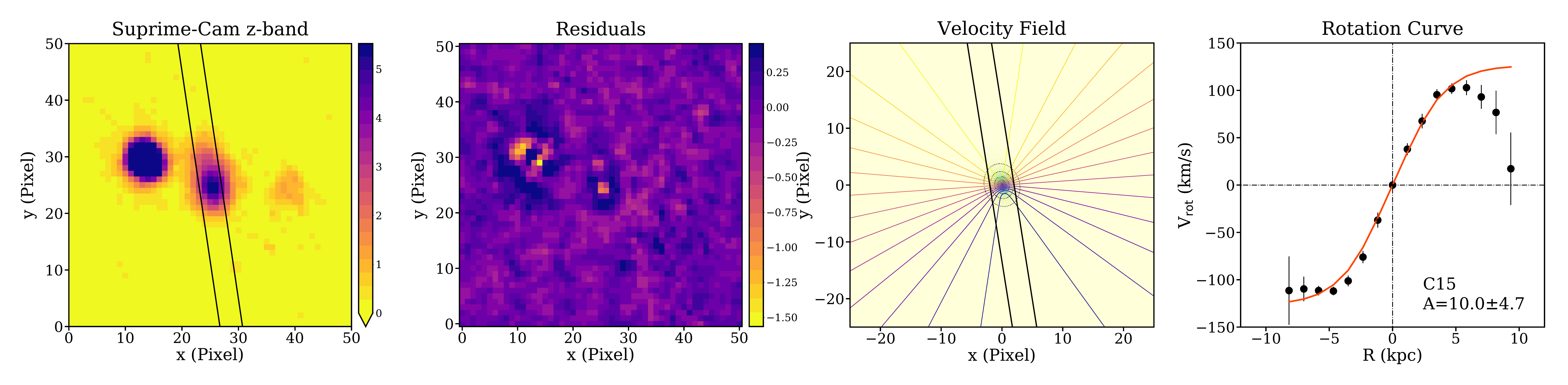}
\caption[]{(Continued)}
\ContinuedFloat

\end{figure*}

\begin{figure*}[h!]
\centering
\includegraphics[width=\textwidth]{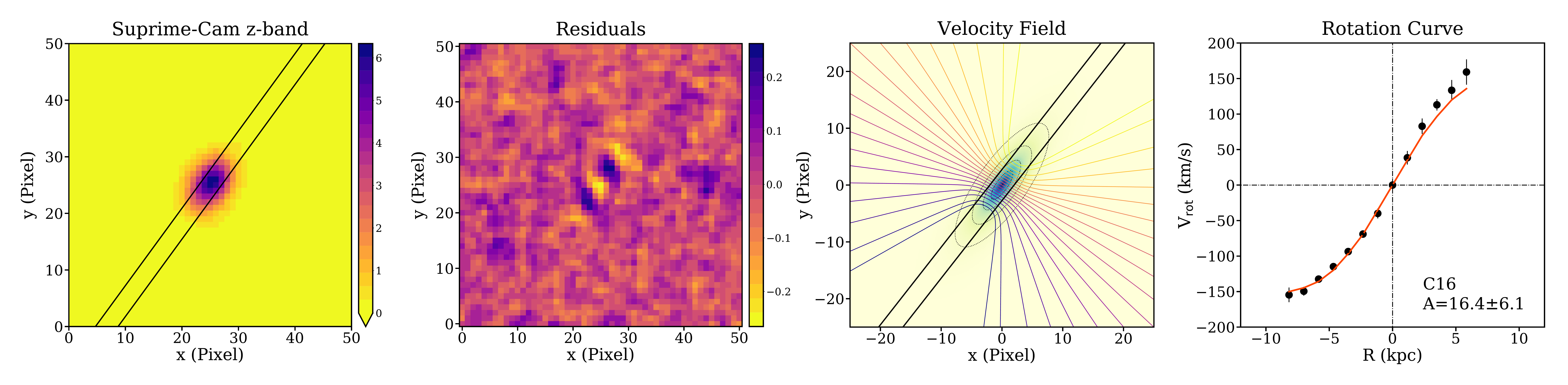}
\includegraphics[width=\textwidth]{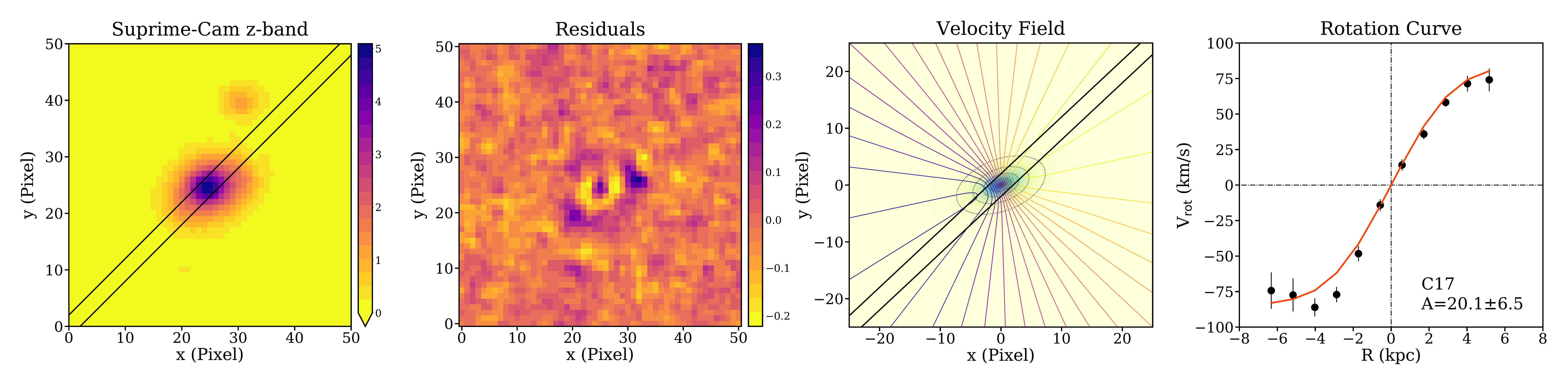}
\includegraphics[width=\textwidth]{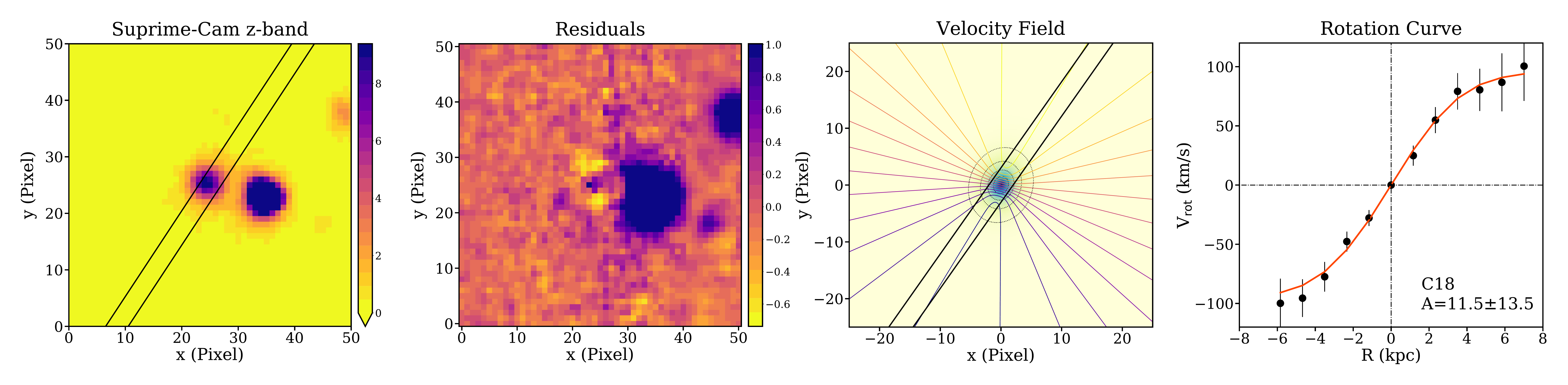}
\includegraphics[width=\textwidth]{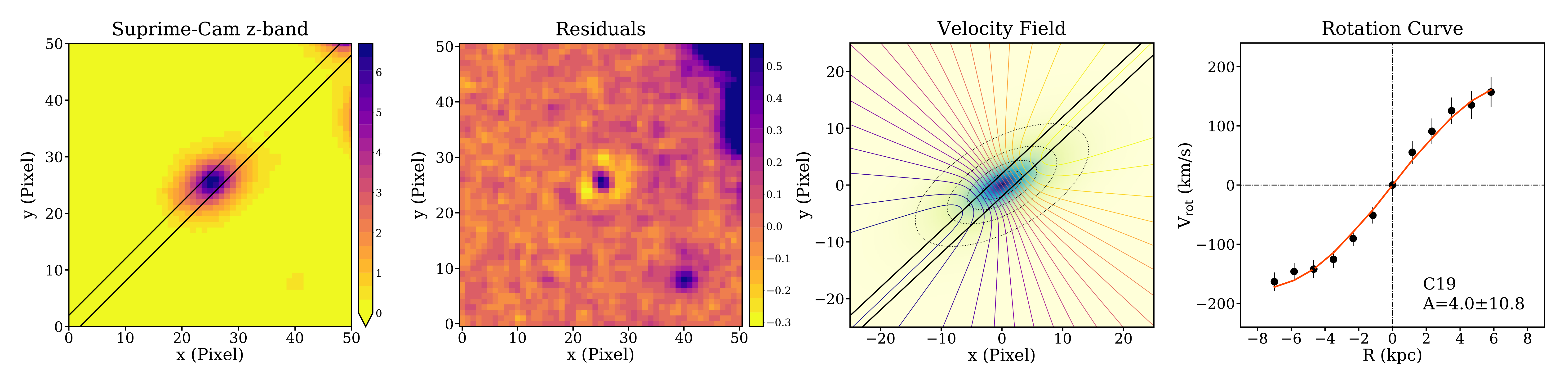}
\caption[]{(Continued)}
\ContinuedFloat

\end{figure*} 

\begin{figure*}[h!]
\centering
\includegraphics[width=\textwidth]{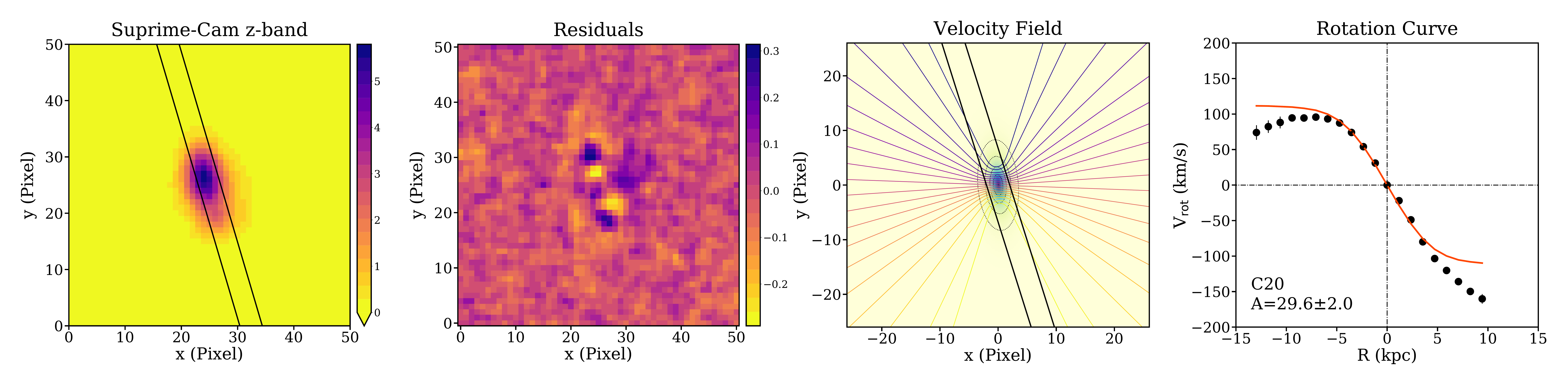}
\includegraphics[width=\textwidth]{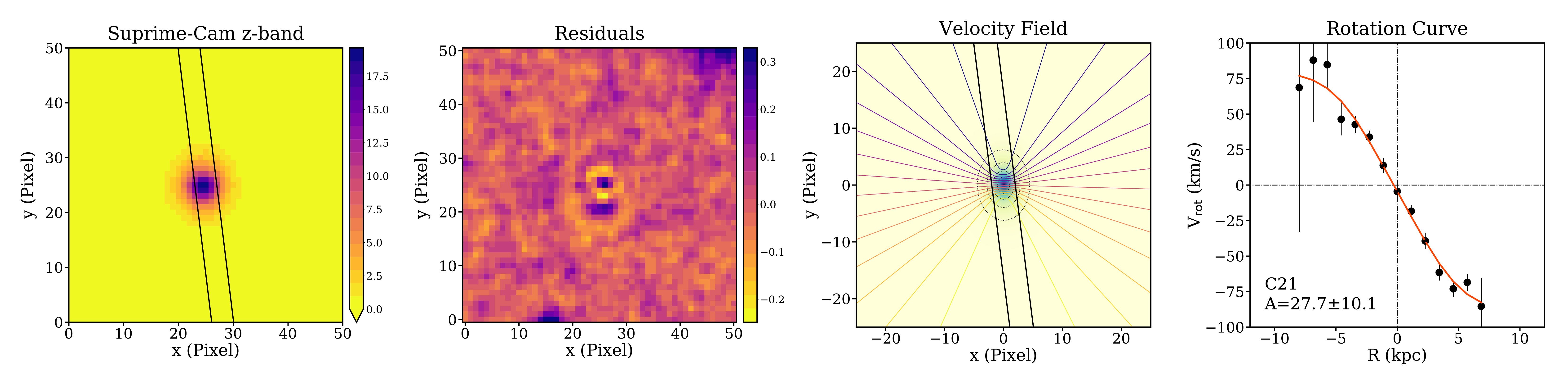}
\includegraphics[width=\textwidth]{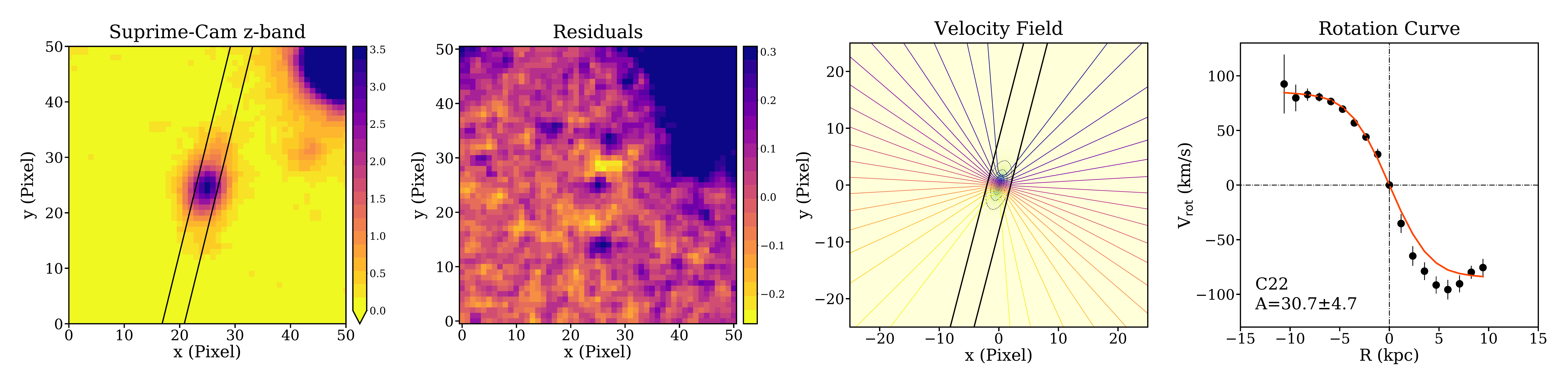}
\includegraphics[width=\textwidth]{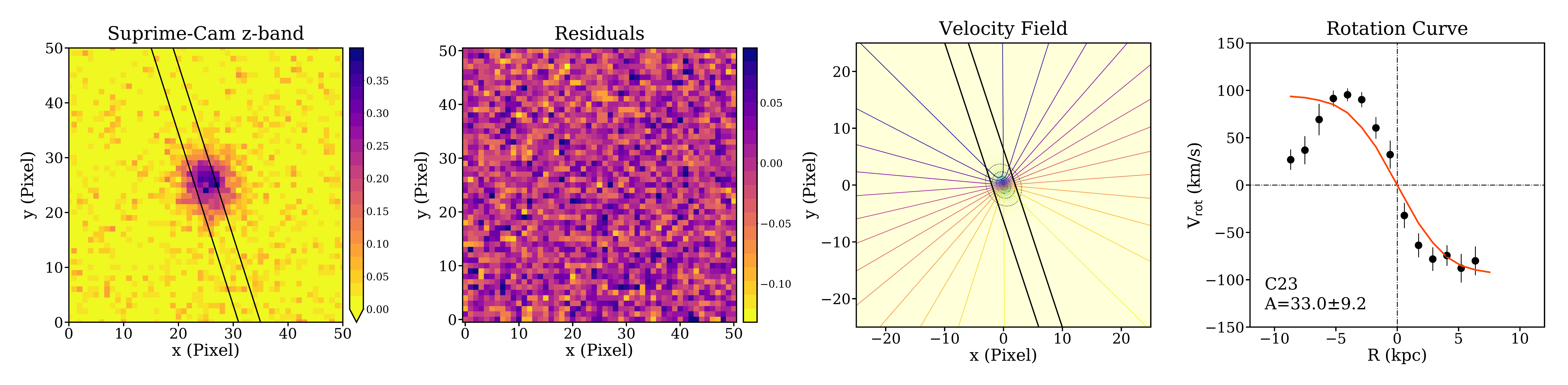}
\caption[]{(Continued)}

\end{figure*} 

\begin{figure*}[h!]
\centering
\includegraphics[width=\textwidth]{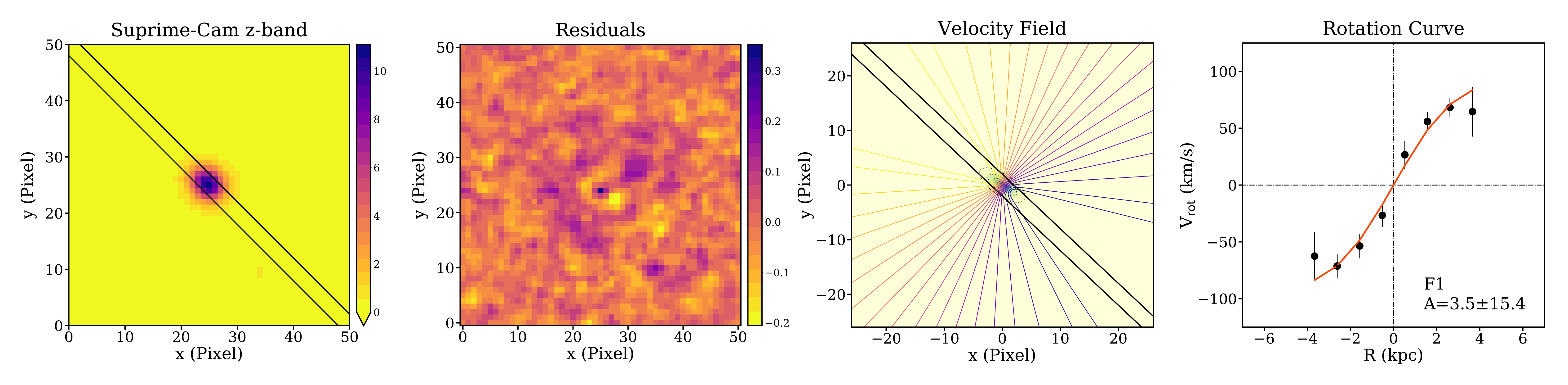}
\includegraphics[width=\textwidth]{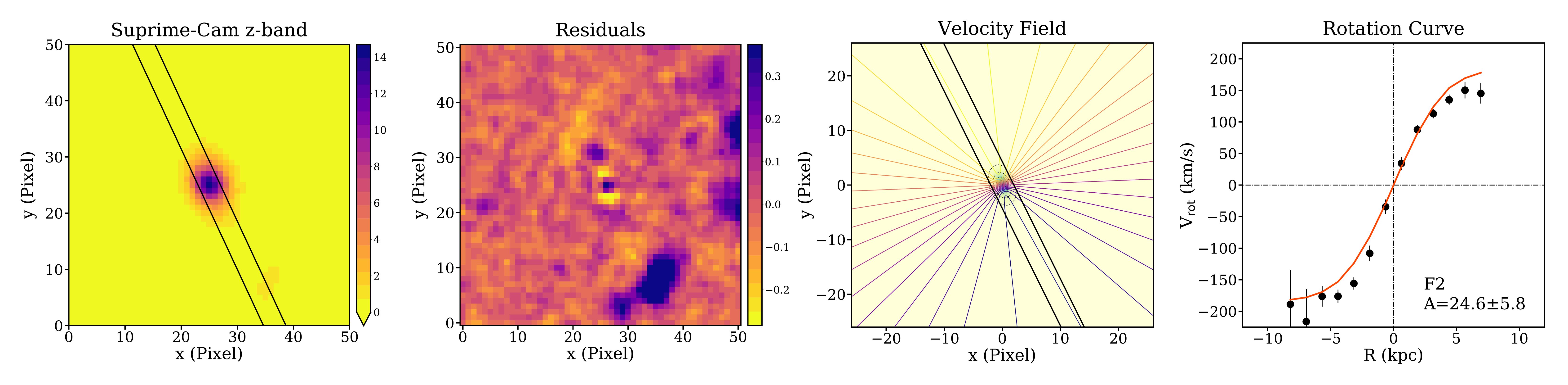}
\includegraphics[width=\textwidth]{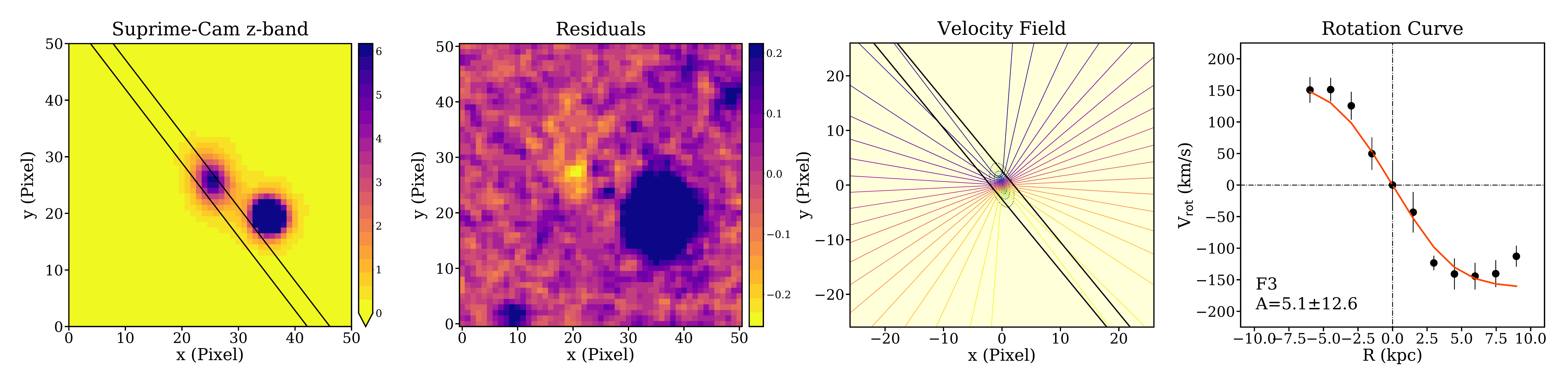}
\includegraphics[width=\textwidth]{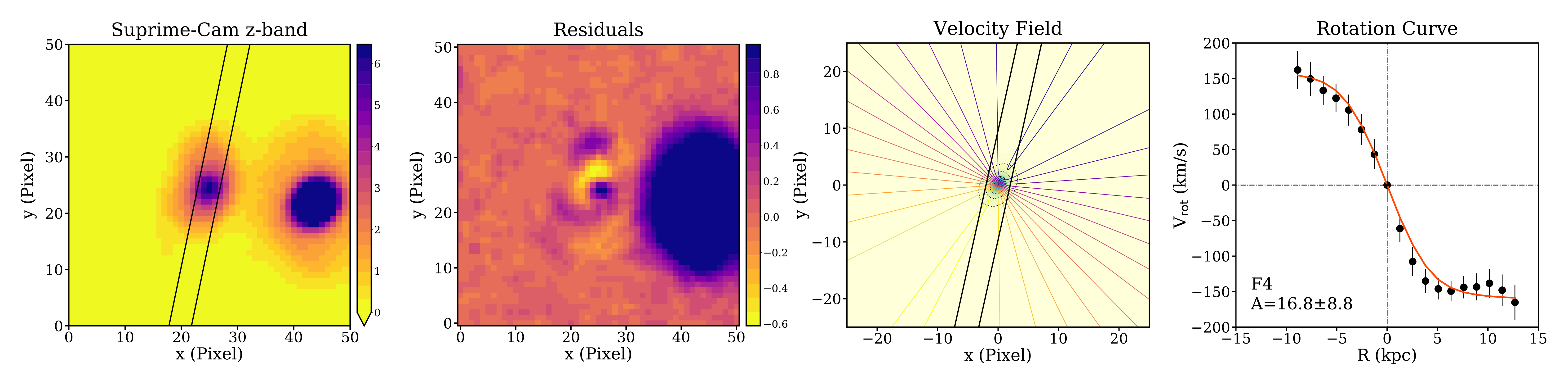}
\includegraphics[width=\textwidth]{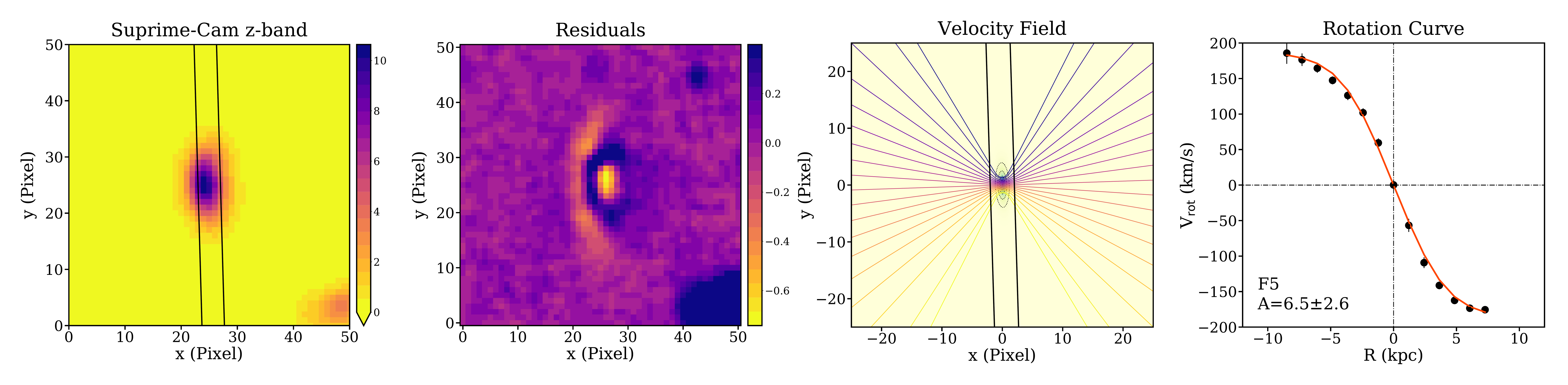}
\caption{Our sample of field galaxies studied following the methods explained in Sect. \ref{SS:Methods} and presented in the same order as in Table \ref{T:tabtop2}. The columns have the same meaning as in Fig. \ref{foot}.}
\ContinuedFloat
\label{foot2}
\end{figure*}

\begin{figure*}[h!]
\centering
\includegraphics[width=\textwidth]{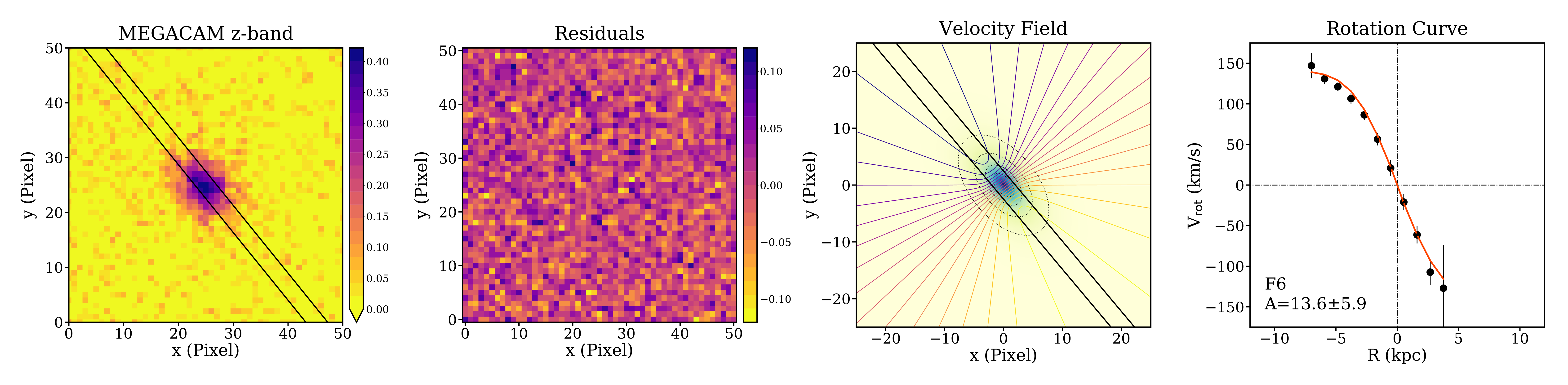}
\includegraphics[width=\textwidth]{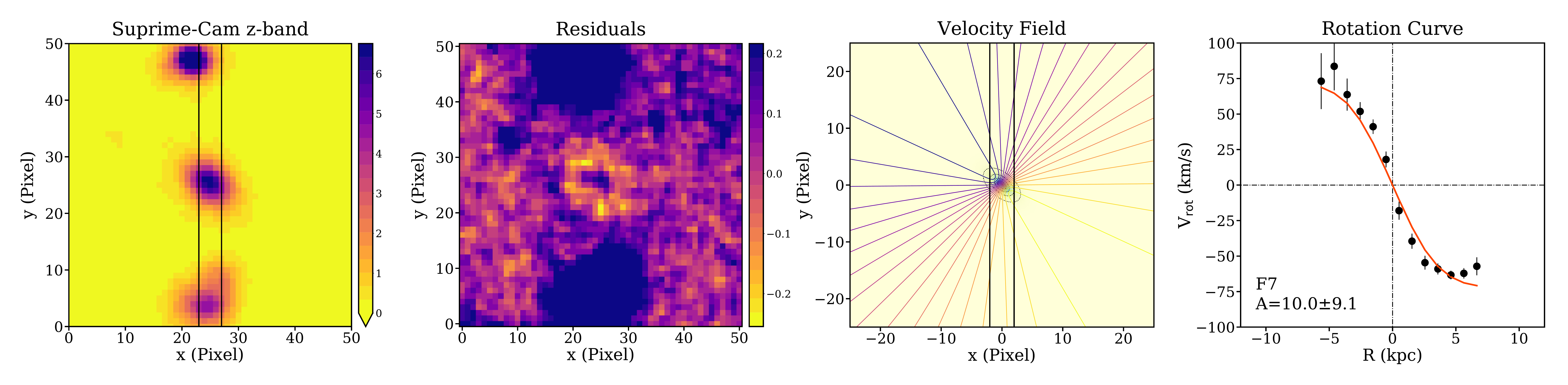}
\includegraphics[width=\textwidth]{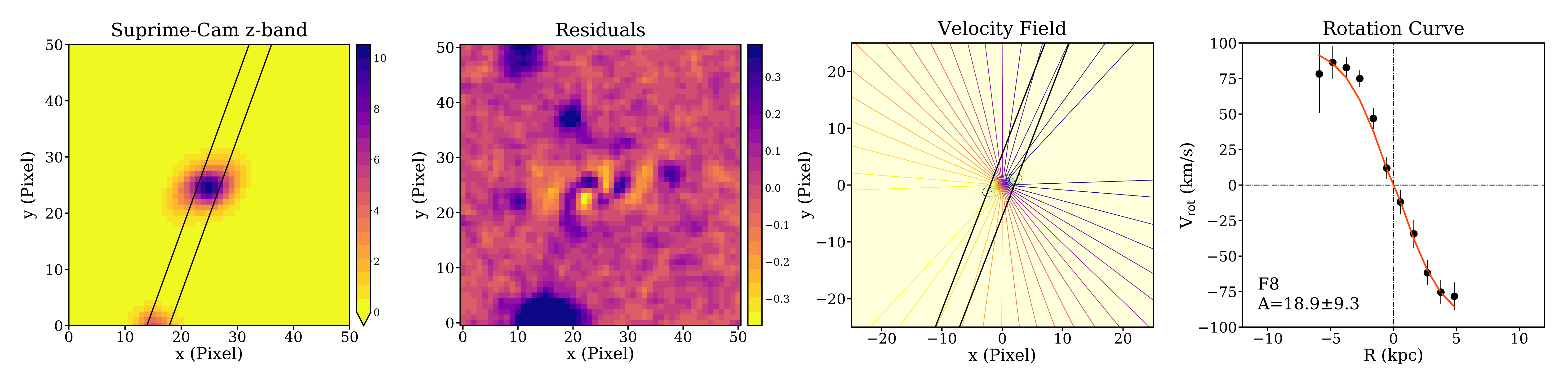}
\includegraphics[width=\textwidth]{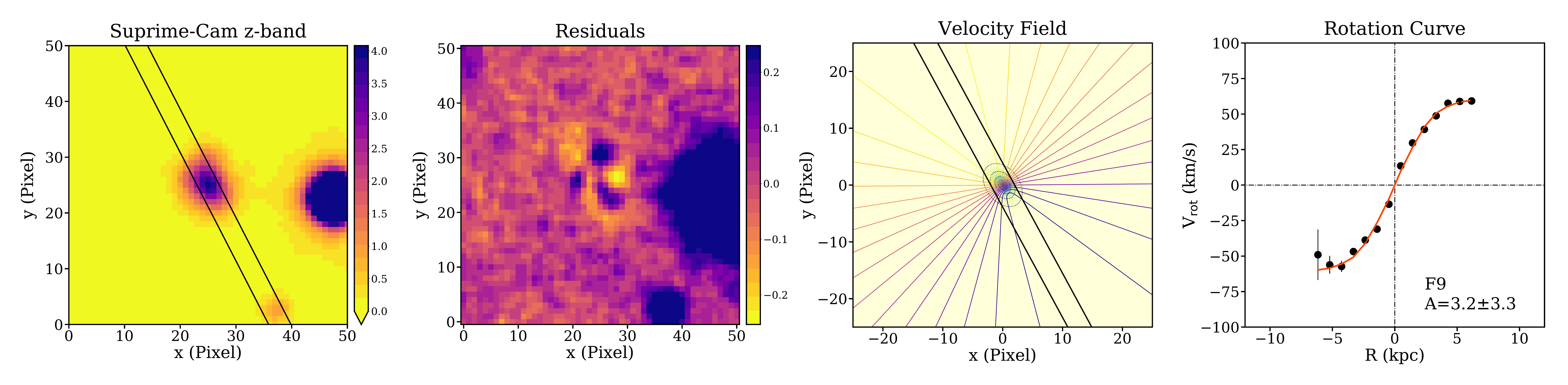}
\includegraphics[width=\textwidth]{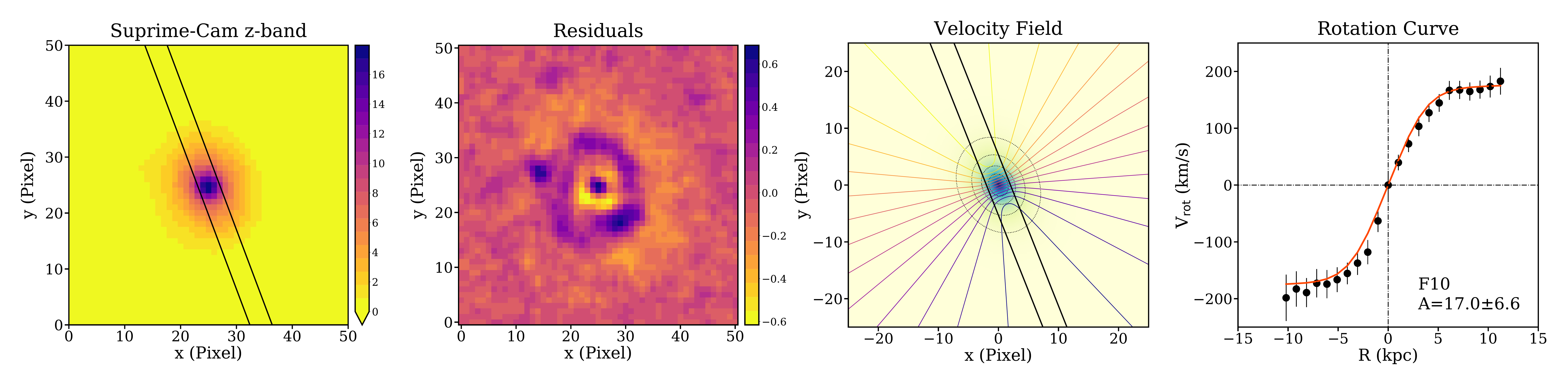}
\caption[]{(Continued)}
\ContinuedFloat
\end{figure*} 

\begin{figure*}[h!]
\centering
\includegraphics[width=\textwidth]{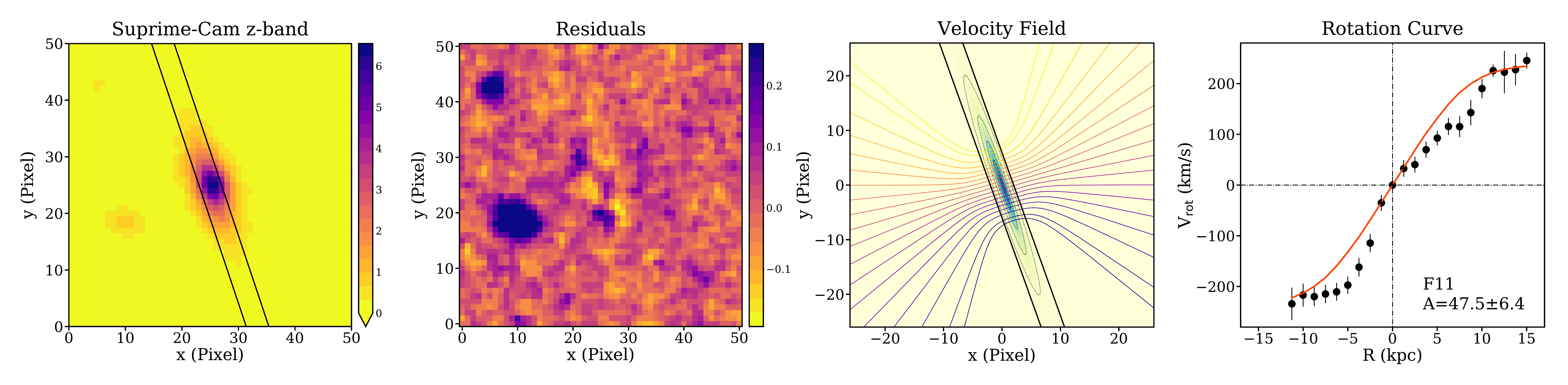}
\includegraphics[width=\textwidth]{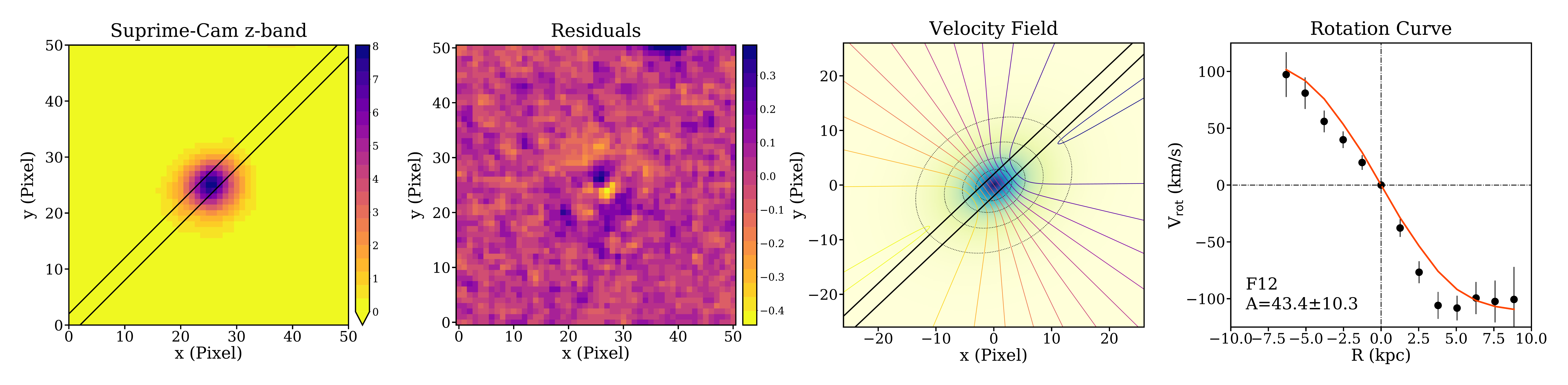}
\includegraphics[width=\textwidth]{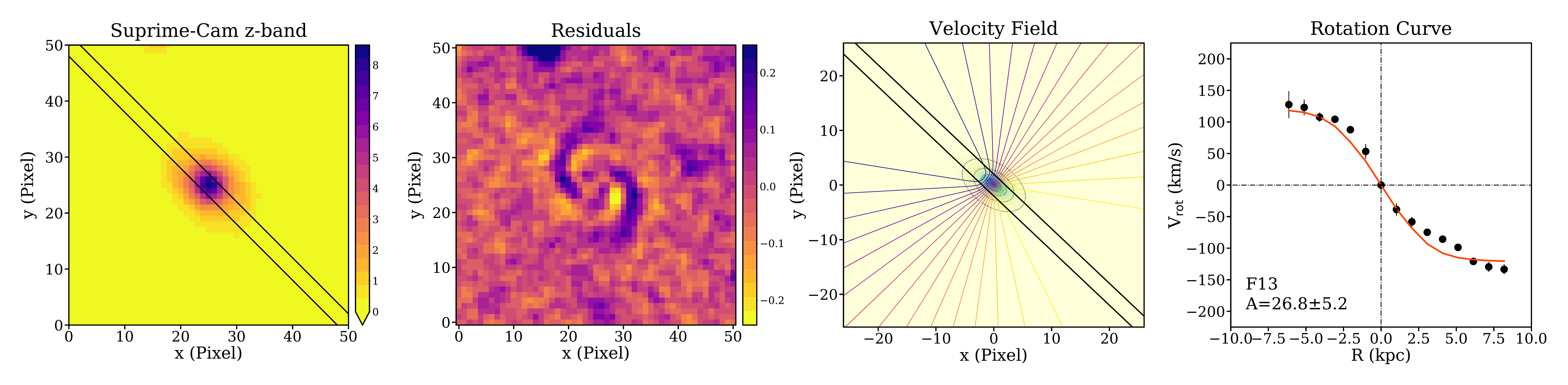}
\includegraphics[width=\textwidth]{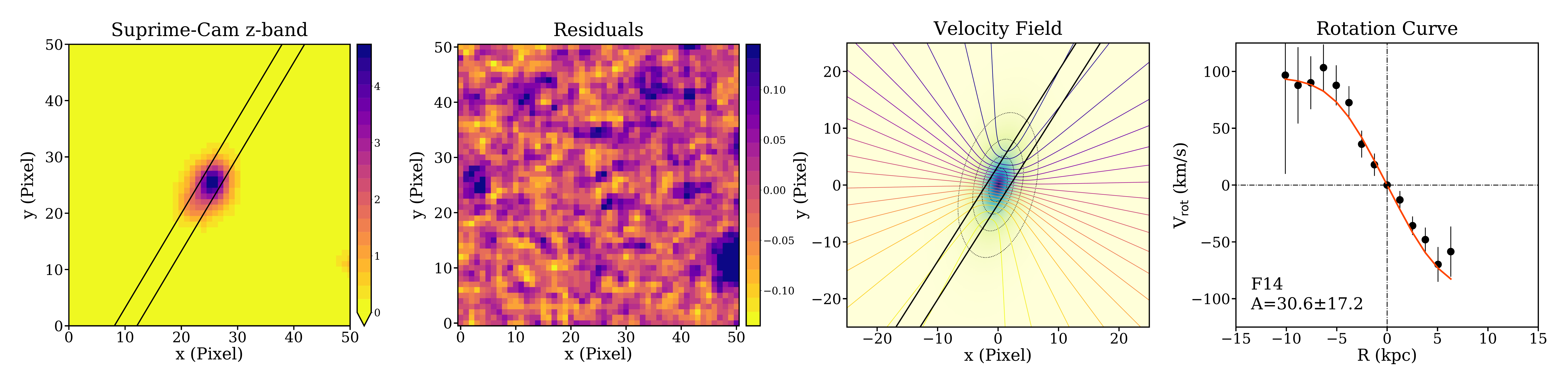}
\includegraphics[width=\textwidth]{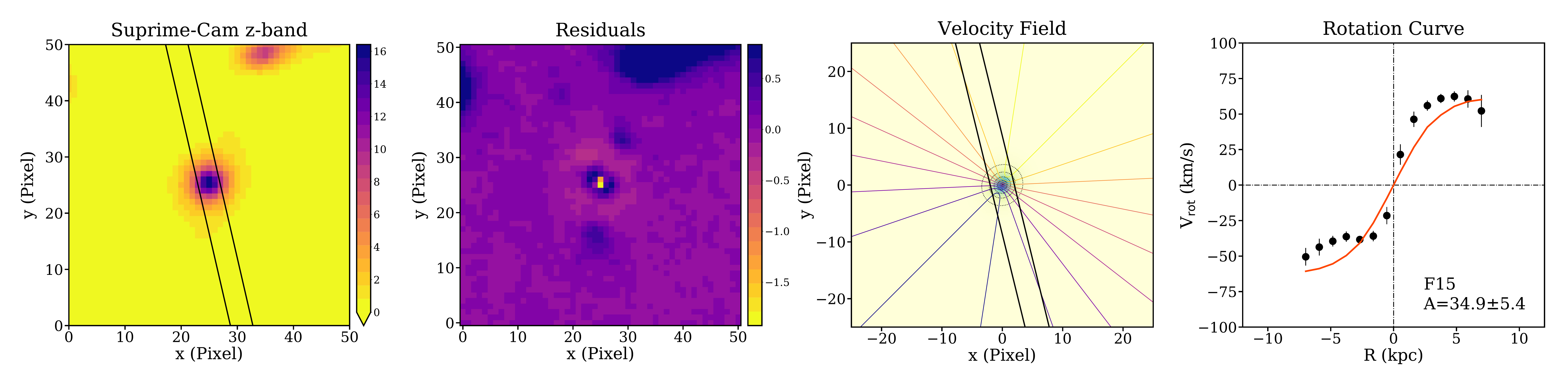}
\caption[]{(Continued)}
\ContinuedFloat
\end{figure*} 

\begin{figure*}[h!]
\centering
\includegraphics[width=\textwidth]{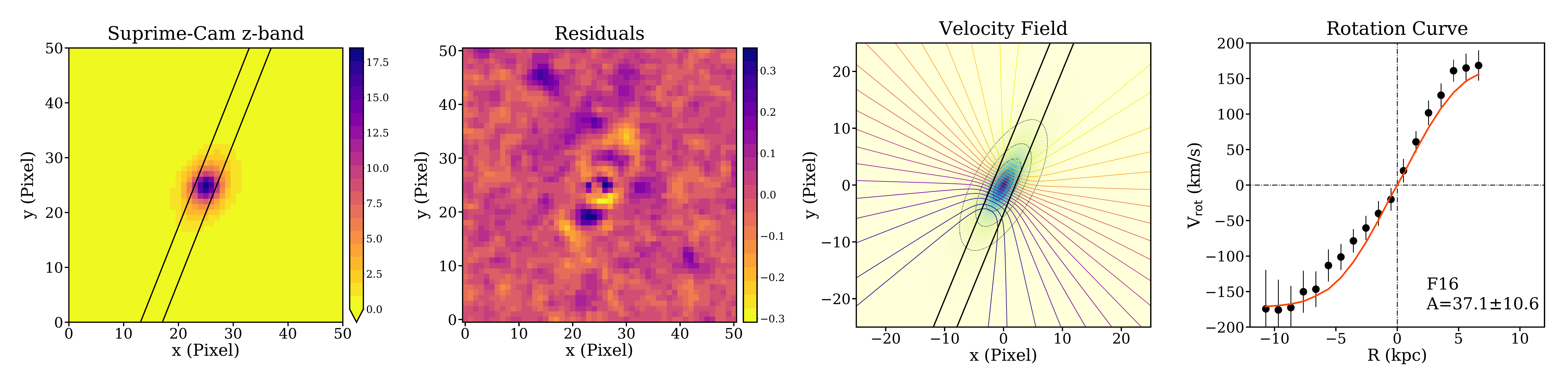}
\includegraphics[width=\textwidth]{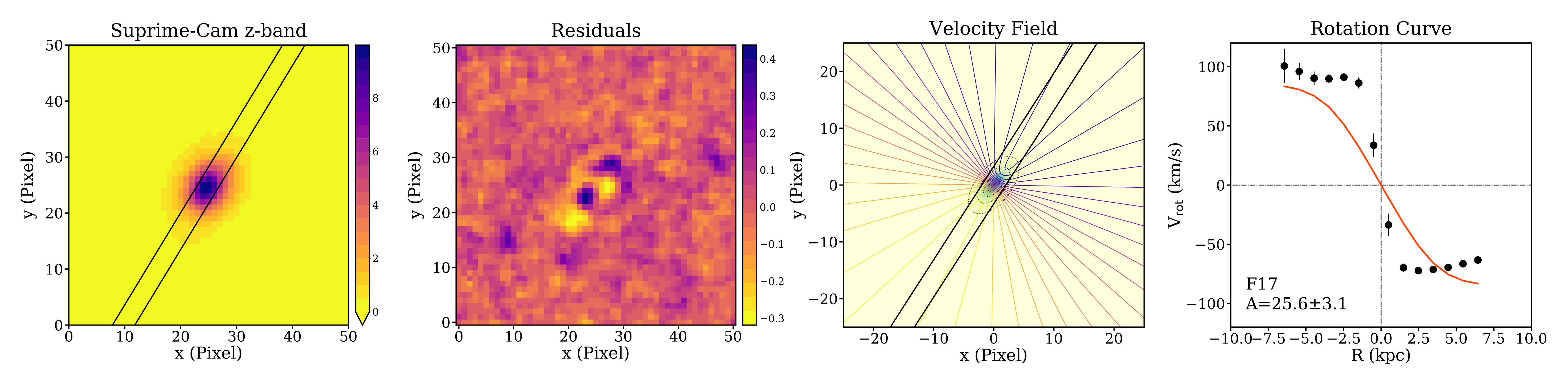}
\includegraphics[width=\textwidth]{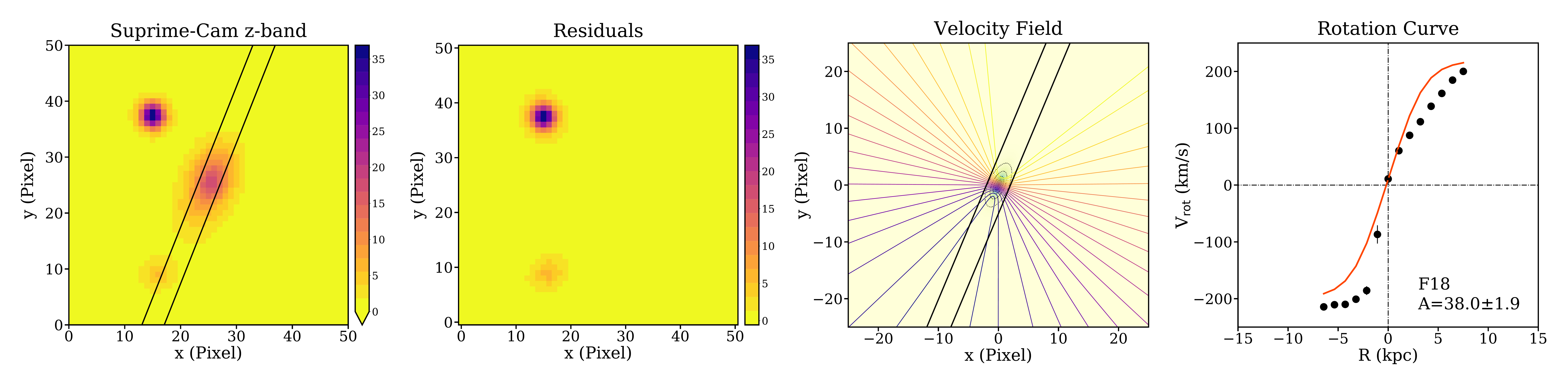}
\caption[]{(Continued)}
\end{figure*}

\begin{figure*}[h!]
\centering
\includegraphics[width=\textwidth]{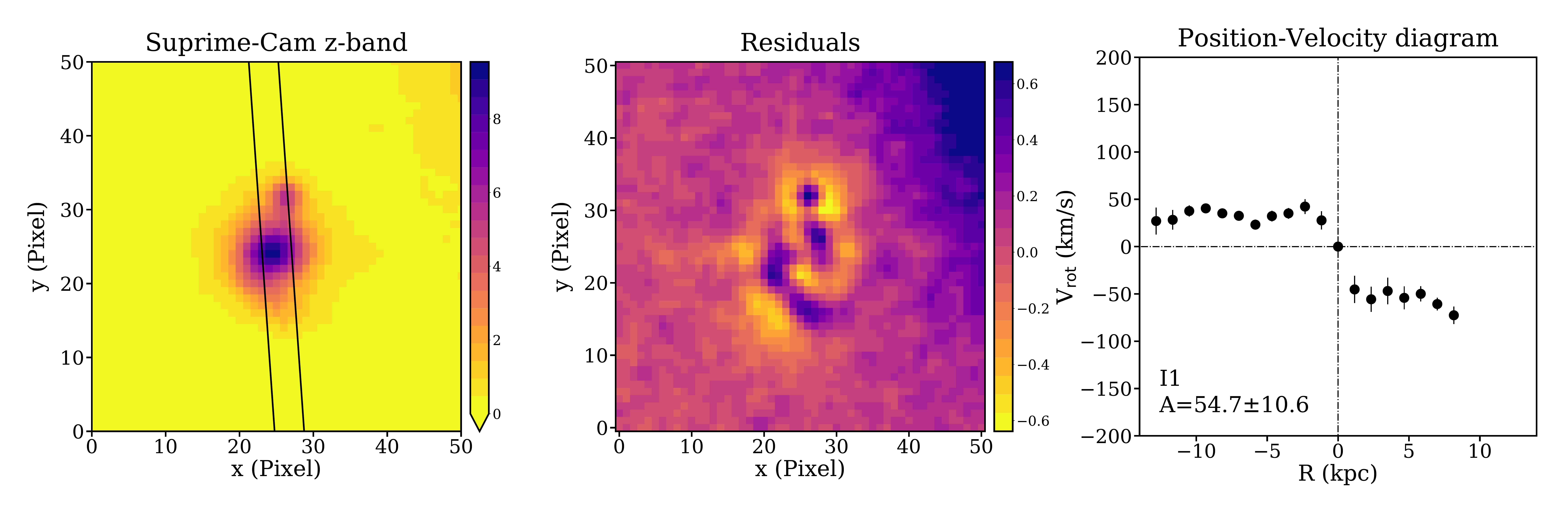}
\includegraphics[width=\textwidth]{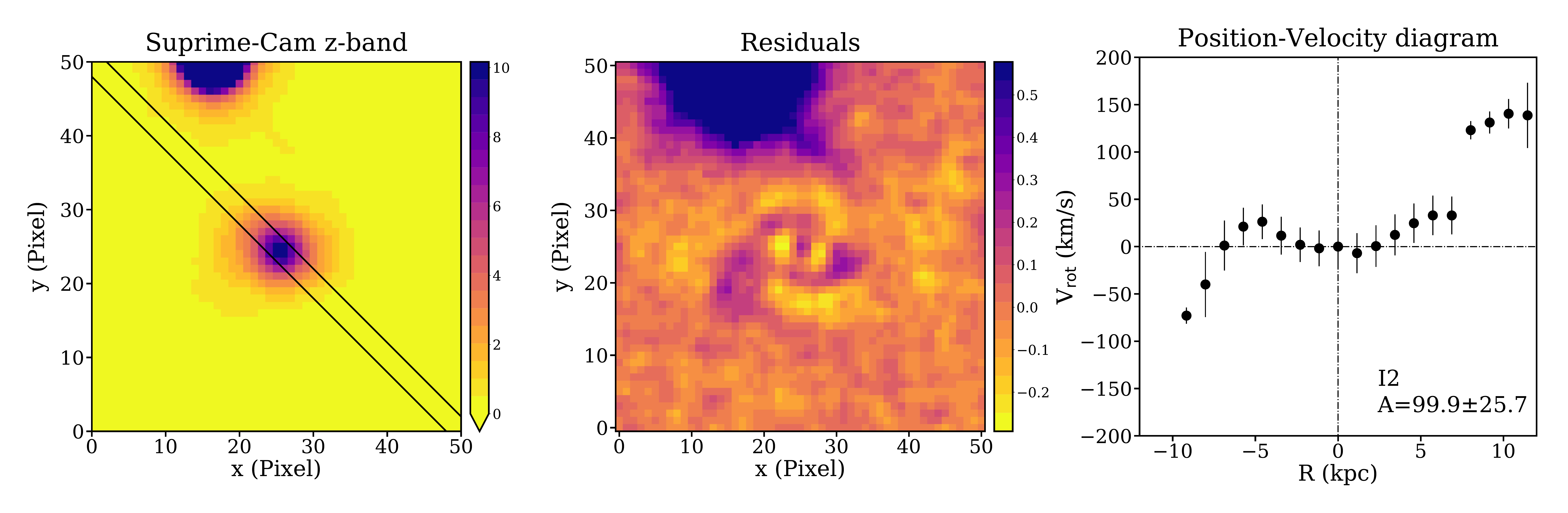}
\includegraphics[width=\textwidth]{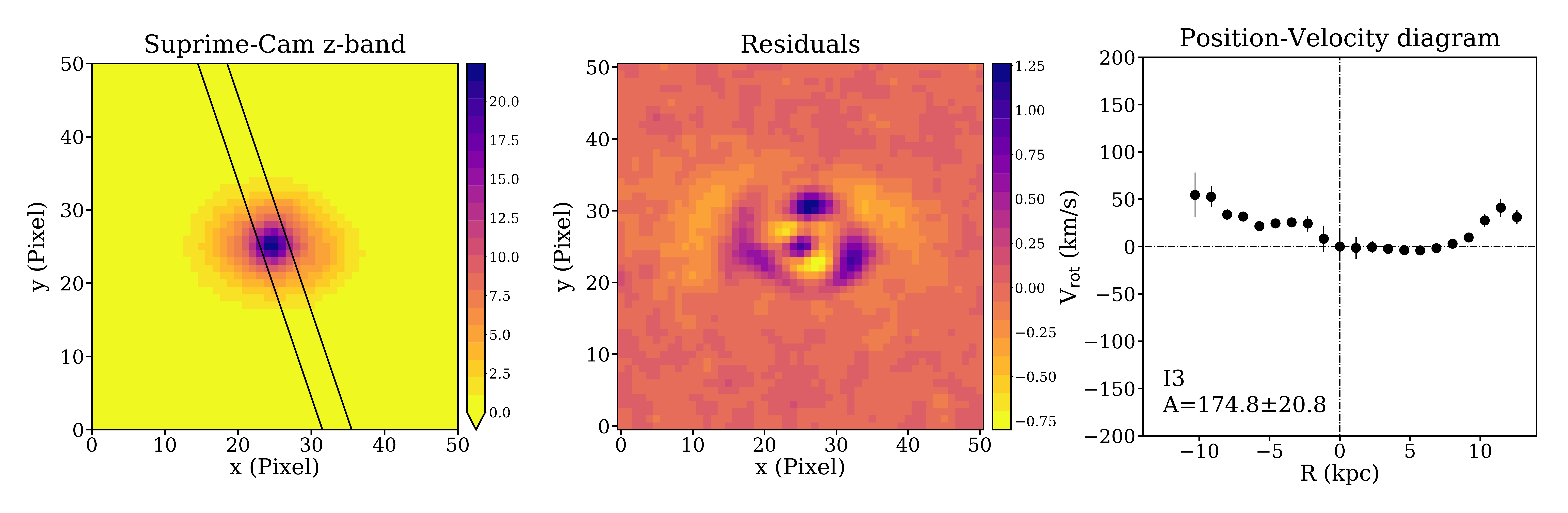}
\caption{Three examples of irregular cluster galaxies ($A\geq50$) according to the asymmetry classification stated in Sect. \ref{SS:Asymmetry}. The columns have the same meaning as in Figs. \ref{foot} and \ref{foot2}.}

\label{foot3}
\end{figure*}

\end{appendix}
\end{document}